\documentclass[times,twocolumn,preprint, longtitle]{elsarticle}
\usepackage{medima}
\usepackage{framed,multirow}
\usepackage{amssymb}
\usepackage{latexsym}
\usepackage{url}
\usepackage{xcolor}
\usepackage{hyperref}
\usepackage{graphicx}
\usepackage{cite}
\usepackage{amsmath,amssymb,amsfonts}
\usepackage{algorithmic}
\usepackage{textcomp}
\usepackage{booktabs}
\usepackage{xspace}
\usepackage{svg}
\usepackage{soul}
\usepackage{amsmath}
\usepackage{footnote}
\usepackage{makecell}
\usepackage{multirow}
\usepackage{enumitem}
\usepackage{array}
\usepackage[nolist,nohyperlinks]{acronym}
\usepackage{subcaption}
\makesavenoteenv{tabular}
\usepackage{caption, subcaption}
\usepackage[figuresleft]{rotating}
\usepackage{multirow}
\usepackage{colortbl}
\usepackage[acronym]{glossaries}
\makeglossaries
\acrodef{GI}{Gastrointestinal}
\acrodef{CRC}{Colorectal Cancer}
\acrodef{CADx}{Computer-Aided Diagnosis}
\acrodef{CNN}{Convolutional Neural Network}
\acrodef{DNN}{Deep Neural Network}
\acrodef{DL}{Deep Learning}
\acrodef{ML}{Machine Learning}
\acrodef{RNN}{Residual Neural Network}
\acrodef{GAN}{Genearative Adversarial Network}
\acrodef{Medico}{Multimedia  for  Medicine  Task}
\acrodef{AI}{Artificial Intelligence}
\acrodef{FPS}{Frame Per Second}
\acrodef{CGAN}{Conditional Generative Adversarial Network}
\acrodef{CADe}{Computer Aided Detection}
\acrodef{BLI}{Blue laser imaging}
\acrodef{COD}{Camouflaged Object Detection}
\acrodef{GRU}{Gated Recurrent Neural Network}
\acrodef{FICE}{Flexible spectral Imaging Color Enhancement}
\acrodef{DSC}{Dice coefficient}
\acrodef{FPS}{Frames per second}
\acrodef{mIoU}{mean Intersection over union}
\acrodef{SOTA}{state-of-the-art}
\definecolor{newcolor}{rgb}{.8,.349,.1}
\usepackage{hyperref}

\journal{Medical Image Analysis}

\begin{document}

\begin{frontmatter}


\title{Validating polyp and instrument segmentation methods in colonoscopy through Medico 2020 and MedAI 2021 Challenges}

\author[1]{Debesh Jha}
\cortext[cor1]{Corresponding author}
\ead{debesh.jha@northwestern.edu}
\author[4]{Vanshali Sharma}
\author[5]{Debapriya Banik}
\author[7]{Debayan Bhattacharya}
\author[5]{Kaushiki Roy}
\author[2]{Steven A. Hicks}
\author[1]{Nikhil Kumar Tomar}
\author[2]{Vajira Thambawita}
\author[11]{Adrian Krenzer}
\author[17]{Ge-Peng Ji}
\author[15]{Sahadev Poudel}
\author[19]{George Batchkala}
\author[26]{Saruar Alam}
\author[9]{Awadelrahman M. A. Ahmed}
\author[21]{Quoc-Huy Trinh}
\author[20]{Zeshan Khan}
\author[12]{Tien-Phat Nguyen}
\author[23]{Shruti Shrestha}
\author[24]{Sabari Nathan}
\author[25]{Jeonghwan Gwak}
\author[1]{Ritika K. Jha}
\author[1]{Zheyuan Zhang}
\author[7]{Alexander Schlaefer}
\author[5]{Debotosh Bhattacharjee}
\author[4]{M.K. Bhuyan}
\author[4]{Pradip K. Das}
\author[29]{Deng-Ping Fan}
\author[21]{Sravanthi Parasa}
\author[16]{Sharib Ali}
\author[2,3]{Michael A. Riegler}
\author[2,3]{P{\aa}l Halvorsen}
\author[27,28]{Thomas de Lange}
\author[1]{Ulas Bagci}

\address[1]{Machine \& Hybrid Intelligence Lab, Department of Radiology, Northwestern University, Chicago, USA}
\address[2]{SimulaMet, Oslo, Norway}
\address[3]{Oslo Metropolitan University, Oslo, Norway}
\address[5]{Jadavpur University, Kolkata, India}
\address[4]{Indian Institute of Technology, Guwahati, India}
\address[7]{Institute of Medical Technology and Intelligent Systems, Technische Universität Hamburg, Germany}
\address[8]{Jaeyong Kang, Information Systems Technology and Design, Singapore University of Technology and Design, Singapore}
\address[9]{University of Oslo, Norway}
\address[11]{Julius-Maximilian University of Würzburg, Germany}
\address[12]{Faculty of Information Technology, University of Science, VNU-HCM, Vietnam}
\address[13]{Vietnam National University, Ho Chi Minh City, Vietnam}
\address[14]{Department of Colorectal Surgery, the Second Affiliated Hospital of Zhejiang University School of Medicine, Zhejiang, China} 
\address[15]{Department of IT Convergence Engineering, Gachon University, Seongnam 13120, South Korea}
\address[16]{School of Computing, University of Leeds, LS2 9JT, Leeds, United Kingdom}
\address[17]{College of Engineering, Australian National University, Canberra, Australia}
\address[18]{School of Computer Science, Wuhan University, Hubei, China}
\address[19]{Department of Engineering Science, University of Oxford, Oxford, UK}
\address[20]{National University of Computer and Emerging Sciences, Karachi Campus, Pakistan} 
\address[21]{Swedish Medical Center, Seattle, USA}
\address[22]{Vietnam National University, Ho Chi Minh City, Vietnam}
\address[23]{NepAL Applied Mathematics and Informatics Institute for Research (NAAMII), Kathmandu, Nepal} 
\address[24]{Couger Inc, Tokyo, Japan}
\address[25]{Department of Software, Korea National University of Transportation, Chungju-si, South Korea}
\address[26]{University of Bergen, Bergen, Norway} 
\address[27]{Department of Medicine and Emergencies - Mölndal Sahlgrenska University Hospital, Region Västra Götaland, Sweden} 
\address[28]{Department of Molecular and Clinical Medicin, Sahlgrenska Academy, University of Gothenburg, Sweden}
\address[29]{Computer Vision Lab (CVL), ETH Zurich, Zurich, Switzerland}

\received{x xxxx xxxx}
\finalform{x xxxx xxxx}
\accepted{x xxxx xxxx}
\availableonline{x xxxx xxxx}
\communicated{xxxx xxxx}

\begin{abstract}
Automatic analysis of colonoscopy images has been an active field of research motivated by the importance of early detection of precancerous polyps. However, detecting polyps during the live examination can be challenging due to various factors such as variation of skills and experience among the endoscopists,  lack of attentiveness, and fatigue leading to a high polyp miss-rate. Therefore, there is a need for an automated system that can flag missed polyps during the examination and improve patient care. Deep learning has emerged as a promising solution to this challenge as it can assist endoscopists in detecting and classifying overlooked polyps and abnormalities in real time, improving the accuracy of diagnosis and enhancing treatment. In addition to the algorithm's accuracy, transparency and interpretability are crucial to explaining the whys and hows of the algorithm's prediction. Further, conclusions based on incorrect decisions may be fatal, especially in medicine. Despite these pitfalls,  most algorithms are developed in private data, closed source, or proprietary software, and methods lack reproducibility. Therefore, to promote the development of efficient and transparent methods, we have organized the \textit{{``Medico automatic polyp segmentation (Medico 2020)''}} and \textit{``MedAI: Transparency in Medical Image Segmentation (MedAI 2021)"} competitions. {The Medico 2020 challenge received submissions from 17 teams, while the MedAI 2021 challenge also gathered submissions from another 17 distinct teams in the following year.} We present a comprehensive summary and analyze each contribution, highlight the strength of the best-performing methods, and discuss the possibility of clinical translations of such methods into the clinic. Our analysis revealed that the participants improved dice coefficient metrics from 0.8607 in 2020 to 0.8993 in 2021 despite adding diverse and challenging frames (containing irregular, smaller, sessile, or flat polyps), which are frequently missed during a routine clinical examination. For the instrument segmentation task, the best team obtained a mean Intersection over union metric of 0.9364. For the transparency task, a multi-disciplinary team, including expert gastroenterologists, accessed each submission and evaluated the team based on open-source practices, failure case analysis, ablation studies, usability and understandability of evaluations to gain a deeper understanding of the models’ credibility for clinical deployment. The best team obtained a final transparency score of 21 out of 25. Through the comprehensive analysis of the challenge, we not only highlight the advancements in polyp and surgical instrument segmentation but also encourage {subjective} evaluation for building more transparent and understandable AI-based colonoscopy systems. Moreover, we discuss the need for multi-center and out-of-distribution testing to address the current limitations of the methods to reduce the cancer burden and improve patient care.  

\end{abstract}

\begin{keyword}
\vspace{-15pt}
\KWD Colonoscopy  \sep polyp segmentation  \sep Transparency \sep polyp challenge \sep computer-aided diagnosis  \sep medicine
\end{keyword}
\end{frontmatter}

\section{Introduction}
\label{section:introduction}
{Gastrointestinal (GI)} cancer is a very important global health problem and the second most common cause of mortality in the United States. According to the recent 2023 estimates, there will be approximately  1,958,310 new cancer incidences and 609,820 cancer deaths in the United States~\citep{siegel2023cancer}. Among various types of cancer, the highest number of deaths occur from lung, prostate, and colorectum in men and lung, breast, and colorectum cancer in women. As colorectal cancer is prevalent among both men and women, it is the second leading cause of cancer related death overall. One of the key indicators of colon cancer is the development of polyps in the colon and rectum. The 5-year survival rate for colon cancer is 68\%, and 44\% for stomach cancer~\citep{asplund2018survival}. If colorectal polyps are detected and removed early, the survival is close to 100.~\citep{levin2008screening}. Thus, regular screening is crucial for early detection of these polyps, as it allows for earlier diagnosis and prompt treatment. 

Endoscopic procedures, such as colonoscopy, are considered the gold standard for detecting and treating mucosal abnormalities in the GI tract (such as polyps) and cancer~\citep{moriyama2015advanced}. However, manual screening for polyps is susceptible to error and is also time-consuming. {This has driven the development of} \ac{CADe} {and} \ac{CADx} {systems that can be integrated into the clinical workflow}~\citep{riegler2016eir} {and potentially contribute to the prevention of colorectal cancer.} In the past, traditional machine learning-based \ac{CADx} systems~\citep{inproceedings2,inproceedings3} were popular.  With the recent advancement in the hardware capabilities, such as powerful GPUs and the emergence of deep learning~\citep{article}, the research has shifted towards deep learning-based \ac{CADx} systems~\citep{fan2020pranet,jha2019resunet++}. These algorithms have shown superior performance compared to traditional \ac{CADx} solutions.

\begin{figure*} [!t]
    \includegraphics[width=\linewidth, height=320pt]{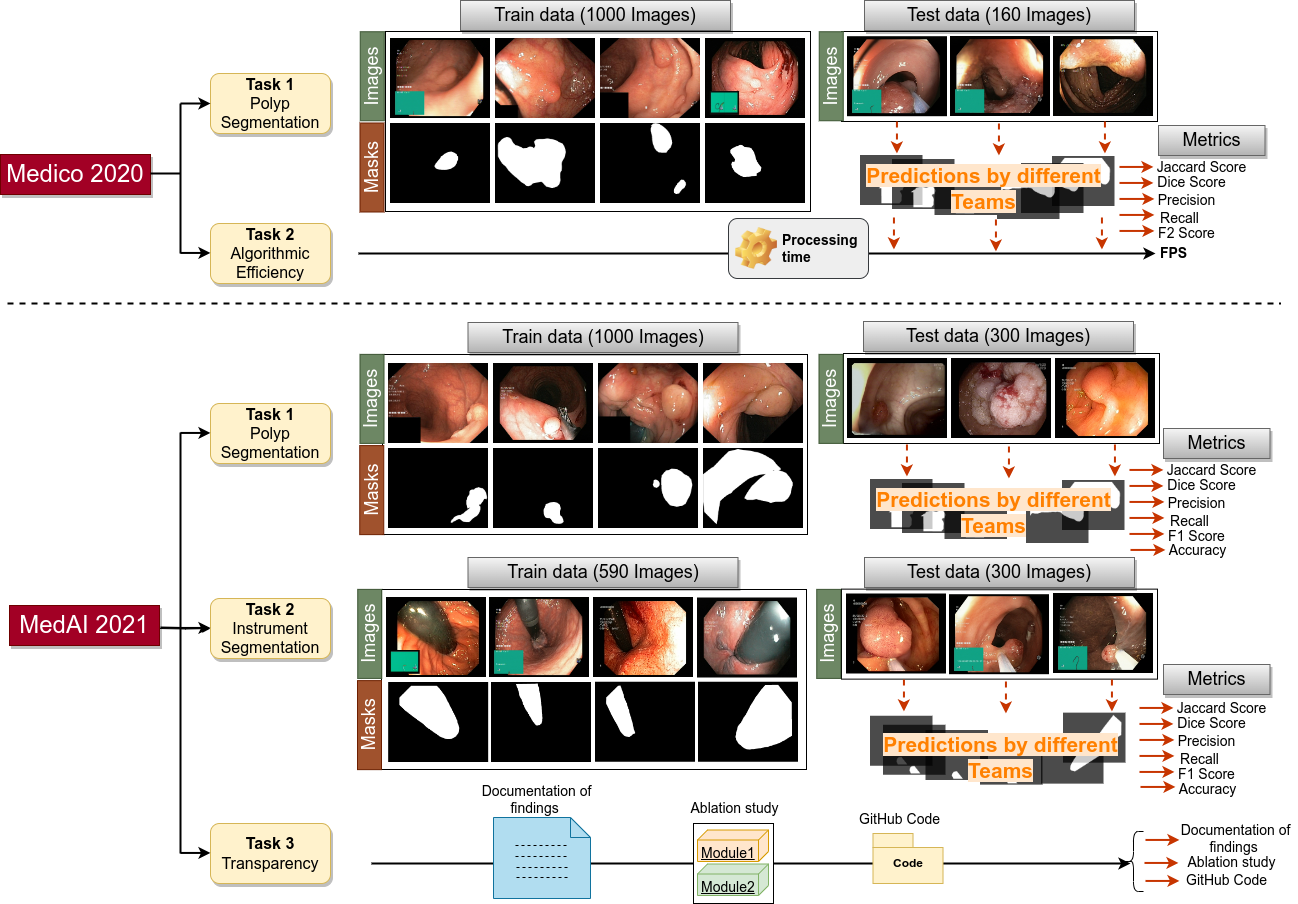}
    \caption{The overview of the ``\textbf{\textit{Medico 2020 Polyp}}"  and ``\textbf{\textit{MedAI 2021 Transparency }}" challenges. We describe each task along with the number of training and testing datasets and the evaluation metrics used in the tasks.}
    \label{GIChallengeoverviewfigure}
\end{figure*}

However, despite their superior performance, deep learning-based \ac{CADx} systems are still considered a ``black box'', meaning their inner workings are not fully understood or there is a lack of transparency in understanding the predictions made by the model. Because of the complexity of multiple layers and interconnected nodes in the convolutional neural network, it is challenging to interpret the decision or understand the features contributing to the outcome. For these systems to be widely adopted in clinical settings, they must be rigorously evaluated on benchmark datasets. They must demonstrate the ability to handle patient and recording device variability, provide explainability and robustness and process data in real-time. Only by carefully evaluating these systems, we can ensure the reliability and effectiveness of detecting and diagnosing cancer and its precursors (such as polyps) in a clinical setting.

In this paper, we present a comprehensive analysis of the results of the two prominent challenges in the field of automatic polyp segmentation, namely, \textit{{``Medico automatic polyp segmentation (Medico 2020)~\footnote{\url{https://multimediaeval.github.io/editions/2020/tasks/medico/}}''}} challenge and the \textit{``MedAI: Transparency in Medical Image Segmentation (MedAI 2021)''~\footnote{\url{https://github.com/Nordic-Machine-Intelligence/MedAI-Transparency-in-Medical-Image-Segmentation}}} challenge. {These challenges aimed to explore the potential of \ac{CADx} solutions on the same shared datasets, focusing on developing novel \ac{SOTA} methods in terms of high-performance metrics, efficiency,  transparency, and explainability, aiming to evaluate the relevance of such algorithms in clinical workflows.} The challenges  posed four distinct tasks:


\begin{itemize}
    \item \textbf{Accurate polyp segmentation task} to develop novel algorithms to enhance the early detection and treatment of colon cancer (Medico 2020, MedAI 2021).
    \item \textbf{Algorithm efficiency task} to develop efficient methods with the {highest} frames-per-second (FPS) on predetermined hardware (Medico 2020).
    \item \textbf{Surgical instruments segmentation task} to enable tracking and localization of essential tools in endoscopy and help to improve targeted biopsies and surgeries in complex GI tract organs (MedAI 2021).
    \item \textbf{Transparency task} to {evaluate different models from a transparency perspective, focusing} on explanations of the training procedure, {failure analysis, and model's predictions interpretation by interdisciplinary team}) (MedAI 2021).
\end{itemize}

These tasks were focused on the development of \ac{SOTA} algorithms for polyp and instrument segmentation in a variety of settings, including performance evaluation, resource utilization (efficiency), and transparency. By analyzing the results of these challenges, we can better understand the field's current state, identify the strength and weaknesses of different methods and find the most effective method for our problem.  It is also useful to identify the research gap and areas for future innovation in the field of polyp, instrument and medical image segmentation. Figure~\ref{GIChallengeoverviewfigure} provides an overview of both challenges along with the total number of images used for training and testing in each task. Ground truth samples with their corresponding original images are also presented for the segmentation tasks. In addition, task-specific metrics are presented (for example, FPS for ``Algorithm efficiency''). 

In short, the main contributions are the following: (i) We present a comprehensive and detailed analysis of all participant results; (ii) we provide an overview and comparative analysis of the developed methods; (iii) we obtain and discuss new insights into the current state of AI in the field of GI endoscopy including open challenges and future directions; and (iv) finally, we provide a detailed discussion of issues such as {trust, safety, interpretability, transparency,} generalizability issues and multi-center in context to current limitations of CADx systems.


\section{Challenge description}
\subsection{{Medico 2020 Automatic Polyp Segmentation Challenge}}
The ``Medico Automatic Polyp Segmentation Challenge" was an international benchmarking challenge hosted through the MediaEval platform {(Multimedia Evaluation Workshop). Medico 2020 is the fourth iteration of the Medico Multimedia Tasks series, following the pattern established in previous years.} This challenge aimed to benchmark automated polyp segmentation algorithms using the same dataset and develop methods to detect difficult-to-detect polyps (such as flat, sessile, and small or diminutive polyps). Researchers from medical image analysis, machine learning, multimedia, and computer vision were invited to submit their results for this challenge, which included two tasks. {The members from the organizer's institute were allowed to participate but were ineligible for receiving any recognition certificates. Participants could use any method, focusing on creating automated solutions. Below, we provide the task description of each sub-task}. 

a) \textbf{Automatic Polyp Segmentation Task}: In this task, the participants were asked to develop innovative algorithms for segmenting polyps in colonoscopic images. The focus was on developing efficient systems that could accurately segment the maximum polyp area in a frame while being fast enough for practical use in a clinical setting. This task addresses the need for robust \ac{CADx} solutions for colonoscopy.

To participate in the challenge, participants were required to train their segmentation models on an available training dataset. Once the test dataset was released, participants could test their models and submit their predicted segmentation maps to the organizers in a zip file with the name of each segmentation map image matching the colonoscopy image in the test dataset. 

{b) \textbf{Algorithmic Efficiency Task}}: CADx systems for polyp segmentation that operate in real-time can provide valuable feedback to clinicians during colonoscopy examinations, potentially reducing the risk of missing polyps and incomplete removal. However, real-time deep learning-based CADx solutions often have fewer parameters and may therefore have lower segmentation accuracy compared to more computationally intensive CADx solutions. In order to address this trade-off between accuracy and speed, the efficiency task of the challenge was designed to encourage the development of lightweight segmentation models that are both accurate and fast.

To participate in this task, participants were asked to submit docker images of their proposed algorithms. These algorithms were then evaluated on a dedicated Nvidia GeForce GTX 1080 graphics card, and the results were used to rank the teams. A \acf{mIoU} threshold was set for considering a solution to be a valid efficient segmentation solution, and teams were ranked according to their \ac{FPS}. By focusing on developing efficient CAD solutions, this task aimed to foster the creation of real-time systems that can provide valuable feedback to clinicians while maintaining high accuracy. A detailed description of the challenge, tasks, and evaluation metrics can be found in~\citep{jha2020medico}. 

\subsection{{MedAI: Transparency in Medical Image Segmentation Challenge}}

MedAI: Transparency in Medical Image Segmentation challenge (MedAI 2021) was held for the first time at the Nordic AI Meet~\footnote{\url{https://nordicaimeet.com}} 2021 {(Nordic young researchers symposium)} that focused on medical image segmentation and transparency in \ac{ML} based \ac{CADx} systems. This challenge proposed three tasks to address specific endoscopic GI image segmentation challenges, including two separate segmentation scenarios and one scenario on transparent ML systems. The latter task emphasized the need for explainable and interpretable \ac{ML} algorithms in the field of medical image analysis. {Similar to the other challenge, participants were granted the flexibility to use any method, focusing on developing automated solutions. The members from the organizer's institute were permitted to participate but were not considered for awards.}

{To participate in this challenge, participants were given a training dataset to use for their {ML} models}. These models were then tested on a concealed test dataset, allowing participants to evaluate their performance. The focus on transparency underscores the importance of developing \ac{ML} algorithms that provide not only accurate and efficient results but also provide interpretable and explainable predictions. By addressing these specific challenges, this challenge aimed to foster the development of innovative and effective \ac{CADx} solutions for GI endoscopy. {Below, we present each challenge sub-task}. 

\textbf{a) Automatic Polyp Segmentation Task}: In this task, participants were invited to submit segmentation masks of polyps from colonoscopic images of the large bowel. They were provided with a training dataset to develop their models, and a hidden test dataset was later released to them without the ground truth segmentation masks. Participants were required to submit a zip file containing their predicted masks in the same resolution as the input images, with the filenames of each mask matching the corresponding input image and using the ``.png" file format. The objective of this task was similar to Medico 2020. By using a hidden test dataset, the results of this task were reliable and provided a valuable benchmark for the field.

\begin{table*}[t!h!]
\centering
\caption{Overview of {\textbf{GI image analysis challenges with a specific focus on polyp detection, segmentation, localization, and WCE lesion detection and segmentation between 2015 and 2021.} Here, WL = White Light Endoscopy, NBI = Narrow Band Imaging \& WCE = Wireless Capsule Endoscopy. The total number of images and videos offered at different tasks are summed and presented in the 'Size' class.}}
\label{tab:challengeoverview}
\resizebox{0.99\linewidth}{!}{
\begin{tabular}{p{10.7cm}|p{1.2cm}|p{1.5cm}|p{4.5cm}|p{5cm}|p{2.5cm}}
\hline
\bf{Challenge Name} & \bf{Organ} & \bf{Modality}  & \bf{Findings} & \bf{Size} & \bf{Dataset} \newline \bf{Availability} \\ 
\hline

Automatic Polyp Detection in Colonoscopy videos 2015~\citep{bernal2017comparative} & Colon & WL & Polyps 
&808 images \& 38 videos & By request \\ \hline


GIANA 2017~\citep{gianadataset2017} & Colon & WL & Polyps \& angiodysplasia & $3,462$ images\newline \& $38$ videos & By request \\ \hline 

GIANA 2018~\citep{angermann2017towards,bernal2018polyp} & Colon & WL, WCE  & Polyps \& small bowel lesions & $8,262$ images \newline \& $38$ videos  & By request \\ \hline 






EndoCV 2021~\citep{ali2022assessing,ali2021polypgen} &Colon & NBI, WL &  Polyps & $3,446$ images & Open academic\\   \hline


\textbf{{Medico 2020~\citep{jha2020medico}} (Ours)} & {Colon} & {WL} & {Polyps} & 160 images (test) \& 1000 images (train)  & {Open academic}\\ \hline

\textbf{MedAI Transparency challenge 2021~\citep{medaireview} (Ours)}  & Colon, bladder &WL &Polyps, Instrument, Normal frames &  600 images (test) \& (1000 +590) images (train)  & Open academic \\\hline

\hline
\end{tabular}}
\end{table*}


\textbf{b) Automatic Instrument Segmentation Task}: The instrument segmentation task required the development of algorithms that could generate segmentation masks for GI accessory instruments such as biopsy forceps or polyp snares used during live endoscopy procedures. This task aimed to create segmentation models that enable tracking and localization of essential tools in endoscopy that could aid endoscopists during interventions (such as polypectomies) by providing a precise and dense map of the instrument. Like the polyp segmentation task, participants were given a training dataset to develop their models. The submission procedure for this task was similar to that of the polyp segmentation task, with participants required to submit zip files containing their predicted masks in the same resolution as the input images and with filenames matching the corresponding input images. 

\begin{figure*} [!t]
     \centering
         \subfloat[{Examples samples from the test data of Medico 2020 (first three columns) and MedAI 2021 (last three columns) for the polyp segmentation task.}\label{fig:polyp20_21}]{%
      \includegraphics[width=0.8\linewidth]{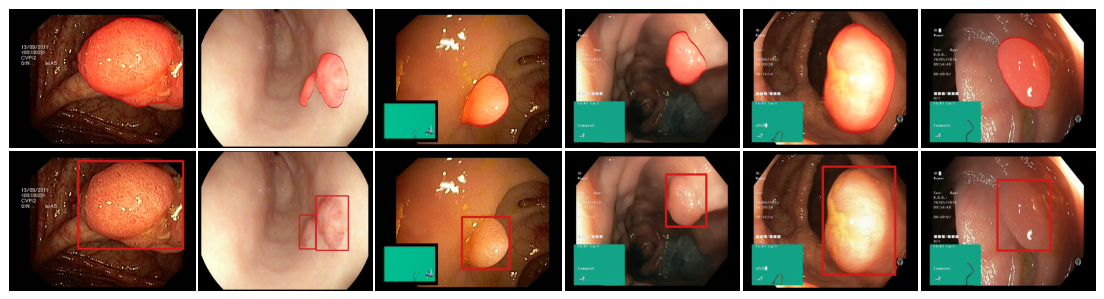}
        }
         \hfill
     \subfloat[Example samples from the MedAI 2021 Instrument segmentation task. \label{fig:inst21}]{%
       \includegraphics[width=0.8\linewidth]{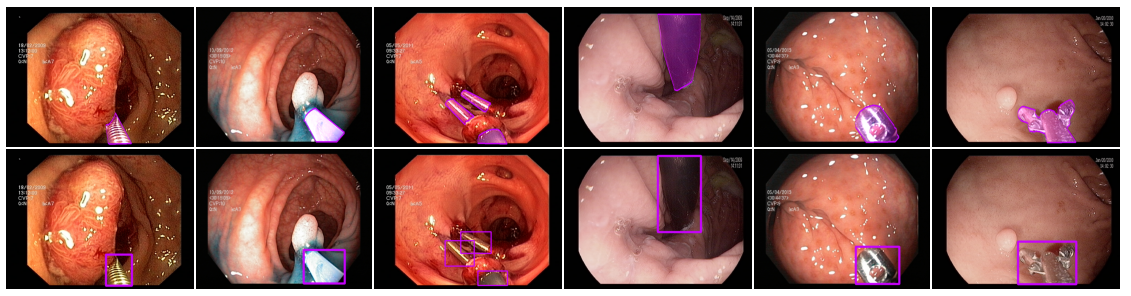}
        }
   \caption{Example of the test datasets from the Medico 2020 and MedAI 2021 datasets.}
\end{figure*}

\textbf{c) Transparency Task}:
The transparency task focused on the importance of transparent research in medical artificial intelligence (AI). The main goal of this task was to evaluate systems from a transparency perspective, which included detailing the training procedure of the algorithms, the dataset used for training, the interpretation of the model's predictions, the use of explainable AI methods, etc. To participate in this task, researchers were encouraged to perform ablation studies, conduct a thorough failure analysis, and share their code in a GitHub repository with clear steps for reproducing the results. {We allowed the participants to submit, considering the transparency and left them to decide what to deliver for the task.}

In addition, participants were required to submit a one-page document summarizing their findings from the transparency task. {We encouraged the participants to list package dependencies and architecture code (with instruction for building, compiling, and training) and share trained model weights in a standardized format. Additionally, we encouraged participants to include the code for model evaluation and provide repository licensing information to enable others to use the code and the trained model responsibly. Moreover, we suggested that the participants explain model predictions using intermediate heatmaps, statistical analysis and alternatives, such as SHapley Additive exPlanations~\citep{lundberg2017unified}.} By promoting transparency in AI research, this task aimed to foster the development of reliable, interpretable, and trustworthy algorithms for use in medical image segmentation. A detailed description of the challenge can also be found in~\citep{medaireview}.

\section{Related Work}
Polyp detection and segmentation using \ac{ML} has been an active field of research for over a decade but have been previously limited by hand-crafted features ~\citep{bernal2012towards, hwang2007polyp}. Previous methods had limitations in sub-optimal performance, poor generalization to unseen images, and complexity that limited real-world applicability. However, in the recent 5-6 years, with the success of Convolutional Neural Networks (CNNs), the polyp segmentation task has seen a tremendous performance boost, including the winning model in the MICCAI challenge~\citep{bernal2017comparative}. The widespread use of CNNs, particularly the U-Net ~\citep{ronneberger2015u} and its variants, have been successfully applied on several polyp segmentation datasets and discussed in challenge reports. In addition, recent advances in CNN architectures for polyp segmentation have focused on improving convolution operations~\citep{alam2020automatic}, adding attention blocks~\citep{jha2019resunet++, oktay2018attention}, incorporating feature aggregation blocks(~\citep{mahmud2021polypsegnet}) and using self-supervised learning techniques~\citep{bhattacharya2021self}. These modifications and learning strategies have proven effective in improving the accuracy and reliability of polyp segmentation using CNNs. Apart from the contributions of individual research groups, several challenges~\citep{bernal2017comparative,ali2021deep} have been organized to improve the detection and classification of mucosal abnormalities in the GI tract from either single image frames or videos. However, the dataset provided in the challenge and the details of the proposed algorithms are often not publicly available, making it difficult to reproduce and build upon them. Hence, there is a need for open-access benchmarking datasets and reproducible algorithms to facilitate progress in this field. 

Table~\ref{tab:challengeoverview} provides an overview of {GI image analysis challenges} held in the past eight years. The challenge was conducted using images from different modalities with a specific focus on polyp segmentation, detection,  localization and wireless capsule endoscopy lesion detection and localization. In 2015, Bernal et al.~\citep{bernal2017comparative} organized the ``Automatic Polyp Detection in colonoscopy videos'' challenge. Likewise, they organized the GIANA challenge in 2017 and 2018\footnote{\url{https://giana.grand-challenge.org/}} focused on colonoscopy data and included tasks such as detection of lesions in Video capsule Endoscopy (VCE), polyp detection, and polyp segmentation. Recent challenges attempted to address generalizability in polyp detection and segmentation~\citep{ali2022assessing} with single frames and sequence colonoscopy datasets. They demonstrated how variability in images can affect algorithm performances.  Altogether, these challenges have led to many algorithmic innovations in detecting and classifying GI abnormalities {(especially polyp segmentation and detection)}.


Additionally, past challenges have not emphasized on the explainability and reliability of deep learning model predictions. Most challenges also do not focus on open source codes for research and development, making it difficult for proposed algorithms to be adopted in clinical settings due to a lack of transparency. Moreover, the reported methods are not reproducible, which hinders further algorithmic advancement. Thus, we lose track of what are best practices and where we are heading in this field. Through our challenges in Medico 2020 and MedAI 2021, we address reproducibility and open science which are the two most important aspects that can enable experienced and new ML scientists to build upon and advance the field.

\section{Challenge datasets and evaluation metrics}

\begin{figure*}[t!]
     \centering
     \begin{subfigure}[b]{0.4\textwidth}
         \centering
         \includegraphics[width=\textwidth]{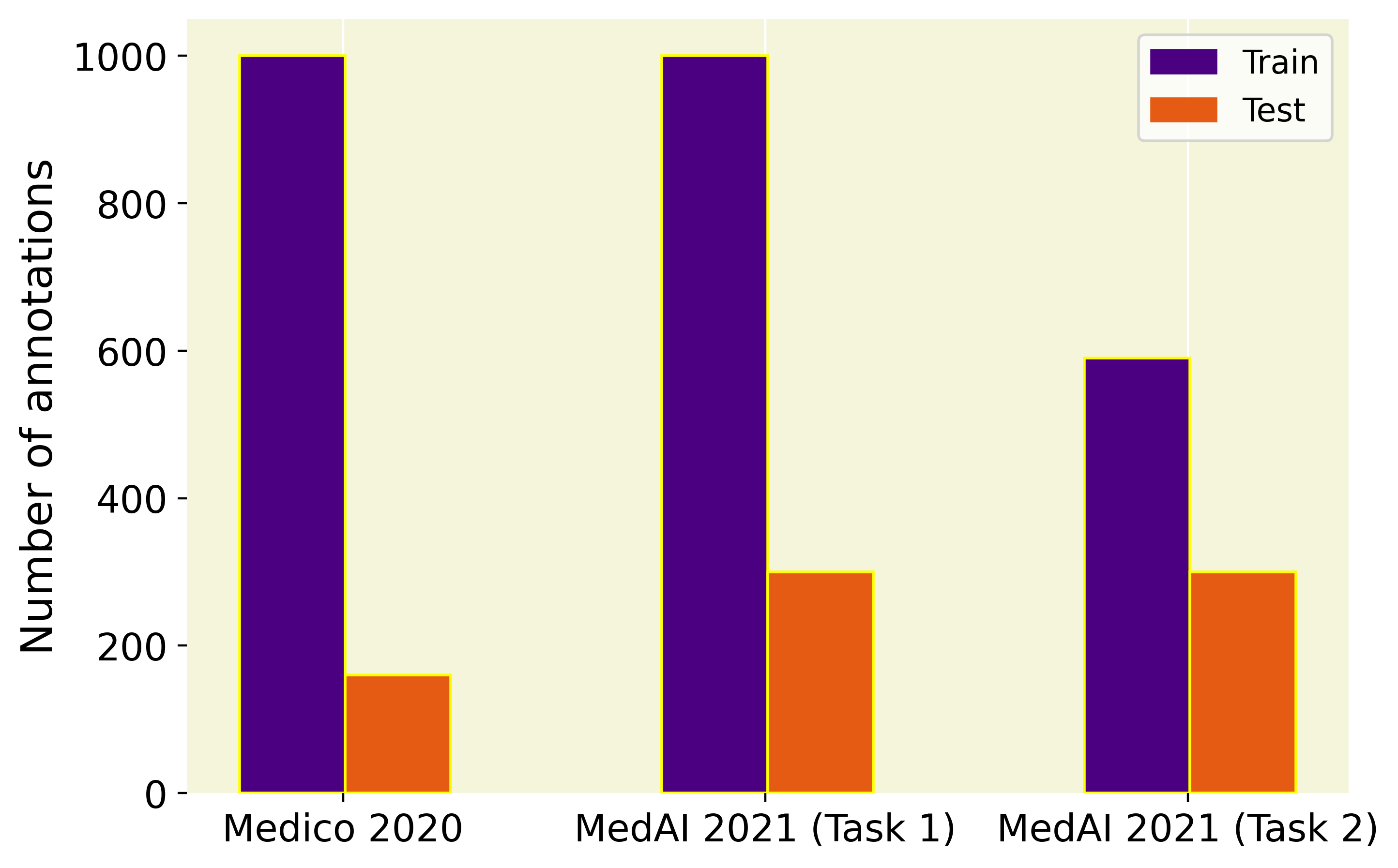}
         \caption{Train and test sample proportion for Medico 2020 and MedAI 2021}
         \label{fig:MedAI2021_inst_train1}
     \end{subfigure}
     \begin{subfigure}[b]{0.4\textwidth}
         \centering
         \includegraphics[width=\textwidth]{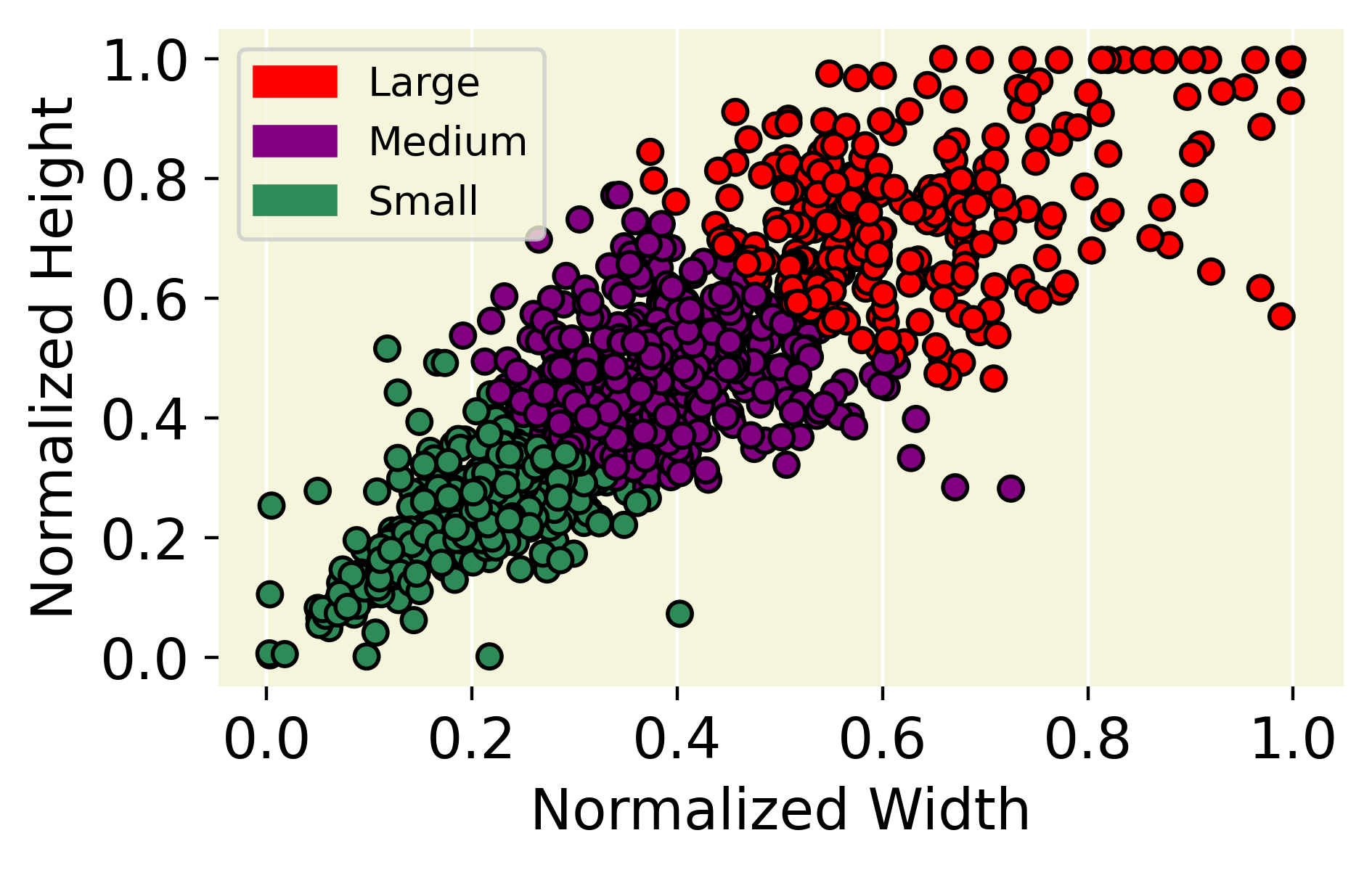}
         \caption{Train data (Medico 2020 and MedAI 2021 Task 1)}
         \label{fig:MedAI2021_inst_train2}
     \end{subfigure}
     \begin{subfigure}[b]{0.4\textwidth}
         \centering
         \includegraphics[width=\textwidth]{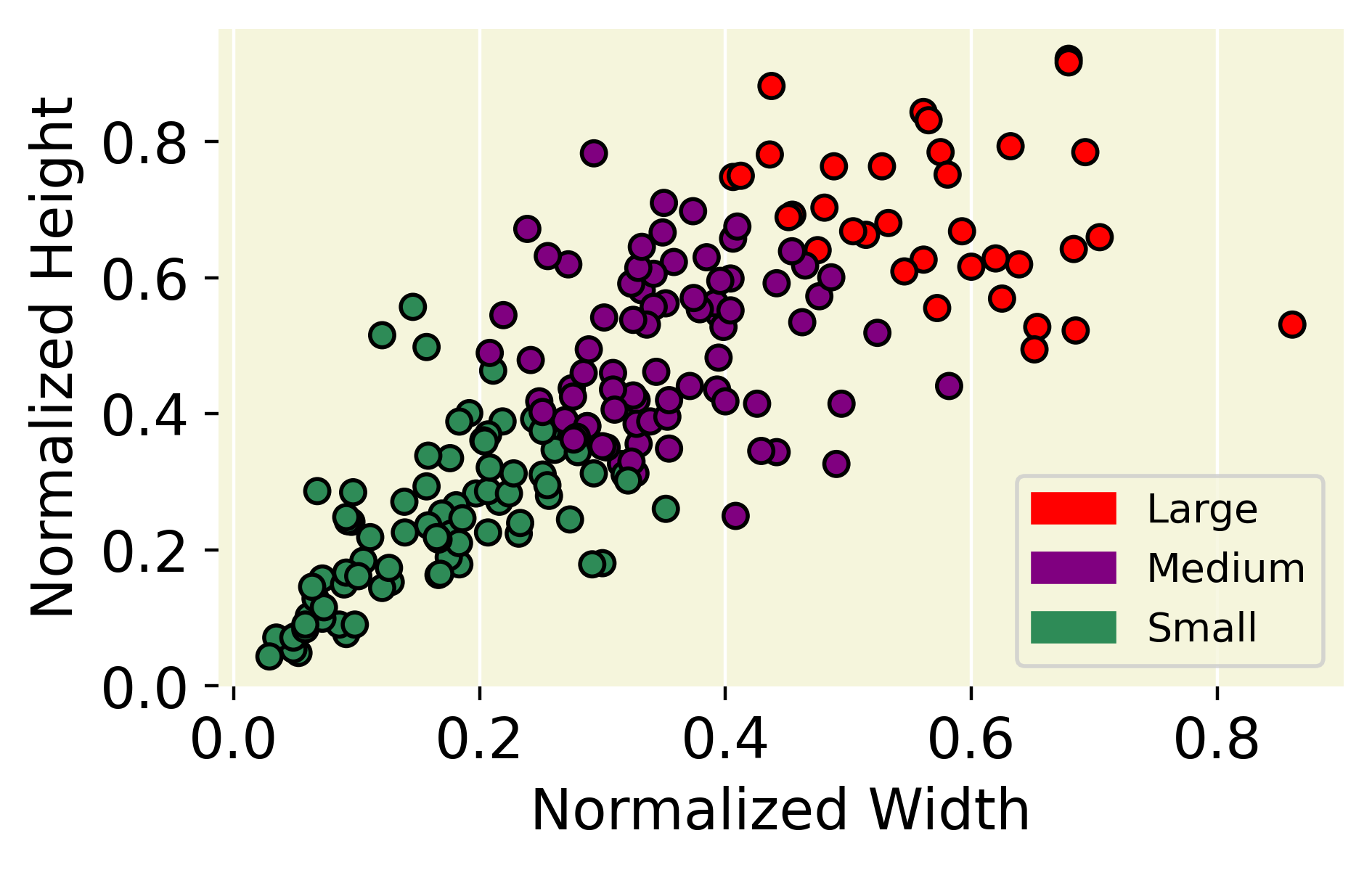}
         \caption{Test data (Medico 2020)}
         \label{fig:MedAI2021_inst_train3}
     \end{subfigure}
     \begin{subfigure}[b]{0.43\textwidth}
         \centering
         \includegraphics[width=\textwidth]{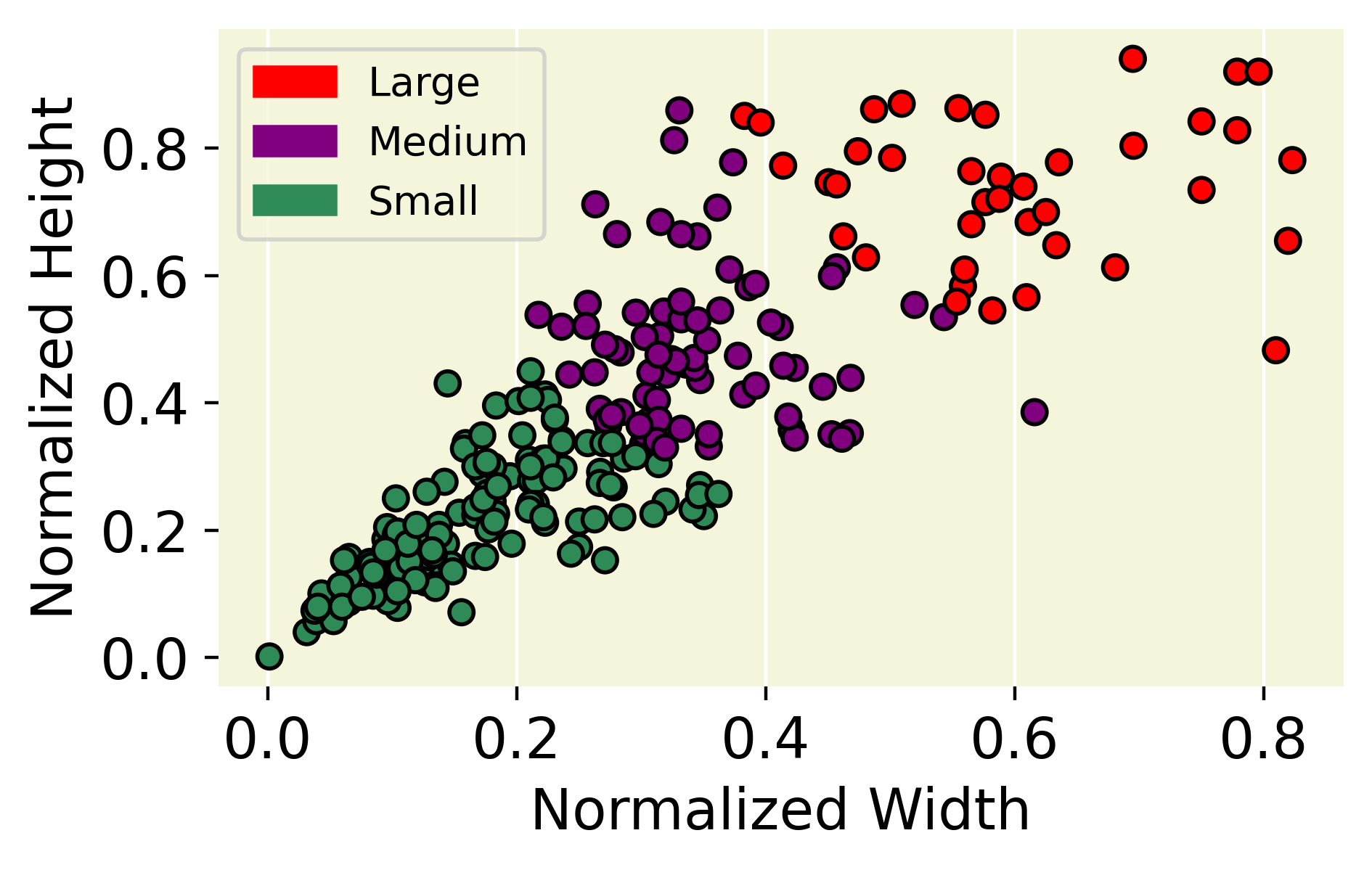}
         \caption{Test data (MedAI 2021 Task 1)}
         \label{fig:MedAI2021_inst_train4}
     \end{subfigure}
     \begin{subfigure}[b]{0.4\textwidth}
         \centering
         \includegraphics[width=\textwidth]{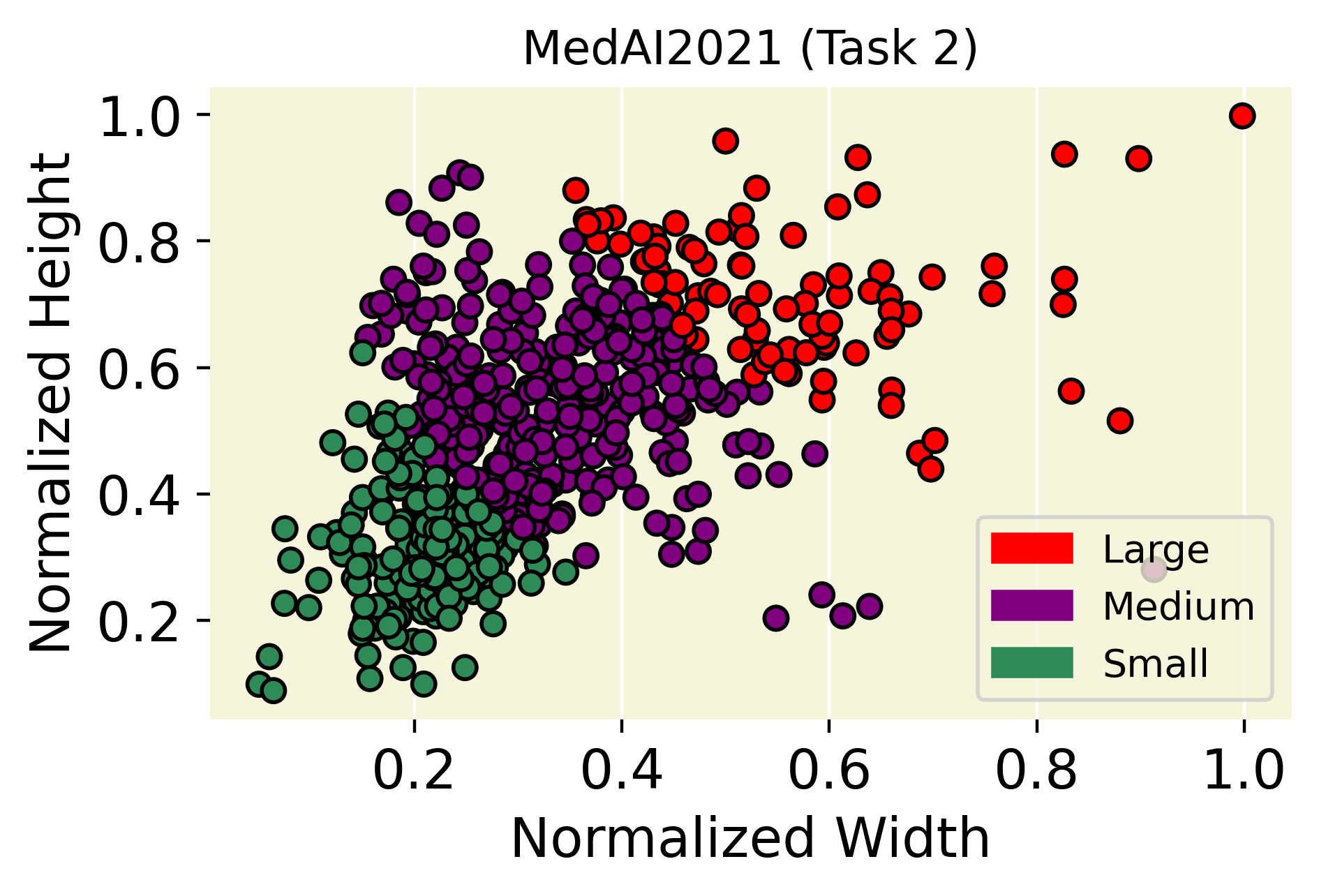}
         \caption{Train data (MedAI 2021 Task 2)}
         \label{fig:MedAI2021_inst_train5}
     \end{subfigure}
     \hspace{0em}
     \begin{subfigure}[b]{0.4\textwidth}
         \centering
         \includegraphics[width=\textwidth]{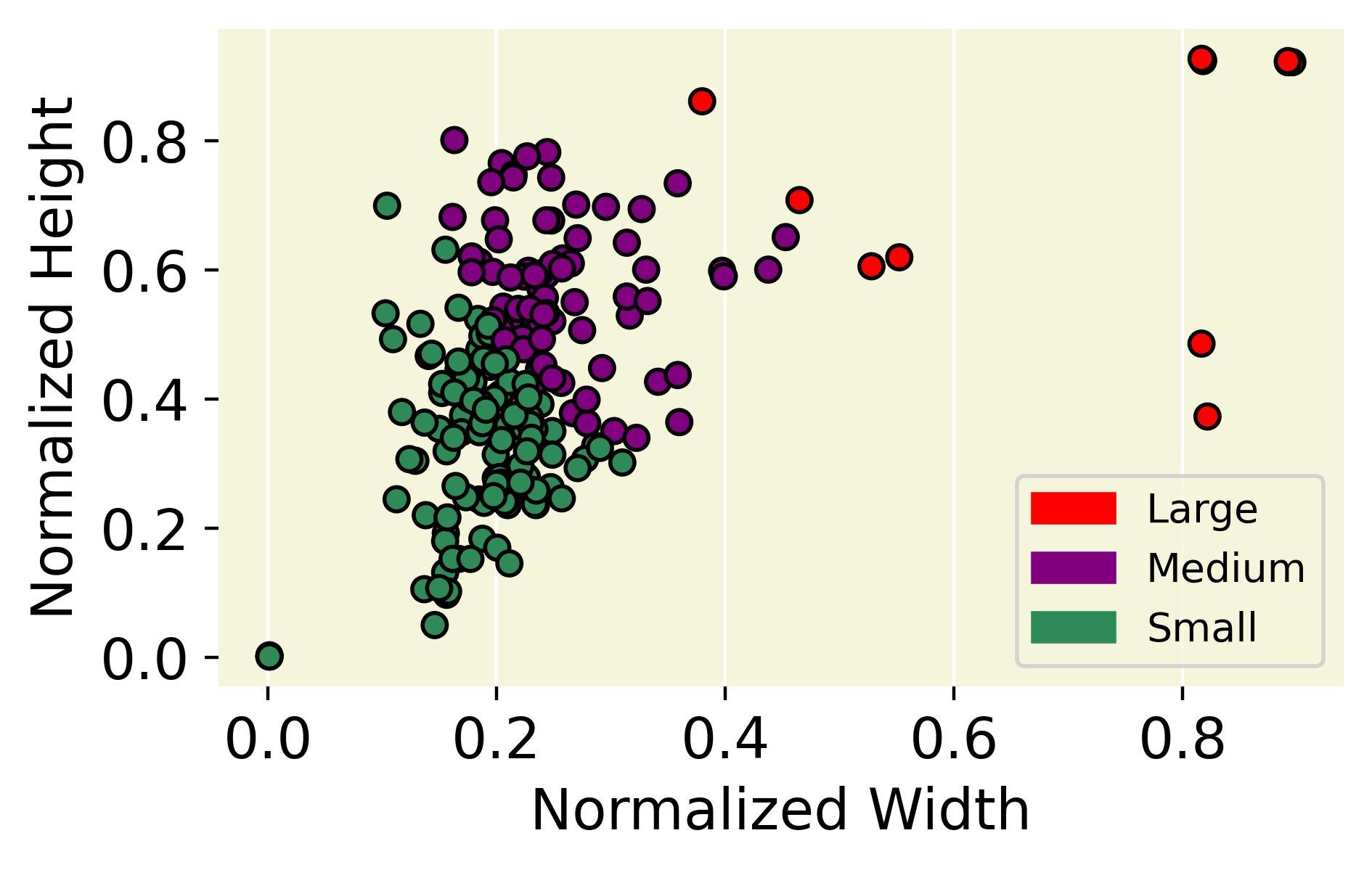}
         \caption{Test data (MedAI 2021 Task 2)}
         \label{fig:MedAI2021_inst_test}
     \end{subfigure}

    \caption{Data distribution details of train and test sets used in Medico 2020 and MedAI 2021 challenges. Large, medium, and small represent the distribution information of regions of interest in the data samples.}
    \label{fig:datadistributuion}
\end{figure*}

\subsection{Medico 2020 dataset}
The training dataset contains 1,000 polyp images and their corresponding ground truth mask taken from Kvasir-SEG~\citep{jha2020kvasir}. {Kvasir-SEG consists of diverse images varying in appearance, such as sizes (for example, diminutive, regular or large), colors (same color as mucosa, or different colors such as reddish), textures (smooth or granular), locations (anywhere in large intestine such as left colon, sigmoid colon or rectum), numbers of polyp per images (for example, one or many), image quality (illumination, artifacts) and shapes (flat, pedunculated, and sessile). The variation ensures that the algorithms trained on this dataset can handle real-world variations in clinical settings. Some samples are shown in Figure~\ref{fig:polyp20_21}.}

{The datasets were acquired from real routine clinical examinations at Vestre Viken Health Trust (VV) in Norway by a team of expert gastroenterologists. The VV is the collaboration of the four hospitals that provide healthcare services to 470,000 people. The resolution of images varies from $332\times487$ to $1920\times1072$ pixels. Some images contain green thumbnails in the lower-left corner of the images showing the position marking from the ScopeGuide (Olympus). After data acquisition, our team categorized the dataset into a polyp class. To extend the dataset to the segmentation class, a team of one experienced engineer, a medical doctor, and an expert gastroenterologist annotated the polyp images using the label box tool. After annotation, we extract the corresponding ground truth and bounding box information. Once the ground truth was created, the images and ground truths were combined to facilitate the review process. These images were sent to a team of expert gastroenterologists for validation through a web-based interface. After validation, we compiled them into training and test datasets. The data proportion for each set followed the general split ratio used in the literature.}

The training dataset has been made publicly available as open access and is widely available at\footnote{\url{https://datasets.simula.no/kvasir-seg/}}. {The test dataset contains unique polyp images encompassing a wide range of diverse clinical scenarios with different polyp characteristics, varying lighting conditions and image resolution, low-quality images, as well as complex polyp images (for example, with instruments and residual stool) that the model has never encountered before. {Only the organizers had access to the test case labels. Currently, the test data can be downloaded from}\footnote{\url{https://drive.google.com/file/d/1uP2W2g0iCCS3T6Cf7TPmNdSX4gayOrv2}}}. 



\subsection{MedAI Transparency challenge 2021 dataset}
{We utilize our Kvasir-SEG}~\citep{jha2020kvasir} {as the development dataset for the polyp segmentation task. Similarly, Kvasir-Instrument~\citep{jha2021kvasir} was used as the training dataset for the instrument segmentation task. It can be downloaded from} \footnote{\url{https://datasets.simula.no/kvasir-instrument/}}. {We followed the same data acquisition and annotation protocol for test dataset creation as the Medico 2020 challenge.} Some sample images for polyp segmentation and instrument segmentation tasks are presented in Figure~\ref{fig:polyp20_21} and Figure~\ref{fig:inst21}. Figure~\ref{fig:datadistributuion} shows the data distribution of the train and test datasets used in Medico 2020 and MedAI 2021. {We have categorized the images into ``small", ``medium'' and ``large'' according to the size of regions of interest using a randomly selected threshold of 0.3 and 0.1 and plotted the normalized height versus normalized width of each data point.} This is to visualize the dimension of each data point and observe the diversity and complexity of the dataset used in the study. The information about the size categories and the dataset's dimensions is crucial for assessing the performance and robustness of the proposed algorithms.

\subsection{Metrics for polyp and instrument segmentation tasks}

{We used mean Intersection over Union ({mIoU}) as a primary evaluation metric for the polyp and instrument segmentation tasks. If the teams achieved the same mIoU values, their ranking was further evaluated based on the higher value of the \ac{DSC}.}  We also recommend calculating other important standard evaluation metrics that hold significant relevance in clinical settings such as Accuracy (Acc), Recall (Rec),  Precision (Pre), F-2 score, and Frames per second ({FPS}) to ensure a detailed evaluation.

\subsection{Metrics for efficiency tasks}   
{Efficiency is crucial in colonoscopy as it directly impacts the models' feasibility and practicality in real-world scenarios. Endoscopists often need to analyze numerous frames in real-time during routine colonoscopy, and lag (latency) in the analysis could lead to suboptimal results. Our approach to FPS calculation was based on the time taken to process a single image, averaged over the entire dataset, and then extrapolated to a per-second rate.} Therefore, we strongly recommend calculating processing speed in terms of \ac{FPS}.

\subsection{Metrics for transparency tasks}
{The transparency task aimed to assess the transparency and understandability of algorithms for medical AI by utilizing a qualitative approach in the evaluation metrics. We evaluated transparency tasks using a more quantitative approach than polyp and instrument segmentation. A multi-disciplinary team assessed each submission and evaluated the transparency and understandability of the proposed solutions. Each team was scored based on the three criteria: open source code, model evaluation and clinical evaluation. The open source code was evaluated based on the presence of a publicly available repository, code quality and quality of the readme file. The model evaluation included failure analysis, ablation study, explainability of the method, and metrics used. Evaluation by clinical experts considered the usefulness of the technique and its understandability. With these three criteria, we aimed to measure the transparency of the provided solutions. A detailed score distribution under different criteria is shown in Table~\ref{tab:transscore}, which was part of our Task 3. Ultimately, this task aimed to promote the development of more transparent and interpretable AI systems.}

\vspace{-4mm}

\section{Participating Research Teams}
\vspace{-2mm}
\subsection{Methods used in Medico 2020}
Table~\ref{table:challenge_summary_medico} summarizes all the teams participating in the ``Medico 2020" challenge. {It can be seen from Table}~\ref{table:challenge_summary_medico}   {that all 17 teams participated in Task 1, whereas only 9 teams participated in Task 2.}

\begin{table} [!t]
\caption{Summary information of participating teams in Medico 2020. Here, `$\surd$'  = Team participated, `--' = No participation, \textbf{Task 1 =} Polyp segmentation task and \textbf{Task 2 =} Algorithm efficiency task.}
\label{table:challenge_summary_medico}
\centering
\begin{tabular}{p{0.7cm}|p{3.5cm}|p{1cm}p{1cm}}
\toprule
\textbf{Chal.} & \centering \textbf{Team Name} & \textbf{Task 1} &\textbf{Task 2}\\
\midrule

\multirow{17}{*}{\rotatebox{90}{\textbf{Medico 2020}}}
              & FAST-NU-DS &$\surd$  &  $\surd$   \\ 
              & AI-TCE & $\surd$  & -- \\ 
              & ML-MMIVSARUAR & $\surd$ &--    \\ 
              & UiO-Zero  & $\surd$ &--\\ 
              & HBKU\_UNITN\_SIMULA & $\surd$ &--  \\ 
              & AI-JMU & $\surd$ &$\surd$  \\ 
              & SBS & $\surd$ & $\surd$  \\ 
              & AMI Lab & $\surd$ &$\surd$ \\ 
              & UNITRK & $\surd$& $\surd$  \\ 
              & MedSeg\_JU & $\surd$ &-- \\ 
              & IIAI-Med & $\surd$ &-- \\ 
              & HGV-HCMUS & $\surd$ & $\surd$   \\ 
              & GeorgeBatch & $\surd$& $\surd$ \\ 
              & PRML2020GU & $\surd$ & $\surd$  \\ 
              & VT & $\surd$ &  -- \\ 
              & IRIS-NSYSU & $\surd$ & -- \\ 
              & NKT & $\surd$& $\surd$  \\ 
              \bottomrule
\end{tabular}
\vspace{-6mm}
\end{table}

\begin{sidewaystable*}
    \footnotesize
    \centering
   \caption{Summary of the participating teams algorithm for Medico 2020. {Here,  ``Aug." = augmentation used, ``SGD" = Stochastic gradient descent,  ``GAN" = generative adversarial network, \\  ``ASPP" = Atrous Spatial Pyramid Pooling, and ``AP" = Average precision.}}
    \label{table:challenge_summary2020} 
    \renewcommand{\arraystretch}{1.2}
    \begin{tabular}{|p{3.3cm}|p{2.6cm}|p{1.8cm}|p{2.6cm}|p{2cm}|p{0.5cm}|p{2cm}|p{1cm}|}
        \toprule
        \textbf{Team Name}  & \textbf{Algorithm} & \textbf{Backbone} & \textbf{Nature} & \textbf{Choice basis} & \textbf{Aug.} & \textbf{Loss} & \textbf{Optimizer} \\\midrule 
        
FAST-NU-DS~\citep{ali2020depth} & Depth-wise separable convolution and ASPP & ResUNet++ & Cascade of depth-wise separable convolutions &  mIoU and DSC & Yes & IoU & Adam\\ \hline

AI-TCE~\citep{nathan2020efficient}  & Multi-Supervision Net  & EfficientNetB4 & Encoder-Multi Supervision Decoder & Acc and DSC & Yes & Categorical cross-entropy + Dice loss & Adam \\ \hline

ML-MMIV SARUAR~\citep{alam2020automatic} & Encoder-decoder based architecture based on ResNet50 &  ResNet50 & Cascade of residual blocks & mIoU and DSC & Yes & Cross-entropy& Adam\\ \hline

UiO-Zero~\citep{ahmed2020generative} & GAN & None & GAN with CNN based generator and discriminator & Image-to-image translation & No &Standard conditional GAN adversarial loss &Adam \\ \hline

{HBKU UNITN SIMULA}~\citep{trinh2020hcmus} & Residual module, Inception module, Adaptive CNN with U-Net and PraNet  & U-Net and ResNet-50 & Cascade of residual blocks and inception module  & mIoU and DSC & Yes &  {Bce + Dsc loss}  & Adam\\ \hline

AI-JMU~\citep{krenzer2020bigger}  & Cascade Mask R-CNN & ResNeSt backbone, 
Cascade Architecture & Deep CNN & DSC and mIoU & Yes & Binary cross-entropy & SGD\\ \hline

 SSB~\citep{shrestha2020ensemble} & U-Net	& ResNet-34, EfficientNet-B2 &	Ensemble & DSC and mIoU &	Yes	& Tversky loss	& Adam \\ \hline

AMI LAB~\citep{kang2020kd}	& Knowledge distillation on ResUNet++ &	ResUNet++ &	Ensemble &	mIoU and DSC & Yes & Distillation loss & Adam \\ \hline

UNITRK~\citep{khadka2020transfer} &	Knowledge transfer using UNet & Pre-trained U-Net model & Encoder-decoder & {mIoU and DSC} & Yes	& Compound loss of DSC and BCE &	Adam \\ \hline

MedSeg\_JU~\citep{banik2020deep} & Conditional GAN (cGAN)  & {None}	& Encoder-decoder & mIoU and DSC & Yes & Weighted loss of MSE and BCE & Adam \\ \hline

IIAI-Med~\citep{ji2020automatic} &PraNet &	Res2Net &	Encoder-decoder	& mIoU, DSC and FPS &	No	& Weighted IoU loss + BCE loss &	Adam \\ \hline

HGV-HCMUS~\citep{nguyen2020hcmus}& PraNet and ResUNet++ with triple path &	ResUNet++	& Encoder-decoder &	mIoU	& Yes &	Categorical crossentropy &	Adam \\ \hline

{GeorgeBatch}~\citep{batchkala2020real}& U-Net &	None	& Encoder-decoder &	Acc and Speed	& Yes &	Non-Binarized IoU &	Adam \\ \hline 

PRML20202GU~\citep{poudel2020automatic} &	Efficient-UNet +Channel-Spatial Attention + Deep Supervision & Variants of EfficientNet	& Encoder-decoder	& mIoU and DSC &	Yes &	BCE + DSC loss &	Adam \\ \hline

VT~\citep{thambawita2020pyramid} & U-Net coupled with PYRA  &None & Encoder-decoder& mIoU and DSC &	Yes	& {BCEWithLogits} {Loss}	& RMSprop\\ \hline

IRISNSYSU~\citep{maxwell2020temporal} & Temporal-Spatial Attention Model &	Faster-RCNN & Hybrid attention interface & AP  &	Yes	& Cross entropy &	Adam \\ \hline

NTK~\citep{tomar2021automatic}	& Residual blocks combined with SE network &	None	& Encoder-decoder	& DSC, mIoU and FPS &	No & BCE + DSC loss &	Adam\\ 
        \bottomrule
    \end{tabular}
\end{sidewaystable*}

\textbf{FAST-NU-DS:} 
Team FAST-NU-DS~\citep{ali2020depth} explored the advantage of using depth-wise separable convolution in the atrous convolution of the ResUNet++\citep{jha2019resunet++} architecture.  Modifications were made to get the lightweight image segmentation.  Deep atrous spatial pyramid pooling was also implemented on the ResUNet++ architecture.  The purpose of this architectural design was to provide good performance on the image segmentation evaluation metrics and inference time.  {To get the lightweight model architecture, changes were made to the atrous bridge in ResUNet++ architecture.  The convolution layer in the atrous bridge was replaced with depthwise separable convolution.  Depth-wise separable convolution first applies channel-wise filters, followed by a $1\times1$ pointwise convolution, to maintain performance while streamlining computations. } The implementation of depth-wise separable convolution resulted in less number of parameters and giga-floating point operations (GFLOPs).  

\textbf{AI-TCE:} 
Team AI-TCE~\citep{nathan2020efficient} proposed an efficient supervision network that uses EfficientNet~\citep{tan2019efficientnet} and an attention Unit. The proposed network had the properties of an encoder-decoder structure with supervision layers. An EfficientNet-B4 was used as a pre-trained architecture in the encoder block. The decoder block combined dense block and Concurrent Spatial and Channel Attention  block. Both the encoder and decoder were connected by Convolution Block Attention Module (CBAM). All the outputs of the decoder layer were supervised, i.e., individual decoder output was taken and upsampled with the output layer and supervised by the loss function. Also, all upsampled outputs were concatenated and fed into CBAM. In the upsampling, the convolution transpose layer was used.  

\textbf{ML-MMIV SARUAR}: Team ML-MMIV SARUAR~\citep{alam2020automatic} used the U-Net with pre-trained ResNet50 on the ImageNet dataset as the encoder for the polyp segmentation task. The use of a pre-trained encoder helped the model to converge easily. The input image was fed into the pre-trained ResNet50 encoder, consisting of a series of residual blocks as their main component. These residual blocks helped the encoder extract the important features from the input image, which were then passed to the decoder. Skip connections between the encoder and decoder branch help the model to get all the low-level semantic information from the encoder, which allowed the decoder to generate the desired feature maps.

\textbf{UiO-Zero}: Team UiO-Zero~\citep{ahmed2020generative}  used the generative adversarial networks (GAN) framework for solving the automatic segmentation problem. Perceiving the problem as an image-to-image translation task, conditional GANs were utilized to generate masks conditioned by the images as inputs. The polyp segmentation GAN-based model consists of two networks, namely a generator and discriminator, that were based on convolution neural networks. A generator takes the images as input and tries to produce realistic-looking masks conditioned by this input and a discriminator, which was basically a classifier that had access to the ground truth masks and tried to classify whether the generated masks was real or not. To stabilize the training, the images were concatenated with the masks (generated or real) before being fed to the discriminator.

\textbf{HGV-HCMUS}: The {HGV-HCMUS}~\citep{trinh2020hcmus} team proposed methods combining the Residual module, Inception module, Adaptive CNN with U-Net~\citep{ronneberger2015u} model, and PraNet~\citep{fan2020pranet} for semantic segmentation of various types of polyps in endoscopic images. The team submitted five different runs considering five different solutions. In the first approach, a simple U-Net architecture was adopted to parse masks of polyps. Second, the regular ReLU was replaced with Leaky ReLU to deal with dead neurons. Third, to further boost the result, an Inception module was introduced to extract better features. Fourth, a pre-trained model with the ResNet-50 backbone was used to build ResUNet, yielding better obtained results. Last, PraNet was employed for polyp segmentation in colonoscopy images.  {This solution provided the best outcome and was used to generate the results.}

\textbf{AI-JMU:} Team AI-JMU~\citep{krenzer2020bigger} explored various image segmentation models, specifically the Cascade Mask R-CNN~\citep{cai2019cascade} and Mask R-CNN~\citep{he2017mask} with ResNet~\citep{he2016deep} as well as the ResNeSt~\citep{zhang2022resnest} architectures was used as the backbone. Additionally, the team investigated the effect of varying the depth of both the ResNet and ResNeSt architectures. Depths of 50, 101, and 200 were evaluated for the ResNeSt model, and depths of 50 and 101 for the ResNet model. {They reported that the best outcome was obtained using ResNeSt-101 when combined with Cascade Mask R-CNN.}

\textbf{SBS}: Team SBS~\citep{shrestha2020ensemble} exploited ResNet 34~\citep{he2016deep} and EfficientNet-B2~\citep{tan2019efficientnet} backbones in the U-Net. The team introduced two different models: Single Model and Ensemble Model. The ResNet-34 was used in the single model. The weights saved after the training phase was loaded in the network, and test data were fed to get the predicted polyp masks. However, in the case of the ensemble model, both ResNet-34 and EfficientNetB2 were used to predict the masks. Then the individual prediction was ensembled using bitwise multiplication between the two predicted masks. The ensemble model provided better evaluation results as compared to the single model, as when multiple algorithms were ensembled predictive power increases and error rate decreases. {Hence, the final results are reported using the ensemble model using ResNet-34 and EffiecientNetB2 as backbones in the U-Net architecture.}


\textbf{AMI Lab:} Team AMI Lab~\citep{kang2020kd} utilized the knowledge distillation technique to improve ResUNet++~~\citep{jha2019resunet++}, which performs well on automatic polyp segmentation. First, the data augmentation module was used to generate augmented images for the input. Second, the obtained augmented images were fed to both the student model and the teacher model. Third, the distillation loss between the outputs of student and teacher models was calculated. Similarly, the loss between the output of the student model and the ground truth label was computed to train the student model.

\textbf{UNITRK:} Team UNITRK~\citep{khadka2020transfer} employed the UNet model pre-trained on the brain MRI dataset. The notion of knowledge transfer has been the key motivating factor to choose a simple pre-trained model. The model was fine-tuned with the polyp dataset. The fine-tuning of the pre-trained model helped to converge faster without the requirement of a large number of training examples. The additive soft attention mechanism was integrated with the pre-trained UNet architecture. The key benefit of this attention UNet structure in comparison to multi-stage CNNs was that it does not require training of multiple models to deal with object localization and thus reduces the number of model parameters. It helps to focus on relevant regions in the input images.

\textbf{MedSeg\_JU}: Team MedSeg\_JU~\citep{banik2020deep} proposed an approach for polyp segmentation based on deep conditional adversarial learning. The proposed framework consists of two interdependent modules: a generator network and a discriminator network. The generator was an encoder-decoder network responsible to predict the polyp mask while the discriminator enforces the segmentation to be as similar to the ground truth segmented mask.  The training process of the network alternates between training the generator and the discriminator, with the generator trained to produce a predicted synthetic mask by freezing the discriminator and the discriminator trained while freezing the generator.

\begin{figure*} [!t]
    \centering
    \includegraphics[width=1\textwidth]{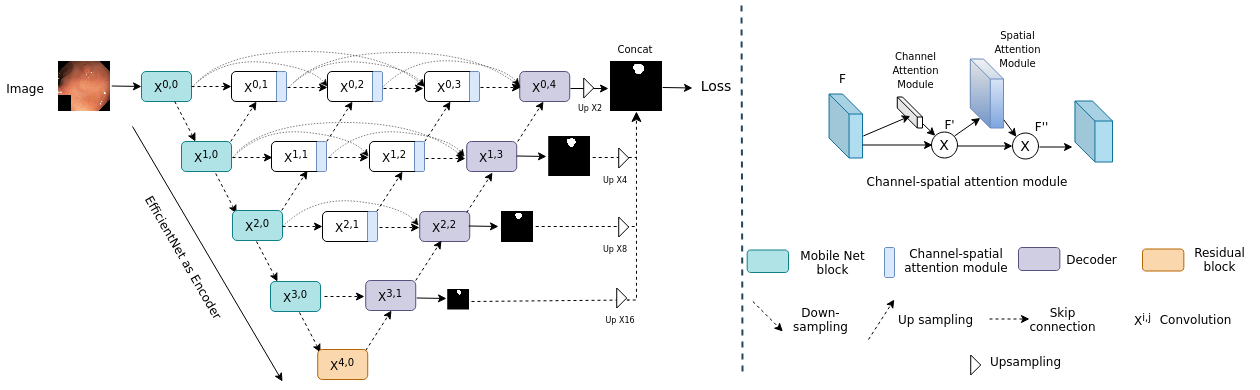}
     \caption{Overview of the winning solution for the Polyp segmentation task (Task 1) from Team \textbf{\textit{PRML2020GU}}. The architecture utilizes pre-trained weights from EfficientNet in the encoder. Additionally, it uses dense skip connections, deep supervision and channel-spatial attention for fast convergence and better performance. }
     \label{bestmethod2020}
\end{figure*}

\textbf{IIAI-Med}: Team IIAI-Med team~\citep{ji2020automatic} presented a novel deep neural network, called the Parallel Reverse Attention Network (PraNet)~\citep{fan2020pranet}, for the task of automatic polyp segmentation at MediaEval 2020. The network first aggregated features in high-level layers using a parallel partial decoder (PPD). This combined feature was then used to generate a global map as the initial guidance area for the following components. Additionally, the network mines boundary cues using a reverse attention (RA) module which establishes the relationship between areas and boundary cues. Thanks to the recurrent cooperation mechanism between areas and boundaries, the PraNet was able to calibrate misaligned predictions, improving segmentation accuracy and achieving real-time efficiency (nearly 30fps). The code and results are available at https://github.com/GewelsJI/MediaEval2020-IIAI-Med.


{\textbf{HBKU\_UNITN\_SIMULA}} Team HBKU\_UNITN\_SIMULA~\citep{nguyen2020hcmus} proposed two different approaches leveraging the advantages of either ResUNet++ or PraNet model to efficiently segment polyps in colonoscopy images, with modifications on the network structure, parameters, and training strategies to tackle various observed characteristics of the given dataset. For the first approach, PraNet was used, which is a parallel reverse attention network that helps to analyze and use the relationship between areas and boundary cues for accurate polyp segmentation. The PraNet with Training Signal Annealing strategy was used to improve segmentation accuracy and effectively train from scratch on the given small dataset.  For the second approach, ResUNet++ was used, which takes advantage of residual blocks, squeeze and excitation blocks, atrous spatial pyramid pooling, and attention blocks. The input path was modified and integrates a guided mask layer to the original structure for better segmentation accuracy.  {They used the two approaches to experiment with different runs. The best polyp segmentation outcome was achieved when the results from three PraNet and five ResUNet++ models, trained on different train-val dataset splits, were averaged. }

\textbf{GeorgeBatch}: Team GeorgeBatch~\citep{batchkala2020real}  used the standard U-Net architecture for the binary segmentation task, and experiments were conducted using the intersection-over-union loss (IoU loss) instead of the commonly used binary cross-entropy (BCE) loss. They also experiment with a combination of both losses in the training process. The motivation behind this approach was to strike a balance between accuracy and speed for using automated systems during colon cancer surveillance and surgical removal of polyps. This balance is considered while experimenting with other parameters like loss function and data augmentation to boost performance. The reported outcomes show that using IoU loss results in enhanced segmentation performance, with a nearly 3\% improvement on the \ac{DSC} metric while maintaining real-time performance (close to 200 FPS). The code and results are available at https://github.com/GeorgeBatch/kvasir-seg.

\textbf{PRML2020GU:} An overview of the approach proposed by team PRML2020GU~\citep{poudel2020automatic} is shown in Figure~\ref{bestmethod2020}. The team employed an EfficientNetB3 as an encoder backbone with a U-Net decoder and leveraged the concept of U-Net++ of redesigning the skip connections to use multi-scale semantic details. The densely connected skip connections to the decoder side enable flexible multi-scale feature fusion both horizontally and vertically at the same resolution. Besides, the proposed method is powered by deep supervision, where all the outputs after deep supervision is averaged, and the final mask is generated. Further, channel-spatial attention enables significantly better performance and fast convergence. Moreover, integrating the channel and spatial attention modules restrains irrelevant features and allows only useful spatial details.


\textbf{VT:} Team VT~\citep{thambawita2020pyramid} proposed a simple but efficient idea of using an augmentation method called pyramid focus-augmentation (PYRA) that uses grids in a pyramid-like manner (large to small) for polyp segmentation. The method has two main steps: data augmentation with PYRA using pre-defined grid sizes followed by training of a DL model with the resulting augmented data. PYRA can be used to improve the performance of segmentation tasks when there is a small dataset to train the DL models or if the number of positive findings is small. The method shows a large benefit in the medical diagnosis use case by focusing the clinician’s attention on regions with findings step-by-step.

\textbf{IRISNSYSU:} Team IRISNSYSU~\citep{maxwell2020temporal} proposed a local region model with attentive temporal-spatial pathways for automatically learning various target structures. The attentive spatial pathway highlights the salient region to generate bounding boxes and ignores irrelevant regions in an input image. The proposed attention mechanism allows efficient object localization, and the overall predictive performance is increased because there are fewer false positives for the object detection task for medical images with manual annotations.

\textbf{NKT:} Team NKT~\citep{tomar2021automatic} proposed a full convolution network following an encoder-decoder approach. It combines the strength of residual learning and the attention mechanism of the squeeze and excitation (SE) network. The encoding network consists of 4 encoder blocks with 32, 64, 128, and 256 filters. The decoding network also consists of 4 decoder blocks with 128, 64, 32, and 16 filters. Both the encoder and decoder block consist of a residual block as their core component. The residual block helps in building deep neural networks by solving the vanishing gradient and exploding gradient problem.

Additionally, in Table~\ref{table:challenge_summary2020}, we provide an elaborate summary of all the research teams who participated in the ``Medico 2020" challenge. It gives a detailed overview of the algorithms, backbone, nature, choice basis,  data augmentation used, loss function, and optimizer used by each participating teams. 

\begin{table} [!t]
\caption{Summary information of participating teams in MedAI 2021. Here, `$\surd$'  = Team participated, `--' = No participation,  \textbf{Task 1 =} Polyp segmentation task, \textbf{Task 2 =} Instrument segmentation task, and \textbf{Task 3 =} Transparency task.  {A total of 16 teams participated in polyp segmentation and instrument segmentation and 14 teams participated in the Transparency tasks in the challenge.}}
\label{table:challenge_summary_MedAI}
\centering
\begin{tabular}{p{0.7cm}|p{3.2cm}|p{1cm}p{1cm}p{1cm}}
\toprule
\textbf{Chal.} & \centering \textbf{Team Name} & \textbf{Task 1}  &\textbf{Task 2} &\textbf{Task 3}\\
\midrule
\multirow{16}{*}{\rotatebox{90}{\textbf{MedAI 2021}}}
              & The Segmentors & $\surd$   & $\surd$ &$\surd$\\ 
              & The Arctic & $\surd$    & $\surd$ &$\surd$ \\ 
              & mTEC & $\surd$ &$\surd$ &$\surd$ \\ 
              & MedSeg\_JU  &-- &  $\surd$ &--\\ 
              & MAHUNM & $\surd$ &  $\surd$ &$\surd$ \\ 
              & IIAI-CV\&Med & $\surd$   &  $\surd$  &$\surd$  \\ 
              & NYCity & $\surd$ &  $\surd$ &$\surd$ \\ 
              & PRML & $\surd$ &   $\surd$ &$\surd$ \\ 
              & leen & $\surd$ &   $\surd$ &$\surd$ \\ 
              & CV\&Med IIAI & $\surd$ &  $\surd$ &$\surd$ \\ 
              & Polypixel & $\surd$ &  $\surd$ &$\surd$ \\ 
              & agaldran & $\surd$ &   $\surd$ &$\surd$ \\ 
              & TeamAIKitchen & $\surd$ &$\surd$  & $\surd$  \\ 
              & CamAI& $\surd$ &  $\surd$ &$\surd$ \\ 
              & OXGastroVision & $\surd$ &  $\surd$ &$\surd$ \\
              & Vyobotics & $\surd$  & -- & --\\ 
              & NAAMII & $\surd$ &  $\surd$ & --\\ 
\bottomrule
\end{tabular}
\end{table}

\begin{sidewaystable*}
    \footnotesize
    \centering
    \caption{Summary of the participating teams algorithm for MedAI 2021.}
    \label{table:challenge_summary2021}
    \begin{tabular}{|p{2.7cm}|p{1.6cm}|p{2.2cm}|p{2cm}|p{2cm}|p{1.8cm}|p{1.5cm}|p{1.7cm}|p{1.3cm}|}
        \toprule
        \textbf{Team Name} & \textbf{Segmentation Task} & \textbf{Algorithm} & \textbf{Backbone} & \textbf{Nature} & \textbf{Choice basis} & \textbf{Augmentation} & \textbf{Loss} & \textbf{Optimizer} \\ \midrule 
The Segmentors~\citep{Mirza2021}& 	Polyp, Instrument	& U-Net	& None	& Encoder-decoder& 	DSC  and mIoU	& Yes	& DSC & 	Adam\\\hline
The Arctic~\citep{Somani2021} & Polyp, Instrument  & DeeplabV3plus + ResNet101 & None		& Hybrid	& DSC	& Yes & 	Cross-entropy	& Adam	\\\hline

mTEC~\citep{bhattacharya_betz_eggert_schlaefer_2021} & Polyp, Instrument	& DPRA-EdgeNet	 &  HarDNet & Cascade	& DSC and mIoU &	No	&  {(Dice  + BCE) loss} & {Adam}	\\ \hline

MedSeg\_JU \citep{Banik2021}	  & Instrument	&  EM-Net	&  EfficientNet-B3	&  Encoder-decoder & DSC	& Yes	& DSC & 	Adam	\\ \hline

MAHUNM~\citep{Haithami2021} & 	Polyp, Instrument	&  DeeplabV3 with GRU	& ResNet-50/ResNet-101	& Sequential	& DSC  and mIoU	& No	&  {BCE With Logits Loss} & {Adam}	\\\hline											IIAI-CV\&Med~\citep{Dong2021}& Polyp, Instrument	& Polyp-PVT, Sinv2-PVT and Transfuse- PVT	& Transformer & 	Ensemble & 	Majority voting	& No & 	IoU & 	Adam\\ \hline

NYCity~\citep{Chen2021}	& 	Polyp, Instrument	& HarDNet-85, ResNet-101	&  Transformer & Ensemble & 	Accuracy & 	Yes & IoU & 	Gradient centralization		\\\hline
															
PRML~\citep{Poudel2021} & Polyp, Instrument	&  Ef-UNet & 	EfficientNet	& Encoder-decoder	& DSC  and mIoU	& No	& DSC Loss	& Adam	\\\hline	
														
leen~\citep{ahmed2021explainable}  & 	Polyp, Instrument	& GAN	& None & Encoder-decoder & DSC  and mIoU & 	No	& BCE and L1 loss	& Adam	 \\\hline	
																	
CV\&Med IIAI~\citep{Chou2021} & Polyp, Instrument	& SINetv2	& PVT v2 & Encoder-decoder&  mIoU	& No	&  {Pixel position-aware loss}	& {Adam}		\\ \hline
																
Polypixel~\citep{Tzavara2021} & Polyp, Instrument	& Transfer learning using EfficientNet B1 & 	None & 	CNN	 & DSC and mIoU & 	Yes& 	IoU	& Adam	\\\hline	
															
agaldran \citep{galdran2021polyp} & 	Polyp, Instrument	& Double Encoder-Decoder with temperature scaling	Feature&  Pyramid Network as Decoder and Resnext101 as pretrained decoder	& Sequential	& DSC	& Yes	& DSC	& Sharpness-aware minimization (SAM) + Adam	\\\hline

TeamAIKitchen~\citep{Keprate2021} & 	Polyp, Instrument& 	U-Net	& None & Encoder-decoder	& DSC	& Yes	& DSC & 	Adam\\\hline 

CamAI~\citep{Yeung2021} & Polyp, Instrument	& Transfer learning (Attention U-Net)	& ResNet-152	& Ensemble	& Accuracy	& Yes	& Unified focal loss	& SGD	\\ \hline	

OXGastroVision~\citep{ali2021iterative}  & 	Polyp, Instrument & 	DDANet + FANet & 	None & Encoder-decoder	& DSC	& No	& BCE and DSC loss	& Adam \\\hline
															
Vyobotics~\citep{Rauniyar2021} & Polyp	& DDANet	& None & Encoder-decoder& DSC and mIoU	& Yes	& BCE and DSC loss	& Adam		\\\hline	

NAAMII~\citep{Rauniyar2021} & Polyp, Instrument	& U2Net	& None & Encoder-decoder&  mIoU	& Yes	& Mean Squared Error, Cross-entropy	& Adam		\\	
 \bottomrule
    \end{tabular}
\end{sidewaystable*}

\subsection{Methods used in MedAI 2021}
In this subsection, we briefly summarize the methods used by the participating teams in the MedAI 2021 challenge.
In Table~\ref{table:challenge_summary_MedAI}, we present the research teams who have participated in each of these three tasks. It can be seen from this table that most of the teams participated in all three tasks except for {three} teams, which participated in either one or two of the sub-tasks. {All participating teams have used the same architecture in their submission for polyp segmentation and instrument segmentation subtasks.} However, two teams, namely \textit{Vyobotics}~\citep{Rauniyar2021} and  \textit{MedSeg\_JU}~\citep{Banik2021} have participated in only one of the subtasks. The team \textit{Vyobotics}~\citep{Rauniyar2021} has participated in the polyp segmentation task whereas the team \textit{MedSeg\_JU}~\citep{Banik2021} has participated in the surgical instrument segmentation task. 


\textbf{The Segmentors}: Team Segmentors~\citep{Mirza2021} 
proposed solution is a UNet-based algorithm designed for segmenting polyps in images taken from endoscopies. The primary focus of this approach was to achieve high segmentation metrics on the supplied test dataset, which was a crucial requirement for accurate and reliable polyp segmentation. To this end, they experimented with data augmentation and model tuning to achieve satisfactory results on the test sets.

\textbf{The Arctic}: Team Arctic~\citep{Somani2021}
utilized a unique hybrid optimization technique that combined the power of DeepLabV3+~\citep{chen2018encoder} and ResNet101~\citep{he2016deep} to address the specific challenges of GI image segmentation effectively. In order to ensure the accuracy of their results, the team employed a 5-fold cross-validation approach, with a learning rate of 0.0001 and a batch size of 12. Additionally, towards transparency, they proposed a method of rendering feature attention maps to visualize the attention of the network on individual pixels within the image.

\textbf{mTEC}: Team mTEC~\citep{bhattacharya_betz_eggert_schlaefer_2021} introduced a new architecture called Dual Parallel Reverse Attention Edge Network (DPRA-EdgeNet) for joint segmentation of polyp masks and polyp edge masks. This architecture utilizes the reverse attention module from PraNet~\citep{fan2020pranet} to perform the segmentation tasks. The team implemented two parallel decoder blocks, with one focused on extracting features for polyp segmentation and the other focused on extracting features for polyp edge segmentation. The polyp mask decoder leverages the features from the edge decoder block to improve the accuracy of the segmentation. Additionally, the team employed deep supervision of both edge and polyp features to stabilize the optimization process of the model.

\textbf{MedSeg\_JU}: Team MedSeg\_JU~\citep{Banik2021} proposed EM-Net, encoder-decoder-based architecture inspired by the M-Net~\citep{7950555} architecture. In their approach, the encoder branch of the network utilized  EfficientNet-B3~\citep{DBLP:journals/corr/abs-1905-11946} as its backbone. The network also employed a multi-scale input method, where the input image was downsampled at rates of 2, 4, and 8 at each level of the encoder branch, providing a multi-level receptive field. The decoder branch was a mirror structure of the encoder, where upsampling was used to increase the size of the feature maps at each level. Skip connections were used to enhance the flow of spatial information lost during downsampling. The final feature maps underwent point-wise convolution and sigmoid activation and were then upsampled to provide deep supervision and a local pixel-level prediction map for each scale of the input image. These maps were then fused to generate the final segmentation mask.

\textbf{MAHUNM}: Team MAHUNM~\citep{Haithami2021} presented an approach for enhancing the segmentation capabilities of DeeplabV3 by incorporating Gated Recurrent Neural Network (GRU). In their approach, the team replaced the 1-by-1 convolution in DeeplabV3 with GRU after the ASSP layer to combine input feature maps. While the convolution and GRU had sharable parameters, the latter had gates that enabled or disabled the contribution of each input feature map. The experimental evaluation conducted on unseen test sets demonstrated that using GRU instead of convolution produced better segmentation results.

\textbf{IIAI-CV\&Med}: Team IIAI-CV\&Med~\citep{Dong2021} 
developed an ensemble of three sub-models, namely Polyp-PVT \citep{dong2021polyp}, Sinv2-PVT, and Transfuse-PVT. The official Polyp-PVT, as designed for polyp segmentation, was adopted without modification and achieved state-of-the-art segmentation capability and generalization performance. Transfuse, also designed for polyp segmentation, was improved by replacing the transformer part with the pyramid vision transformer (PVT)~\citep{wang2022pvt} to enhance its performance. The official Sinv2~\citep{fan2021concealed}, which proposes an end-to-end network for searching and recognizing concealed objects, was employed and its original backbone of Res2Net was replaced with a stronger PVT transformer~\citep{wang2022pvt} to extract more meaningful features.

\textbf{NYCity}: Team NYCity~\citep{Chen2021} presented a novel multi-model ensemble framework. The team first collected a set of SOTA models in this field and further improved them through a series of {refinements. These models include TransFuse~\citep{zhang2021transfuse} and HarDNet-MSEG~\citep{huang2021hardnet}. They improvised TransFuse by replacing its backbone with HarDNet-85~\citep{chao2019hardnet} and placing an additional BiFuse layer. They further modified HarDNet-MSEG by using HarDNet-85 and ResNet-101~\citep{he2016deep} as the backbone. Additionally, they made modifications to the decoder and adopted different receptive fields.} By integrating those fine-tuned models into a more powerful ensemble framework, they were able to achieve improved performance. 

\textbf{PRML}: Team PRML~\citep{Poudel2021} introduced Ef-UNet, a segmentation model that is composed of two main components. First, a U-Net encoder that utilizes EfficientNet~\citep{DBLP:journals/corr/abs-1905-11946} as a backbone, which allows the generation of different semantic details in multiple stages. Second, a decoder integrates spatial information from different stages to generate a final precise segmentation mask. Using EfficientNet as the encoder backbone provides Ef-UNet with the ability to efficiently extract high-level features from the input images while the decoder component effectively integrates these features to produce accurate segmentation results.

\textbf{leen}: Team leen~\citep{ahmed2021explainable} utilized the GANs framework to produce corresponding masks that locate the polyps or instruments on GI polyp images. To ensure transparency and explainability of their models, the team leen adopted the layer-wise relevance propagation (LRP) approach~\citep{bach2015pixel}, which is one of the most widely used methods in explainable artificial intelligence. This approach generated relevant maps that display the contribution of each pixel of the input image in the final decision of the model.

\textbf{CV\&Med IIAI}: Team CV\&Med IIAI~\citep{Chou2021}
proposed a novel dual model filtering (DMF) strategy, which effectively removed false positive predictions in negative samples through the use of a metrics-based threshold setting. To better adapt to high-resolution input with various distributions, the PVTv2~\citep{wang2022pvt} backbone was embedded into the SINetV2~\citep{fan2021concealed} framework. The SINetV2 framework with camouflaged object detection was used for better identification ability, as polyp segmentation is a downstream task. Additionally, extensive experiments have been conducted to study the effectiveness of DMF, and it was found that the method performs well under different data distributions, making it a favorable solution for problems where the training dataset had a different distribution of negative samples compared to the testing dataset.

\textbf{Polypixel}: Team Polypixel~\citep{Tzavara2021} presented a study in which they used both pretrained and non-pretrained segmentation models for the polyp and instrument segmentation task. The team trained and validated both models on the dataset. The model architectures were retrieved from a Python library, ``Segmentation Models"~\url{https://github.com/qubvel/segmentation_models}, that contained different CNN architectures. This library offered models with both untrained and pre-trained weights, which were trained on the ImageNet dataset. To find the optimal fit for their datasets, they experimented and tested their results using EfficientNet, MobileNet, SE-ResNet, Inception, ResNet, and VGG. They achieved the best results with EfficientNetB1 for the polyp segmentation task.

\begin{figure} [!t]
    \centering
    \includegraphics[width=1\linewidth]{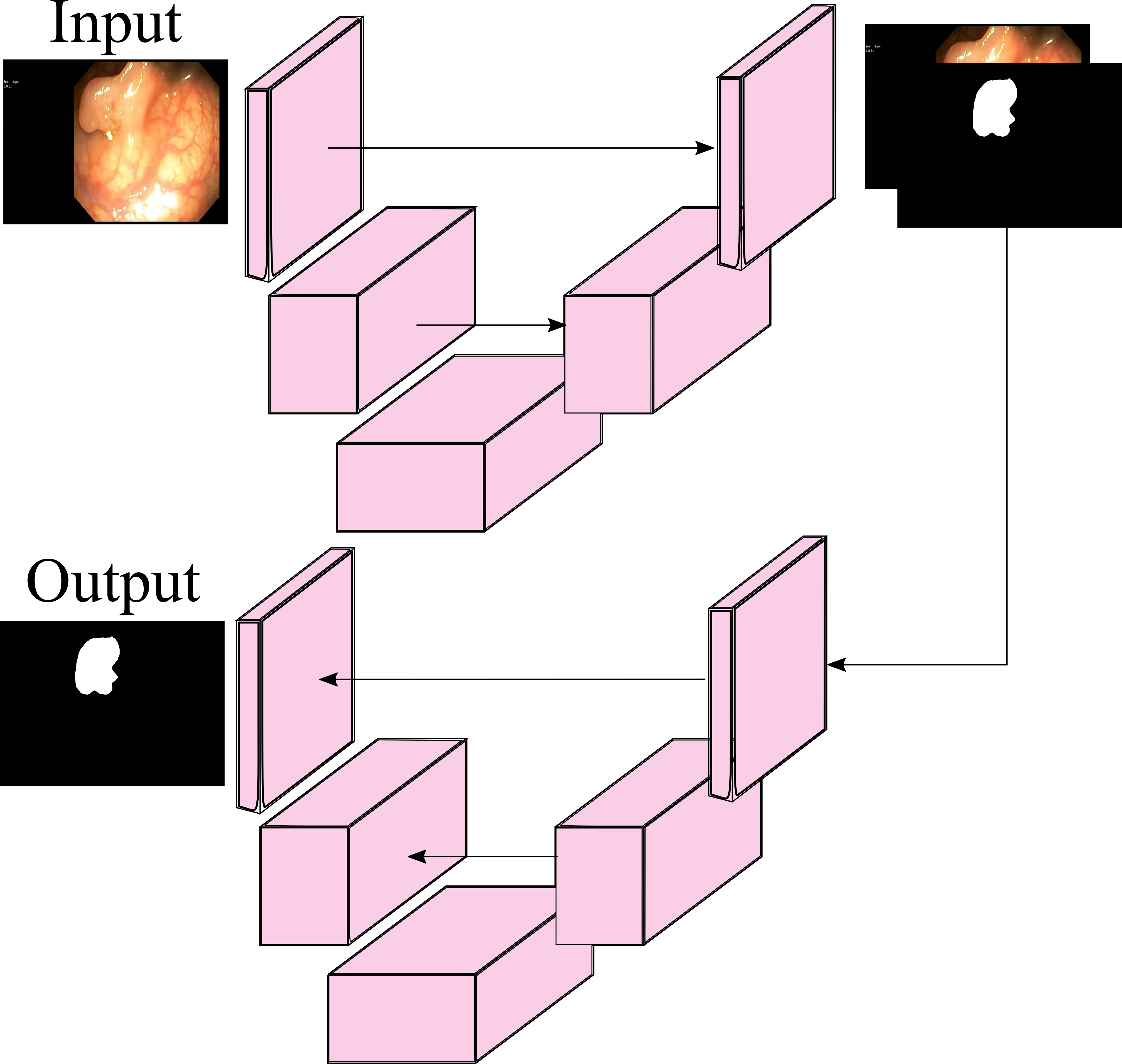}
     \caption{Overview of winning solution of MedAI 2021 proposed by Team \textbf{\textit{agaldran}}. A double encoder-decoder network was used to segment polyps and surgical instruments.}
     \label{fig:bestmethodmedAI2021}
\end{figure}

\textbf{agaldran}: Team agaldran~\citep{galdran2021polyp} utilized a double encoder-decoder structure for polyp and instrument segmentation, which consists of two U-Net like structures arranged sequentially as shown in Figure~\ref{fig:bestmethodmedAI2021}. The first encoder-decoder network processes the original image and produces output fed into the second encoder-decoder network. According to the authors, this setup allows the first network to highlight the important features of the image for segmentation, while the second network further improves the predictions of the first network. {For the architectural design of a double encoder-decoder network, they incorporate Feature Pyramid Network (FPN)~\citep{lin2017feature} architecture as the decoder mechanism, along with Resnext101 that serves as the pretrained decoder~\citep{kolesnikov2020big}. This is done to optimize the feature extraction. To further refine the model's optimization process, they used Sharpness-Aware Minimization (SAM)  along with the ADAM optimizer~\citep{foret2020sharpness}.} The team employed a 4-fold cross-validation approach to train their models, training with four separate models and using temperature sharpening across the ensemble model to produce the final segmentation maps.

\textbf{TeamAIKitchen}: Team TeamAIKitchen~\citep{Keprate2021} presented a methodology for developing, fine-tuning, and analyzing a U-Net-based model for generating segmentation masks for the polyp segmentation task. { They modified the original U-Net architecture to extend it to work with less training samples and to generate the output mask of the same size as the input. ReLU activation function was used in the hidden layers. They further experimented with different batch sizes and selected 8 as the best. Same architecture was used for polyp and instrument segmentation with early stopping criteria.} 

\textbf{CamAI}: Team CamAI~\citep{Yeung2021} presented a deep learning pipeline that is specifically developed to accurately segment colorectal polyps and various instruments used during endoscopic procedures. To improve transparency and interpretability, the pipeline leveraged the Attention U-Net architecture, which enables visualization of the attention coefficients to identify the most salient regions of the input images. This allowed for a better understanding of the model's decision-making process and facilitated the identification of potential errors. To further improve performance, the pipeline incorporated transfer learning using a pre-trained encoder. Additionally, test-time augmentation, softmax averaging, softmax thresholding and connected component labeling were used to further refine predictions and boost performance.

\textbf{OXGastroVision}: Team OXGastroVision~\citep{ali2021iterative} presented a novel solution that utilizes two state-of-the-art deep learning models, namely the iterative FANet~\citep{tomar2022fanet} architecture and DDANet~\citep{tomar2021ddanet}. The FANet is based on a feedback attention network that allows rectifying predictions iteratively. It consists of four encoder and four decoder layers. Similarly, DDANet is based on a dual decoder attention network with one shared encoder at each layer. While the iterative mechanism in the full FANet architecture can lead to larger computational time, DDANet has real-time performance (70 FPS) but sub-optimal output. To overcome these limitations, the team proposes to use the segmentation maps from the DDANet output as input for the FANet iterative network for pruning. This approach aims to achieve a balance between computational efficiency and segmentation accuracy.

\textbf{Vyobotics}: Team Vyobotics~\citep{Rauniyar2021}  presented a solution based on dual decoder attention network (DDANet)~\citep{tomar2021ddanet}, a deep learning model that has been specifically designed to achieve decent performance and real-time speed. The team performed data augmentation and trained a smaller network. This smaller network has a lower number of trainable parameters, which resulted in lower GPU training time. The ultimate goal of this approach was to achieve decent evaluation metrics while maintaining a decent FPS speed, which is crucial for real-time applications.

\textbf{NAAMII}: The team participated in polyp and instrument segmentation tasks. They employed $U^2Net$~\citep{qin2020u2} as the base network. They added a separate learnable CNN network on the decoder part of the U2Net to regress the HoG features of the input images. The output from each decoder block was fed into the HoG regressor and learned the parameters to predict the HoG correctly. They jointly minimized Mean Squared Error (MSE) loss for HoG features and CrossEntropy loss for Segmentation. {However, they only submitted their method description to the organizer and did not publish it as a research paper.}


\begin{table}[!t]
    \centering
    \caption{{Performance comparison on Polyp segmentation task (Medico 2020). `Bold' refers to the best score and `red' color refers to the second best score. We follow this consistently in all the Tables.} $\uparrow$ {indicates a higher value is better.}}\label{table:medicochallengeresults}
    \resizebox{\columnwidth}{!}{
    \begin{tabular}{ l l l l l l }
        \toprule
        \textbf{Team Name} & {\textbf{mIoU}} $\uparrow$ & DSC $\uparrow$ & Recall $\uparrow$ & Precision $\uparrow$  & F2 $\uparrow$ \\ \midrule

PRML2020GU	        & \textbf{0.78975}	&\textbf{0.86076}	& {0.90312}	&0.86731		& {0.87481} \\

\shortstack{HBKU\_UNITN\_\\SIMULA}	& {0.77736}	&0.84768	&0.85034	& {0.88971}		&0.84483 \\

AI-TCE	            &0.77733	& {0.85030}	&\textbf{0.91646}	&0.83897		&\textbf{0.87901} \\

HGV-HCMUS	        &0.76597	&0.84050	&0.89439	&0.84455		&0.85768 \\

IIAI-Med            &0.76195	&0.83854	&0.83049	&\textbf{0.90121}		&0.82837\\

SBS	                &0.75503	&0.83162	&0.83168	&0.88513	    &0.82490\\

\shortstack{ML-MMIV\\ Saruar}	    &0.75168	&0.82289	&0.83908	&0.88228		&0.82492\\

AI-JMU	            &0.73742	&0.81437	&0.82661	&0.87432		&0.81038\\

MedSeg\_JU	        &0.71330	&0.80195	&0.83542	&0.82864		&0.81240\\

VT	                &0.70578	&0.79264	&0.88353	&0.78784		&0.82368\\

NKT	                &0.68473	&0.78012	&0.80771	&0.81264		&0.78546\\

UNITRK	            &0.64379	&0.72878	&0.70989	&0.85726	    &0.71312\\

GeorgeBatch         &0.63511	&0.73276	&0.75003	&0.82294		&0.73615\\

AMI Lab	            &0.61958	&0.70889	&0.72865	&0.79140		&0.71226\\

IRIS-NSYSU	        &0.50353	&0.64173	&0.87915	&0.58498		&0.75089\\

UiO-Zero	        &0.43814	&0.56185	&0.69721	&0.55587		&0.61102\\

FAST-NU-DS	        &0.18344	&0.26691	&0.27447   &0.29184	    &0.26762\\


        \bottomrule
    \end{tabular}}
\end{table}





        


        
        
        
        
        
        
        
       
        
        
         
\begin{table} [!t]
    \centering
    \caption{Algorithm efficiency task for polyp segmentation (Medico 2020). Note that some teams provided the same solution for this task as used in Task 1, whereas others designed different architecture specifically for the efficiency task (Task 2). $\uparrow$ {indicates a higher value is better.}}
    \label{table:Algorithm_efficiency_task}
     \resizebox{\columnwidth}{!}{
    \begin{tabular}{ l l l l l l l  }
        \toprule
        \textbf{Team Name} & {mIoU} $\uparrow$ & DSC $\uparrow$ & Recall $\uparrow$ & Precision $\uparrow$ & F2 $\uparrow$ & {\textbf{FPS}} $\uparrow$ \\ \midrule
       GeorgeBatch	& 0.6351 & 0.7327 & 0.7500 & 0.8229	 & 0.7361 & \textbf{196.79} \\
       UNITRK & 0.6437 & 0.7287 & 0.7098 & {0.8572}  & 0.7131 & {116.79} \\
       NKT	& 0.6847 & 0.7801 & 0.8077 & 0.8126 & 0.7854 & 80.60 \\
       \shortstack{HBKU\_UNITN\_\\SIMULA} & \textbf{0.7364} & {0.8074} & 0.8164 & \textbf{0.8646} & 0.8067 & 33.27 \\
        SBS	& {0.7341} & \textbf{0.8148} & {0.8764} & 0.8145 & \textbf{0.8354} & 26.66 \\
         AMI Lab & 0.6195 & 0.7088 & 0.7286 & 0.7914  & 0.7122 & 107.87 \\
        FAST-NU-DS & 0.6582 & 0.7556 & \textbf{0.8982} & 0.7171  & {0.8109} & 67.51 \\
        AI-JMU & 0.7213 & 0.8017 & 0.8359 & 0.8495 & 0.8056 & 3.36 \\
        PRML2020GU & 0.5083 & 0.6265 & 0.6003 & 0.7870 & 0.6029 & 2.25 \\
        \bottomrule
    \end{tabular}}
\end{table}

\section{Results}
In this section, we present a summary of the evaluated results obtained on the test dataset by all the participating teams in the two challenges: ``Medico 2020” and ``MedAI 2021”. Each challenge consists of tasks with a specific focus and evaluation metrics. There were two tasks for the Medico 2020 challenge, namely \textit{polyp segmentation} and \textit{algorithm efficiency} tasks. In the MedAI 2021, there were three tasks, namely \textit{polyp segmentation}, \textit{endoscopic accessory instrument segmentation} and \textit{transparency task}. The teams were evaluated based on standard evaluation metrics such as mIoU, DSC, Rec, Pre, Acc, F1, F2, and FPS. {We emphasized mIoU, DSC, and FPS more, whereas we also acknowledge the importance of recall and precision as they are useful metrics in clinical settings.} We have highlighted the best and the second-best scores in boldface and red color, respectively, for all the tasks in the two challenges.

\subsection{Medico 2020 results}
\subsubsection{{Polyp segmentation task}}
\begin{figure*} [!t]
    \centering
    \includegraphics[width =0.8\textwidth]{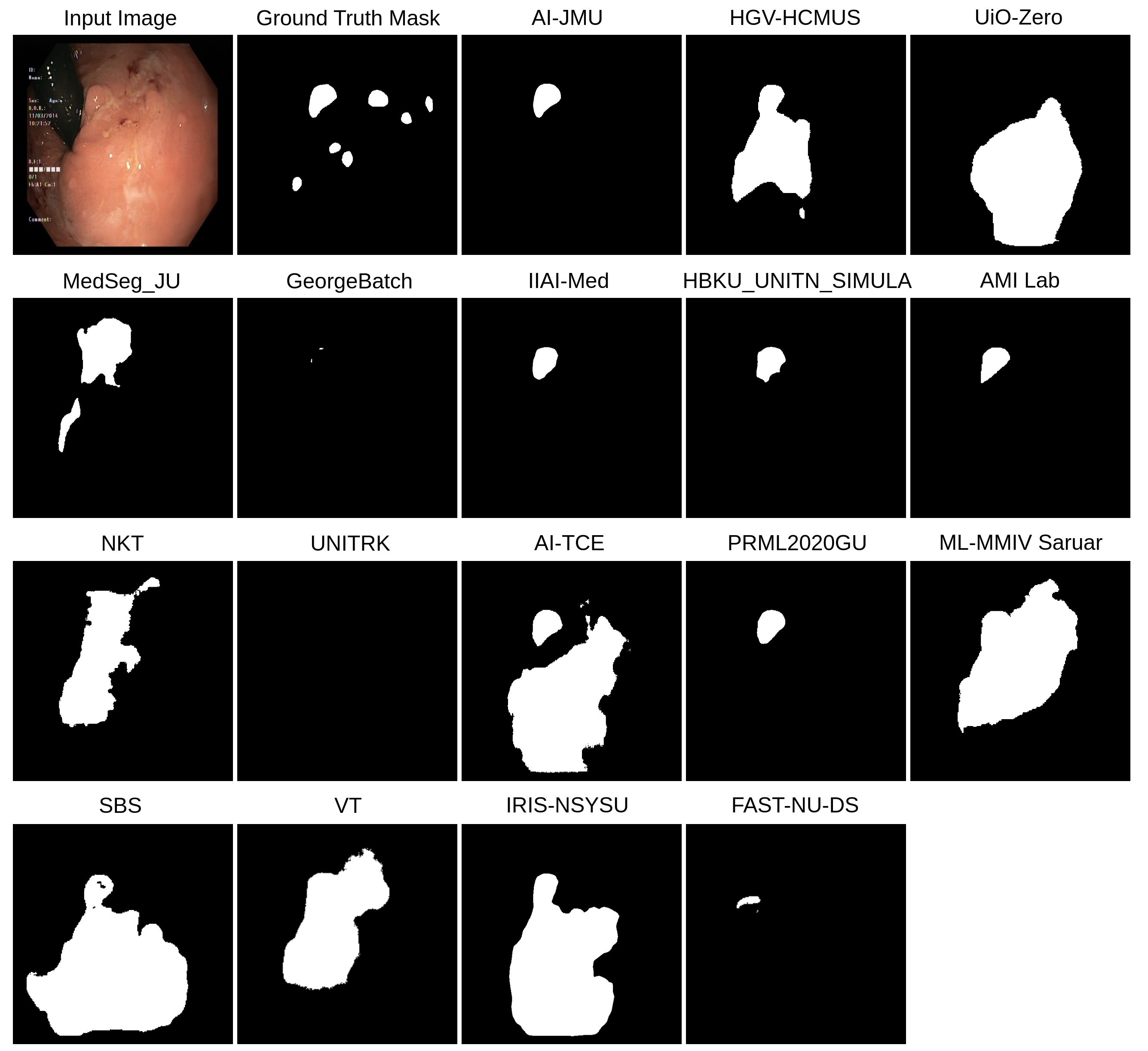}
    \caption{{The figure shows the qualitative results of participating teams for the polyp segmentation task in the Medico 2020 Challenge on challenging scenarios. When each team's predicted mask is compared with its corresponding ground truth, we observe that none of the teams obtained results that fit well with the ground truth.}}
    \label{fig:medico2020results}
\end{figure*}

In Table~\ref{table:medicochallengeresults}, we provide the results for the \textit{polyp segmentation} task. It can be observed that Team ``PRML2020GU'' outperforms other participating teams in the polyp segmentation task. It achieves a mIoU of 0.7897, DSC of 0.8607, recall of 0.9031, precision of 0.8673, and F2 of 0.8748. Team ``HBKU\_UNITN\_SIMULA'' was the second best performing team with mIoU of 0.7773. similarly, ``AI-TCE'' was the third best performing team with mIoU of 0.7773. The best-performing team, ``PRML2020GU,'' used an encoder-decoder structure with EfficientNet as the backbone and a U-Net decoder with channel-spatial attention and deep supervision. This architecture had an improvement of 1.23\% and 1.30\% over the mIoU and DSC achieved by the Team ``HBKU\_UNITN\_SIMULA'', which used an average of three PraNet and five ResUNet++ trained on different training and validation datasets. 



\subsubsection{{Algorithm efficiency task}}
For the second task, as in Table~\ref{table:Algorithm_efficiency_task}, team ``PRML2020GU'' has poor speed performance with a processing speed of only 2.25 fps, which is not desirable for a real-time efficient model. An interesting observation is that Team ``GeorgeBatch'' outperforms other participating teams in the algorithm efficiency task with a processing speed of 196.79 fps, as seen from Table~\ref{table:Algorithm_efficiency_task}. However, it is worth noting that the team obtained a low mIoU of 0.6351 for the polyp segmentation task, even though we are considering it as the winner in this task. {Team ``UNITRK" obtained a second-best fps of 116.79. Similarly, team ``NKT" obtained a balanced mIoU of 0.6847 and a high speed of 80.60 fps, and was ranked third for this task. Despite the two teams, ``UNITRK" and ``GeorgeBatch'', achieving the highest evaluation fps values, there is a trade-off between speed and mIoU.} Low FPS cannot be used for real-time medical processing applications, and low overlap evaluation metrics cannot generate precise segmentation masks. To provide further insight, we have included the qualitative results of all the teams participating in the Medico 2020 challenge in Figure~\ref{fig:medico2020results}. We can see that none of the teams came close to the ground truth mask. Achieving a balance between these metrics is crucial for developing an efficient polyp segmentation model.

\subsection{MedAI 2021 challenge results}

\begin{figure*} [!t]
    \centering
    \includegraphics[width =0.8\textwidth]{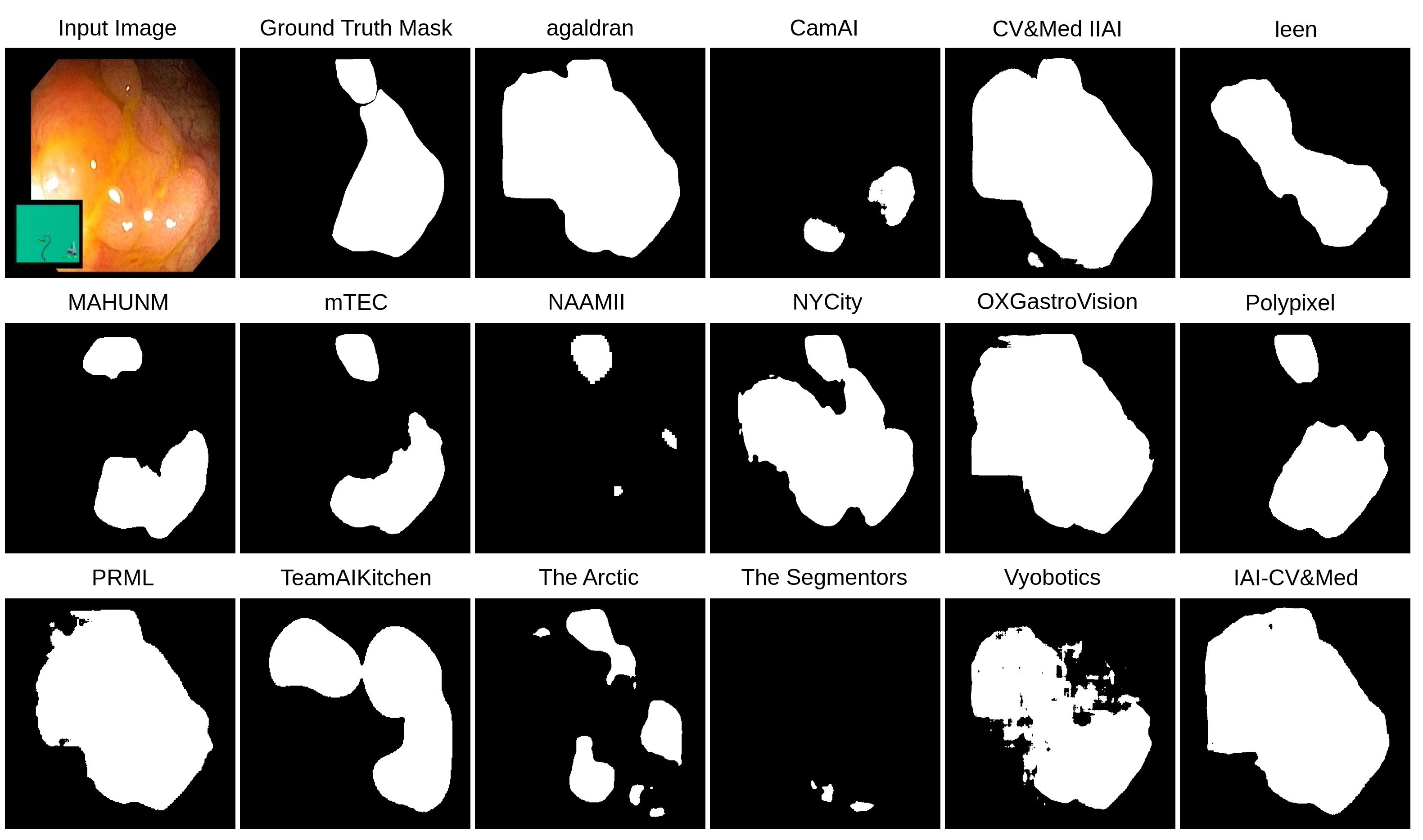}
    \caption{Qualitative results of all the methods participating in polyps segmentation challenge in MedAI2021.}
    \label{fig:medaipolypchallenge2011}
\end{figure*}

\begin{figure*}[!t]
    \centering
    \includegraphics[width =0.8\textwidth]{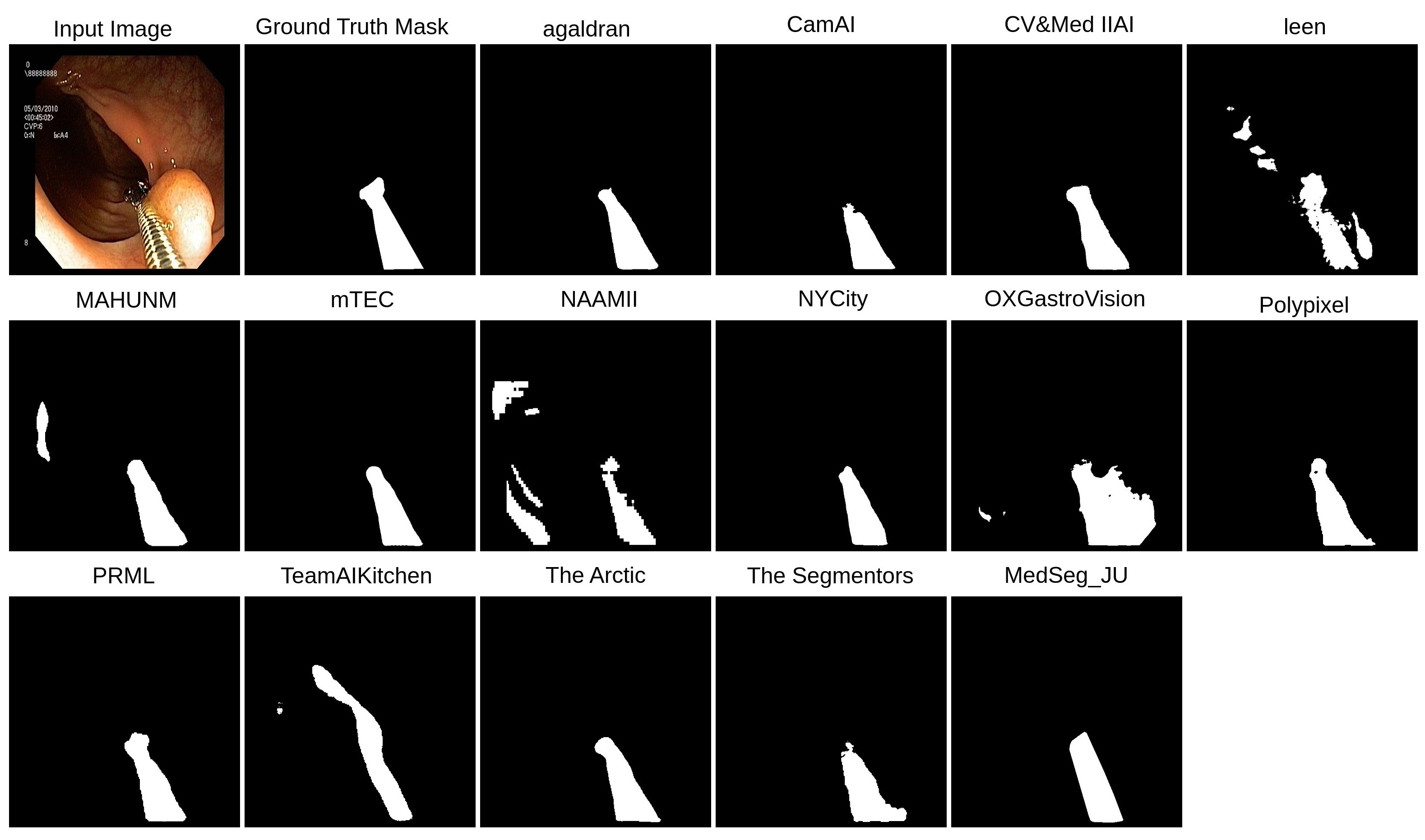}
    \caption{Qualitative results of all the methods participating in surgical instrument segmentation challenge in MedAI2021.}
    \label{fig:medAI_ins_2021}
\end{figure*}

\subsubsection{{Polyp Segmentation Task}}

In Tables~\ref{table:medai}, we tabulated the evaluation results of all the participating teams in MedAI 2021 for  polyp segmentation task. From Table, it can be observed that team ``agaldran” outperforms other teams in the polyp segmentation task with mIoU of 0.8522, {and}  DSC of 0.8965. Team {``CV\&Med IIAI”} also showed good performance and was ranked $2^{nd}$ in the polyp segmentation task with {a mIoU of 0.8484}, a very small difference from the best-performing team. In Figure~\ref{fig:medaipolypchallenge2011}, we present the qualitative results of the participating teams for the polyp segmentation task of MedAI~2021. None of the methods performed well on this challenging image, emphasizing the need for more robust polyp segmentation methods. However, in the overall test set, the predicted segmentation masks from most of the team performed well on regular polyps (see Supplementary materials). Overall, the qualitative masks produced by teams ``agaldran”  and {``CV\&Med IIAI”} were better as compared to the other teams. 





\begin{table}[!t]
    \centering
    \caption{{Performance evaluation for the participating teams for the polyp segmentation task in MedAI 2021 Challenge. $\uparrow$ {indicates a higher value is better.}}}\label{table:medai}
    \begin{tabular}{ l l l l l l l }
        \toprule
        Team Name  		& {\textbf{mIoU}} $\uparrow$  & DSC $\uparrow$ & Recall $\uparrow$ & Precision $\uparrow$ \\ \midrule
       
        agaldran   & \textbf{0.8522} &	{0.8965}  &	0.9009	& {0.9242} \\
        
        CV\&Med IIAI	 &	{0.8484}	& \textbf{0.8993}	& {0.9186}	& 0.9100\\
        
        NYCity	& 0.8418 &	0.8885		&0.8794&	\textbf{0.9319} \\
        IIAI-CV\&Med		& 0.8361	& 0.8927		& \textbf{0.9195}	& 0.8963 \\
        
        mTEC	 &0.8334	&0.8892	&0.9010	&0.9096 \\
        PRML	 &	0.8116&	0.8669	&0.8852	&0.8922 \\
        CamAI		&0.8083&	0.8701		&0.8702	&0.9052 \\
        The Arctic &0.8022	&0.8533		&0.8604&	0.8821 \\
        Polypixel		&0.7997	&0.8567		&0.8868	&0.8659 \\
        MAHUNM 	&0.7495&	0.8189	&0.8397	&0.8568 \\
        OXGastroVision	 &	0.7334&	0.7966&	0.8158&	0.8374 \\
        Vyobotics  & 0.7220	&0.7967	&0.8214&	0.8359 \\
        NAAMII 	& 0.6041&	0.6940&	0.7499	&0.7334 \\
        leen   &0.4595&	0.5531	&0.6389&	0.5860 \\
        The Segmentors  &	0.3789&	0.4205&	0.4178&	0.4640 \\
       TeamAIKitchen  &	0.2904	&0.4100	&0.7152	&0.4910 \\
        \bottomrule
    \end{tabular}
\end{table}

\begin{table}[!htpb]
    \centering
    \caption{{Performance of participating teams for instrument segmentation task of MedAI 2021 Challenge. $\uparrow$ {indicates a higher value is better.}}}\label{table:medai2}
    \begin{tabular}{ l l l l l  }
        \toprule
        TeamName & {\textbf{mIoU}} $\uparrow$  & DSC $\uparrow$  & Recall $\uparrow$ & Precision $\uparrow$  \\ \midrule
         agaldran		& \textbf{0.9364}	&\textbf{0.9635}		&{0.9692}&	\textbf{0.9632} \\
         NYCity		&{0.9326}	& {0.9586} &	\textbf{0.9712}&	{0.9516} \\
         mTEC&	0.9245&	0.9553		&0.9687	&0.9490 \\
         PRML&0.9178&	0.9528 &	0.9687&	0.9441 \\
         IIAI-CV\&Med&	0.9148&	0.9490&		0.9612	&0.9473\\
         CV\&Med IIAI&	0.9136	&0.9512&		0.9605	&0.9500 \\
         Polypixel		&0.9114&	0.9478&	0.9591&	0.9438 \\
         CamAI&0.9085&	0.9437 &	0.9454&	0.9514 \\
        The Arctic &	0.9078&	0.9448	&	0.9735&	0.9231 \\
        OXGastroVision&	0.8692	&0.9073&	0.9236&	0.9096 \\
        MAHUNM&	0.8523&	0.9080	&	0.9535&	0.8864 \\
        MedSeg\_JU&	0.8205&	0.8632	&	0.9005&	0.8464 \\
        TeamAIKitchen	&0.7257	&0.7980&	0.7955&	0.8510 \\
        leen		&0.6991&	0.7845 &	0.7963&	0.8232 \\
        NAAMII	&0.6857	&0.7741 &	0.8321	&0.7669 \\
        The Segmentors&	0.3668	&0.3971&	0.3985&	0.4040 \\
        \bottomrule
    \end{tabular}
\end{table}

\subsubsection{{Instrument Segmentation Task}}

From Table~\ref{table:medai2}, it can be observed that the same team, ``agaldran” also outperforms other participating teams in the instrument segmentation task with a high mIoU of 0.9364 and DSC of 0.9635.  Team ``NYCity” was ranked $2^{nd}$ in this task with a mIoU of 0.9326 and DSC of 0.9586. However, Team ``NYCity” obtained the highest recall of 0.9712, which signifies it has low false negative (FN) regions in the predicted segmentation mask compared to team ``agaldran”. Another interesting observation is the team ``agaldran” also achieved higher metric values for the instrument segmentation task as compared to the polyp segmentation task, as instrument segmentation is relatively easier than polyp extraction due to the greater variability of the latter regarding color and appearance. In Figure~\ref{fig:medAI_ins_2021}, we also present the qualitative results of the research teams who participated in the instrument segmentation challenge of MedAI2021. From the qualitative results, it can be observed that the ground truth prediction made by team ``agaldran” is also superior to the other team.  Therefore, it can be concluded from the obtained evaluation metrics for the two tasks that team ``agaldran” proposed a more robust algorithm and was accurately able to segment polyp and instrument at high accuracy.

\begin{table*} [!t]
\caption{Evaluation of the \textbf{`Transparency tasks'} for MedAI 2021 Challenge. For this task, a team of experts accessed the submission based on several criteria and provided a score based on the availability and quality of the source code (for e.g., open access, public availability, and documentation for reproducibility), model evaluation (for e.g., failure analysis, ablation study, explainability, and metrics used) and qualitative evaluation from clinical experts (e.g., usefulness and understandability of the results). {Here, `0' refers to no submissions for the transparency task. Doctor evaluation was only calculated for the team which manuscript were accepted. }}
\label{tab:transscore}
\centering
\resizebox{\linewidth}{!}{
\begin{tabular}{|l|ccc|cccc|ll|c|}
\hline
\multicolumn{1}{|c|}{}  & \multicolumn{3}{c}{\textbf{Open Source}}  & \multicolumn{4}{|c}{\textbf{Model Evaluation}}  & \multicolumn{2}{|c|}{\textbf{Doctor Evaluation}}  & \\ \cline{2-10}
\multicolumn{1}{|c|}{\multirow{-2}{*}{\textbf{Team Name}}} & \multicolumn{1}{c|}{\textbf{\begin{tabular}[c]{@{}l@{}}Publicly \\ available \\ (0 or 1) \end{tabular}}} & \multicolumn{1}{c|}{\textbf{\begin{tabular}[c]{@{}l@{}}Code \\ Quality \\ (0-3) \end{tabular}}} & \textbf{\begin{tabular}[c]{@{}l@{}}Readme \\ (0-3)\end{tabular}} & \multicolumn{1}{c|}{\textbf{\begin{tabular}[c]{@{}l@{}}Failure \\ Analysis \\ (0-3)\end{tabular}}} & \multicolumn{1}{c|}{\textbf{\begin{tabular}[c]{@{}l@{}}Ablation \\ Study \\ (0-3)\end{tabular}}} & \multicolumn{1}{c|}{\textbf{\begin{tabular}[c]{@{}l@{}}Explainability \\ (0-3) \end{tabular}}} & \textbf{\begin{tabular}[c]{@{}l@{}}Metrics \\ Used \\ (0 or 1)\end{tabular}} & \multicolumn{1}{c|}{\textbf{\begin{tabular}[c]{@{}l@{}}Usefulness \\ (0-3)\end{tabular}}} & \multicolumn{1}{c|}{\textbf{\begin{tabular}[c]{@{}l@{}}Understandable \\ (0-5)\end{tabular}}} & \multirow{-2}{*}{\textbf{Final Score}} \\ \hline

agaldran   & \multicolumn{1}{c|}{1} & \multicolumn{1}{c|}{2} & 3 & \multicolumn{1}{c|}{3}    & \multicolumn{1}{c|}{3}  & \multicolumn{1}{c|}{3}   & 1 & \multicolumn{1}{c|}{2}  & \multicolumn{1}{c|}{3}  & \cellcolor[HTML]{57BB8A}21 \\ \hline

mTEC  & \multicolumn{1}{c|}{1}   & \multicolumn{1}{c|}{1}  & 3  & \multicolumn{1}{c|}{3}   & \multicolumn{1}{c|}{1} & \multicolumn{1}{c|}{0}  & 1   & \multicolumn{1}{c|}{3} & \multicolumn{1}{c|}{4}  & \cellcolor[HTML]{7ECBA5}17 \\ \hline

CamAI  & \multicolumn{1}{c|}{1} & \multicolumn{1}{c|}{1}  & 1   & \multicolumn{1}{c|}{2} & \multicolumn{1}{c|}{1}   & \multicolumn{1}{c|}{2}   & 1  & \multicolumn{1}{c|}{2}  & \multicolumn{1}{c|}{5}  & \cellcolor[HTML]{87CFAC}16 \\ \hline

The Arctic  & \multicolumn{1}{c|}{1} & \multicolumn{1}{c|}{2}  & 1  & \multicolumn{1}{c|}{1} & \multicolumn{1}{c|}{0}  & \multicolumn{1}{c|}{3}   & 1   & \multicolumn{1}{c|}{1}  & \multicolumn{1}{c|}{3}  & \cellcolor[HTML]{A4DBC0}13 \\ \hline

IIAI-CV\&Med   & \multicolumn{1}{c|}{1}  & \multicolumn{1}{c|}{1}  & 2  & \multicolumn{1}{c|}{0} & \multicolumn{1}{c|}{0}  & \multicolumn{1}{c|}{0} & 1     & \multicolumn{1}{c|}{1}                  & \multicolumn{1}{c|}{4}   & \cellcolor[HTML]{C1E6D4}10 \\ \hline

Polypixel  & \multicolumn{1}{c|}{1}  & \multicolumn{1}{c|}{1}  & 2  & \multicolumn{1}{c|}{0}  & \multicolumn{1}{c|}{0}  & \multicolumn{1}{c|}{0}  & 1 & \multicolumn{1}{c|}{0}   &\multicolumn{1}{c|}{0}  & \cellcolor[HTML]{F1FAF5}5   \\ \hline

leen  & \multicolumn{1}{c|}{0}  & \multicolumn{1}{c|}{1}  & 0  & \multicolumn{1}{c|}{0}  & \multicolumn{1}{c|}{0}   & \multicolumn{1}{c|}{2}  & 1  &\multicolumn{1}{c|}{0} &\multicolumn{1}{c|}{0} & \cellcolor[HTML]{FBFEFC}4  \\ \hline

MAHUNM  & \multicolumn{1}{c|}{1} & \multicolumn{1}{c|}{1} & 0  & \multicolumn{1}{c|}{0} & \multicolumn{1}{c|}{0} & \multicolumn{1}{c|}{0} & 1 & \multicolumn{1}{c|}{0}  &\multicolumn{1}{c|}{0}   & \cellcolor[HTML]{FBECEB}3  \\ \hline


OXGastroVision  & \multicolumn{1}{c|}{0} & \multicolumn{1}{c|}{2} & 0 & \multicolumn{1}{c|}{0} & \multicolumn{1}{c|}{0} & \multicolumn{1}{c|}{0}  & 1  & \multicolumn{1}{c|}{0} &    \multicolumn{1}{c|}{0}                  & \cellcolor[HTML]{FBECEB}3                                      \\ \hline

CV\&Med IIAI  & \multicolumn{1}{c|}{0}    & \multicolumn{1}{c|}{1}  & 0  & \multicolumn{1}{c|}{1}  & \multicolumn{1}{c|}{0}  & \multicolumn{1}{c|}{0} & 1  & \multicolumn{1}{c|}{0}  & \multicolumn{1}{c|}{0}          & \cellcolor[HTML]{FBECEB}3  \\ \hline

PRML  & \multicolumn{1}{c|}{0} & \multicolumn{1}{c|}{1}  & 0   & \multicolumn{1}{c|}{0}  & \multicolumn{1}{c|}{0}  & \multicolumn{1}{c|}{0} & 1 & \multicolumn{1}{c|}{0} &\multicolumn{1}{c|}{0}   & \cellcolor[HTML]{F4C6C3}2  \\ \hline

TeamAIKitchen  & \multicolumn{1}{c|}{0} & \multicolumn{1}{c|}{1}  & 0  & \multicolumn{1}{c|}{0} & \multicolumn{1}{c|}{0} & \multicolumn{1}{c|}{0} & 1 & \multicolumn{1}{c|}{0}  &  \multicolumn{1}{c|}{0}   & \cellcolor[HTML]{F4C6C3}2  \\ \hline

The Segmentors   & \multicolumn{1}{c|}{0}    & \multicolumn{1}{c|}{0}  & 0   & \multicolumn{1}{c|}{0}  & \multicolumn{1}{c|}{0} & \multicolumn{1}{c|}{0}    & 1    &\multicolumn{1}{c|}{0}       &\multicolumn{1}{c|}{0}   & \cellcolor[HTML]{FBFEFC}1      \\ \hline

NYCity   & \multicolumn{1}{c|}{0}   & \multicolumn{1}{c|}{0}  & 0 & \multicolumn{1}{c|}{0}  & \multicolumn{1}{c|}{0}  & \multicolumn{1}{c|}{0}  & 1  & \multicolumn{1}{c|}{0}  &         \multicolumn{1}{c|}{0}   & \cellcolor[HTML]{EDA19B}1  \\ \hline

\hline
\end{tabular}
}

\end{table*}

\begin{figure} [!t]
    \centering
    \includegraphics[width =\columnwidth]{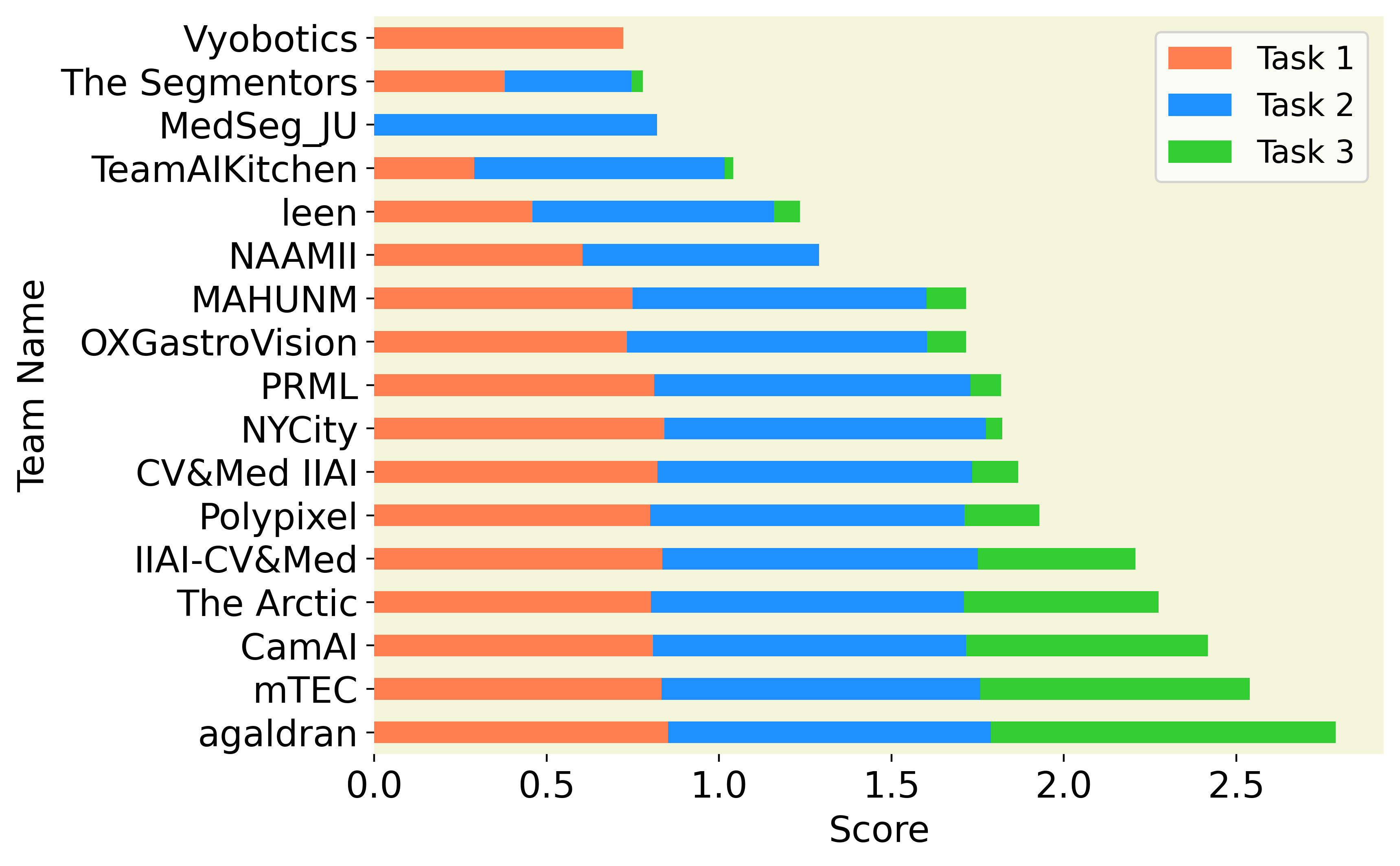}
    \caption{Task-wise scores achieved by participating teams of MedAI 2021 challenge. Team rankings are decided on the basis of overall scores in all three tasks. {Here, we plot the mIoU of Task1 and Task 2, and we have normalized the transparency score to calculate the overall score.}}
    \label{fig:medaipolypchallenge2021}
\end{figure}

\begin{figure*}[p]
     \centering
     
     \begin{subfigure}[b]{0.35\textwidth}
         \centering
         \includegraphics[width=\textwidth]{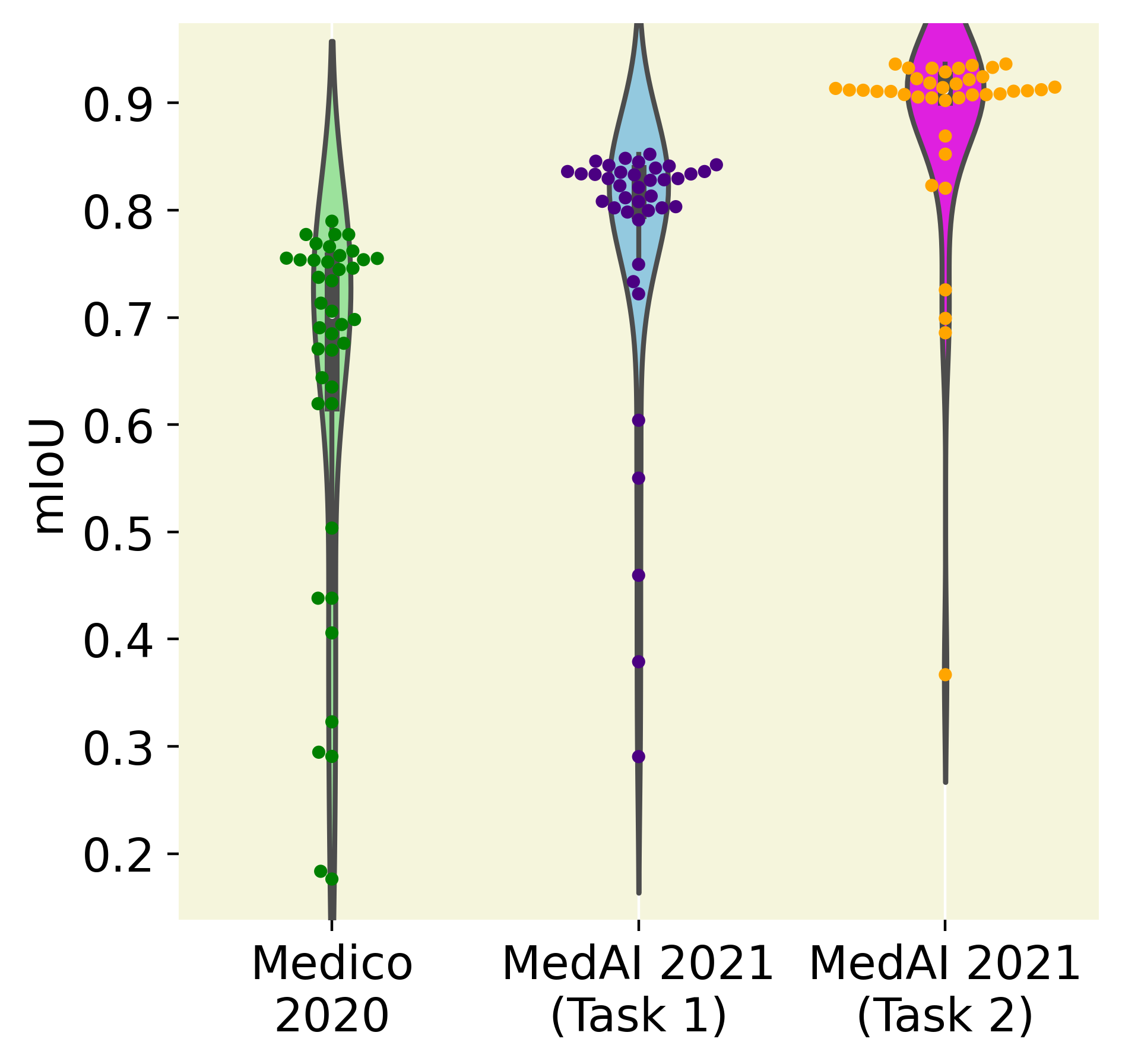}
         \caption{}
         \label{fig:avg_score}
     \end{subfigure}

     \begin{subfigure}[b]{0.95\textwidth}
         \centering
         \includegraphics[width=\textwidth]{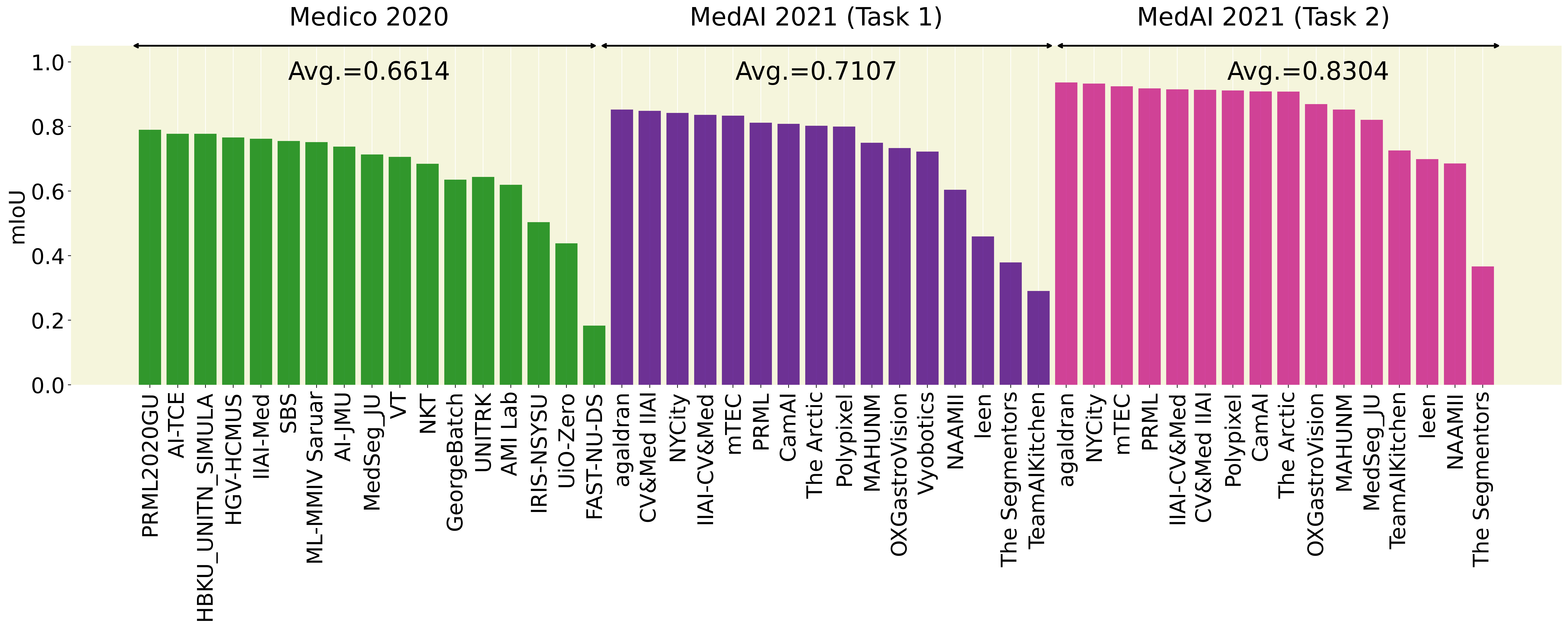}
         \caption{}
         \label{fig:avg_graph}
     \end{subfigure}

      \begin{subfigure}[b]{0.95\textwidth}
         \centering
         \includegraphics[width=\textwidth]{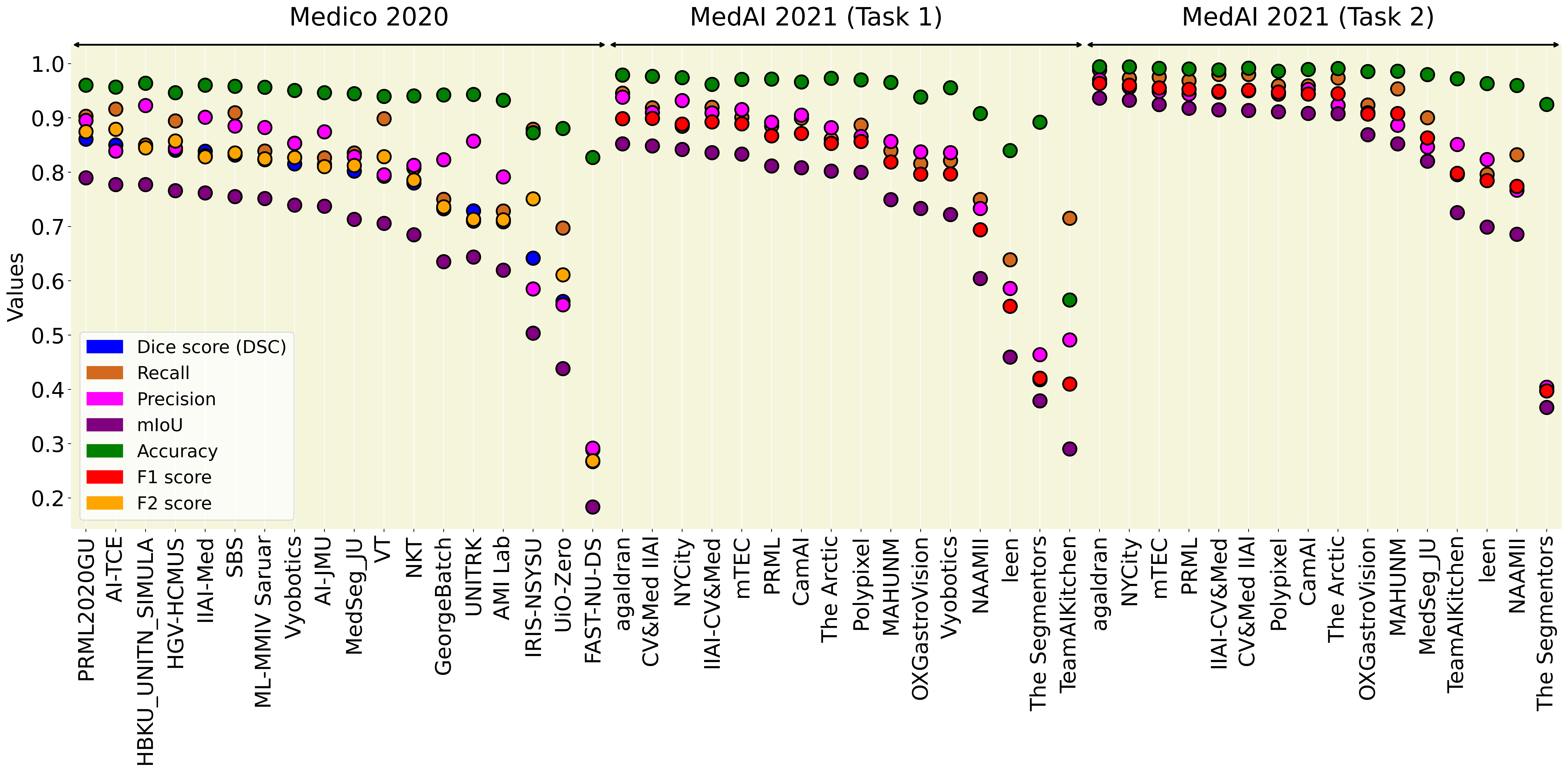}
         \caption{}
         \label{fig:comparison_graph}
     \end{subfigure}
    
   \caption{(a) Violin plots with overlaid swarm plots depicting statistics of submissions received for different tasks for the two challenges, (b) mIoU score comparison of different teams in three tasks of Medico 2020 (polyp segmentation) and MedAI 2021 (Task 1: polyp segmentation and Task 2: instrument segmentation), and (c) Strip plots for all segmentation metrics (Dice score, recall, precision, mIoU score, accuracy, F1 score, and F2 score) reported by different teams in both challenges for all test data samples.}
\end{figure*}


\subsubsection{{Transparency Task}}
We present the transparency results in Table~\ref{tab:transscore}. Team ``agaldran'' outperformed other competitors with a final score of 21 out of 25. Similarly,  ``mTEC'' obtained a score of 17 out of 25 and was ranked $2^{nd}$. Likewise, team ``CamAI'' obtained a score of 16 out of 25 and was ranked third in the transparency task. There were also efforts from teams such as ``The Arctic'', which obtained a score of 13, and ``IIAI-CV\&Med'', which obtained a score of 10. These scores show their effort to provide a transparent solution to the polyp and instrument segmentation tasks. We provide the final ranking and task-wise scores in Figure~\ref{fig:medaipolypchallenge2021}. Notably, team \textit{\textbf{``agaldran''}} outperformed others in all three tasks and overall challenge and emerged as the winner of the MedAI 2021 challenge. Overall, \textit{\textbf{``mTec''}} secured the second position. Following closely behind, \textit{\textbf{``CamAI''}} showcased the third-best solution. {The overall rank was computed by combining the mIoU scores of polyp and instrument segmentation tasks and the Transparency score.}

Figure~\ref{fig:avg_score} illustrates the plot of {mIoU} reported by each team in their submissions in the two challenges with three different tasks. It can be observed that the \textit{polyp segmentation task} from 2020 to 2021 gained improvement with a larger number of submissions achieving a {mIoU of more than 0.80 and the best-performing team with a mIoU of around 0.85}. Similar progress can be observed in Figure~\ref{fig:avg_graph} where an overall {mIoU increased by 4.93\%  when an average score is computed over all participating teams' individual best mIoU in the 2021 polyp segmentation challenge.} We further compared all segmentation metrics, including DSC, recall, precision, mIoU score, accuracy, and F2 score, as shown in Figure~\ref{fig:comparison_graph}.  Notably, the different evaluation metrics scores are consistent with instrument segmentation tasks in the MedAI challenge. However, there is a high variation in the mIoU between the different teams in the polyp segmentation tasks of Medico 2020 and MedAI 2021 challenges.

These values pertain to the best score corresponding to a particular metric the individual team obtained in different executions. It is to be noted that each team was given the opportunity to submit five different submissions, and the best results for the best submission are reported in the Tables here. From here, it can be observed that most teams in the MedAI 2021 challenge reported overall high scores in terms of various segmentation metrics when compared to Medico 2020 outcomes, thus highlighting the improved performance trends in automated systems over time. Furthermore, it can also be visualized that unlike the high variations shown by teams' scores in the polyp segmentation task, better performance and smaller deviations in scores are reported in the instrument segmentation task. The high variations in the polyp segmentation results also show that polyp segmentation is more challenging because of the presence of variations in the size, structure and appearance of the polyps, and the presence of the artifacts and lighting conditions deteriorate the algorithm's performance.


\section{Discussions}
The rapid advancement in the AI-based techniques that support CADe and CADx systems has resulted in the introduction of numerous algorithms in the domain of medical image analysis, including colonoscopy. To assess the performance of these algorithms, it is important to benchmark on the particular set of datasets. It enables the comparison and analysis of different techniques and assists in identifying challenging cases that need to be targeted using improved methodologies. This also includes cases that are misled by the presence of artifacts and occlusion by surgical instruments~\citep{ali2020objective}. Besides developing and analyzing AI-based algorithms, it is crucial to include explainability and interpretability to infuse trust and reliance during the adoption of automated systems in clinical settings. Therefore, the challenges discussed in this paper not only focus on lesion and instrument segmentation but also emphasize the importance of transparency in medical image analysis. This section covers the findings, { limitations, analysis of failing cases, trust, safety and interpretability of the methods, future steps and strategies covering both challenges}, Medico 2020 and {MedAI 2021}.

\subsection{Medico 2020 challenge methods} 
 Most of the methods reported in the Medico 2020 challenge focus on encoder-decoder architecture {(for example, U-Net, ResUNet++, PraNet, Efficient UNet, etc)}. Other networks used include {conditional GAN and Faster R-CNN}. The overview of the methods is provided in Table~\ref{table:challenge_summary2020}. For more detailed architectural information, we have also included the backbone and algorithm used by each team. Further, we also report the nature of the algorithm and the choice basis of evaluation, such as mIoU, DSC or FPS. Additionally, we provide information about the augmentation and hyperparameters, such as loss function and optimizers. {It is noteworthy that all the top three teams ``PRML2020GU", ``HBKU\_UNITN\_SIMULA" and ``AI-TCE"  used the encoder-decoder architecture}. Out of 17 participating teams, only three teams adopted some other architectures. Comparative analysis shows that the highest-scoring encoder-decoder network outperforms the GAN-based approach by a significant margin of 0.3517 in mIoU and 0.2989 in DSC score. Similarly, compared to the R-CNN-inspired networks (team ``IRIS-NSYSU''), the best approach (team ``PRML2020GU'') achieves an improvement of 0.2863 in mIoU score and 0.2191 DSC score.

Medico 2020 challenges provide valuable insight and trends for the {polyp segmentation and biomedical image analysis challenges. Most deep learning frameworks submitted for the challenge used the Adam optimizer to optimize their network. However, a handful of teams used other optimizers, such as SGD or RMSProp. Additionally, most of the teams used data augmentation to boost the number of training samples prior to training their frameworks to improve the performance of their architecture. There have been different preferences in loss function where most of the team used ``BCE + DSC loss", ``binary cross-entropy,"  IoU loss, etc. However, from the results of the top three teams, it can be concluded that ``BCE + DSC loss" is best for this dataset. Similarly, in terms of the backbone for the model architecture, the EfficientNet variant (selected by PRML2020GU) or EfficientNetB4 (selected by AI-TCE) were most favorable.}


\subsection{MedAI 2021 challenge methods}
The summary of the different approaches adopted by the participating teams of the MedAI2021 Challenge is presented in Table~\ref{table:challenge_summary2021}. To provide a brief overview of the general techniques adopted by the different teams, they can be categorized based on the nature of the approach followed, such as ensemble models, encoder-decoder based architectures, CNN, and hybrid CNN models. Almost all the teams presented the same model for both the tasks proposed in the challenge. Most teams explored ensemble modeling, encoder-decoder networks, or a combination of both in the polyp segmentation task. Another criterion of categorization could be CNN or transformed-based approaches. It is observed that the top-ranked team ``agaldran'' utilized two encoder-decoder networks and reported a mIoU score of 0.8522. {Similarly, {``CV\&Med IIAI''} was ranked second, and Team ``NYCity" was ranked third in the polyp segmentation task with a competitive mIoU value of 0.8484 and 0.8418, respectively. Similar to the Medico 2020 polyp segmentation challenge, where GAN-based methods were adopted by teams (for example, Team ``leen") failed to perform well in this challenge for polyp and instrument segmentation tasks. It is to be noted that the winning team, ``agaldran'' used a double encoder-decoder structure with two U-Net, where they incorporated FPN and Resnext101 as the pretrained decoder. They also use SAM and Adam optimizer to optimize the model further. The other competitive team {``CV\&Med IIAI''} used the SINetv2 algorithm with PVTv2 as the backbone, and NYCity used the combination of HarDNet-85 ResNet101.}

In the MedAI2021 instrument challenge, participants mainly focused on either ensemble models or encoder-decoder networks similar to the polyp segmentation task. As the majority of the teams utilized the same model that they proposed for the polyp segmentation problem in this task, the categorization of overall methods remains the same as that of the first task described above. The top rank is secured by Team ``agaldran'', with encoder-decoder architecture, pyramid network as the decoder, and Resnext101 as the pre-trained decoder. The second-ranked model by Team ``NYCity'' is the CNN and transformer based ensemble model, which achieved only a slight difference in the scores from the leading model. {mTec was ranked third in the challenge, which used dual parallel reverse attention edge network (DPRA-EdgeNet)~\citep{bhattacharya_betz_eggert_schlaefer_2021}. The architecture used HardNet~\citep{chao2019hardnet} as the backbone.}

The challenge shows that most of the teams were reluctant to share their method (refer to Table~\ref{tab:transscore}). {From the table, it can be seen that only five teams were qualified for the doctor evaluation.} Additionally, the quality of the code submitted by most of the team was not satisfactory. Most of the participants did not put much effort into the readme file. Additionally, most teams neglected the failure analysis, ablation study and explainability in their submission. {Moreover, based on the doctor's evaluation, only the solution provided by a few teams (for example, ``agaldran'', ``mTEC''  ``CamAI'',  ``The Arctic,"  and ``IIAI-CV\&Med'') was considered useful and understandable.} 




\subsection{Analysis of the failed cases}
We have analyzed the regular and failing cases in polyp and surgical tool segmentation to highlight the limitations of the current methods so that these cases can be considered during further algorithm development. Figure~\ref{fig:medico2020results} and Figure~\ref{fig:medaipolypchallenge2011} show examples of instances where the models fail for most cases. From the results on the test dataset, it was observed that most of the algorithms failed on diminutive and flat polyps located in the left colon. These are the challenging classes in the colon and require effective detection and diagnosis system. Similarly, although most of the methods performed well on the diagnostic and therapeutic surgical tool, there were issues with the images having caps and forceps. Similarly, the performance on the challenging images for polyps and instruments (see Figures~\ref{fig:medico2020results}, \ref{fig:medaipolypchallenge2011}, ~\ref{fig:medAI_ins_2021} {\textbf{and supplementary material}}) {as algorithms could still struggle with difficult and rare cases like sessile polyps, even if they perform well on overall quantitative metrics. Therefore, investigating the cause for misclassification for such samples in the dataset and failure analysis will be critical to focus for future research. This can include generalization performance evaluation on unseen test data from different hospitals. Such investigations can reduce the chances of underperformance on rare cases.}

\subsection{Trust, safety, and interpretability of methods}
Integrating CADe or CADx in clinical settings necessitates addressing factors such as trust, safety, and interpretability to ensure its adoption. The {high variations and potential bias} in the curated datasets used to train such models and the actual scenarios in which they are adopted create a high chance of biases, impacting the generalizability of the method. Such bias ultimately makes it challenging to infuse trust while adopting CADe or CADx tools and questions the safety of patients. {To tackle this issue, we introduced a transparency task in the MedAI2021 challenge that underscores the need for interpretability, reproducibility, and explainability in medical AI research, including polyp and instrument segmentation.}

{Our initiative aimed to light the potential risk that can arise from wrong decisions based on model and algorithmic bias. Our dataset contained polyp cases with varied appearances in terms of shapes, sizes, the presence of artifacts, lightning conditions, textures, and the different numbers of polyps per image that are encountered in real-world clinical settings. Additionally, we have included frames containing surgical instruments to support the cases of occluded endoluminal elements or polyps that could arise in general. Some of the methods adopted by the participating teams include the submission of intermediate heatmaps using approaches like layer-wise relevance propagation that showed visual explanation and highlighted the model decision-making process. Team ``agaldran"  provided detailed ablation studies in support of the predictions obtained. By promoting transparency through subjective analysis and addressing potential biases, the MedAI challenge aimed to foster trust in the presented solution and ensure safety in adopting such methods in the clinic.}

\subsection{Limitation of the Medico 2020 and MedAI 2021}
In our study, we aimed to standardize the challenge of polyp and instrument segmentation by providing the same test sets and evaluation metrics to all participants. To achieve this, we introduced variable polyp cases, including polyps with different sizes, noisy frames with artifacts, blurry images, and occlusion. We also added regular frames to the test set to ensure that participants drew the ground truth manually and did not cheat. However, our study has some limitations. Although we used datasets collected from four medical centers in Norway, these images are from a single country, limiting the ethnicity variance though there is very limited differences if any in the mucosal appearance between ethnicities. Nevertheless, there is a need for a more diverse dataset that includes multiple ethnicities and countries also because the prevalence of various diseases varies between regions. Moreover, the current models should be tested on multi-center datasets to assess their generalization ability.

There was no online leaderboard in our challenge due to the Mediaeval policy. Therefore, we manually calculated the predictions submitted by each team. Each team had limitations of 5 submissions for each task, which restricted further optimization opportunities. Although we have also introduced normal findings from the GI tract to trick the participants and models, our challenge only used still frames and did not incorporate video sequence datasets. {Even when the best performing algorithms are tested on a temporal video sequence dataset, it is possible that the performance can drop.} Most of the images are only from white light imaging. Although our dataset was annotated by one annotator and checked by two gastroenterologists, there is still a possibility of bias in the labels. In the accessory instrument challenge, we had more images from the stomach class than accessory instruments such as biopsy forceps or snares due to the lack of availability of datasets. Finally, despite including diverse cases in the polyp and instrument segmentation challenge, we still had limited flat and sessile polyps, frequently missed during routine colonoscopy examinations. Incorporating multi-center data, video sequences data and addressing label biases will lead to more comprehensive and reliable evaluations of AI-based colonoscopy systems.


\subsection{Future steps and strategies}
In our study, we aimed to promote transparency and interpretability in machine learning models for the GI tract setting. However, more work is needed to understand how decisions are made and identify potential biases or errors in a quantitative manner to build trust in such systems in a clinical setting. To achieve this, we plan to test the best-performing algorithms on large-scale datasets to observe their scalability. {We will consider using more quantitative metrics, such as statistical mixed models, bootstrapping analysis and estimate confidence intervals.} Additionally, we will also include metrics such as Hausdorff distance and normalized surface distance.  

We will emphasize more transparent decision-making methods and visualize interpretability results while focusing on clinical relevance rated by expert clinicians instead of just one objective metric. To achieve this, we have already started collecting large-scale datasets and plan to build a tool if the algorithms are robust enough and verified by our gastroenterologists. {Next, we will propose a challenge to polyp video sequences analysis. We will explore the integration of state space models, such as Video Vision Mamba-based framework~\citep{yang2024vivim}, to capture the temporal information in video sequences that affect the efficiency and accuracy of segmentation tasks. It is worth noting that there has been innovation within hardware (colonoscope) for safer medical colonoscopy devices, such as developing fully flexible automated colonoscopes to offer expanded fields of view rather than 120-170° visualization, which can capture dead spots, improving the lesions' miss-rate. These scopes are currently in the final stage of development. This hardware would require high processing speed to locate potential lesions in real time for a smooth workflow. We believe these solutions from our challenge could help address the complexities with the improved hardware and improved image quality.}


\vspace{-3mm}

\section{Conclusion}
\label{section:conclusion}
Our study aimed to provide a comprehensive analysis of the methods used by participants in the Medico 2020 and MedAI 2021 competitions for different medical image analysis tasks. We designed the tasks and datasets to demonstrate that the best-performing approaches were relatively robust and efficient for automatic polyp and instrument segmentation. We evaluated the challenge based on several standard metrics. In MedAI 2021, we also used a quantitative approach, where a multi-disciplinary team, including gastroenterologists, accessed each submission and evaluated the usefulness and understandability of their results. {Through the qualitative results, we found that even the best-performing method underperforms in rare cases. This highlights the need for further investigation to understand the cause of misclassification.} During the ``performance task'' and ``algorithm efficiency'' tasks, we observed a trade-off between mIoU and inference time when tested across unseen still frames. {For the instrument segmentation challenge, we observed that almost all teams performed relatively well, as segmenting instruments is easier than polyp segmentation.} From the transparency task, we observed that more effort is required from the community to enhance the transparency of the proposed model. Overall, we also observed that several teams demonstrated the use of data augmentation and optimization techniques to improve performance on specific tasks. Our study highlights the need for multi-center dataset collection from larger and more diverse populations, including experts from various clinics worldwide. {More competitions should be held on polyp video sequences to observe the efficiency difference in still frames and video sequences.} Further research should investigate multiple polyp classes that typically fail in clinical settings, multi-center clinical trials, and the emphasis on real-time systems. Additionally, research on transparency and interpretability should be emphasized as it could help build clinically relevant and trustworthy systems.

\section*{Acknowledgment}
D. Jha is supported by the NIH funding: R01-CA246704 and R01-CA240639. V. Sharma is supported by the INSPIRE fellowship (IF190362), DST, Govt. of India. D. Bhattacharya is funded partially by the i$^3$ initiative of the Hamburg University of Technology and by the Free and Hanseatic City of Hamburg (Interdisciplinary Graduate School “Innovative Technologies in Cancer Diagnostics and Therapy”). K. Roy is thankful to DST Inspire Ph.D fellowship (IF170366).

\section*{Authors contribution}
D. Jha conceptualized, initiated, and coordinated the work. He also led the data collection, curation, and annotation processes for Medico 2020 and evaluated the Medico 2020 Challenge. S. Hicks and M.A. Riegler initiated the MedAI 2021 Challenge, organized the data collection together with D. Jha and conducted all evaluations for the challenges, organized the reviews and coordinated with all the authors. V. Sharma analyzed the results and prepared most graphs for technical validation along with N. K. Tomar. She also wrote a part of the results and discussion. D. Banik, D. Bhattacharya and  K. Roy wrote part of the introduction, related work, and participants' methods and provided subsequent feedback on the method's tables. M.A. Riegler and P. Halvorsen facilitated the data and organization for both competitions. Our gastroenterologists, T. de Lange and S. Parasa, reviewed the annotations and provided the required feedback during dataset preparation and evaluation. Challenge participants provided the method details for Medico 2020. All authors read the manuscript, provided substantial input, and agreed to the submission.

\bibliographystyle{vendor/model2-names.bst}\biboptions{authoryear}
\bibliography{refs}

\begin{thebibliography}{84}
\expandafter\ifx\csname natexlab\endcsname\relax\def\natexlab#1{#1}\fi
\providecommand{\url}[1]{\texttt{#1}}
\providecommand{\href}[2]{#2}
\providecommand{\path}[1]{#1}
\providecommand{\DOIprefix}{doi:}
\providecommand{\ArXivprefix}{arXiv:}
\providecommand{\URLprefix}{URL: }
\providecommand{\Pubmedprefix}{pmid:}
\providecommand{\doi}[1]{\href{http://dx.doi.org/#1}{\path{#1}}}
\providecommand{\Pubmed}[1]{\href{pmid:#1}{\path{#1}}}
\providecommand{\bibinfo}[2]{#2}
\ifx\xfnm\relax \def\xfnm[#1]{\unskip,\space#1}\fi
\bibitem[{Ahmed and Ali(2021)}]{ahmed2021explainable}
\bibinfo{author}{Ahmed, A.}, \bibinfo{author}{Ali, L.A.}, \bibinfo{year}{2021}.
\newblock \bibinfo{title}{Explainable medical image segmentation via generative adversarial networks and layer-wise relevance propagation}.
\newblock \bibinfo{journal}{arXiv preprint arXiv:2111.01665} .
\bibitem[{Ahmed and Ali(2020)}]{ahmed2020generative}
\bibinfo{author}{Ahmed, A.}, \bibinfo{author}{Ali, M.}, \bibinfo{year}{2020}.
\newblock \bibinfo{title}{Generative adversarial networks for automatic polyp segmentation}.
\newblock \bibinfo{journal}{arXiv preprint arXiv:2012.06771} .
\bibitem[{Alam et~al.(2020)Alam, Tomar, Thakur, Jha and Rauniyar}]{alam2020automatic}
\bibinfo{author}{Alam, S.}, \bibinfo{author}{Tomar, N.K.}, \bibinfo{author}{Thakur, A.}, \bibinfo{author}{Jha, D.}, \bibinfo{author}{Rauniyar, A.}, \bibinfo{year}{2020}.
\newblock \bibinfo{title}{Automatic polyp segmentation using u-net-resnet50}.
\newblock \bibinfo{journal}{arXiv preprint arXiv:2012.15247} .
\bibitem[{Ali et~al.(2021)Ali, Dmitrieva, Ghatwary, Bano, Polat, Temizel, Krenzer, Hekalo, Guo, Matuszewski et~al.}]{ali2021deep}
\bibinfo{author}{Ali, S.}, \bibinfo{author}{Dmitrieva, M.}, \bibinfo{author}{Ghatwary, N.}, \bibinfo{author}{Bano, S.}, \bibinfo{author}{Polat, G.}, \bibinfo{author}{Temizel, A.}, \bibinfo{author}{Krenzer, A.}, \bibinfo{author}{Hekalo, A.}, \bibinfo{author}{Guo, Y.B.}, \bibinfo{author}{Matuszewski, B.}, et~al., \bibinfo{year}{2021}.
\newblock \bibinfo{title}{Deep learning for detection and segmentation of artefact and disease instances in gastrointestinal endoscopy}.
\newblock \bibinfo{journal}{Medical Image Analysis} \bibinfo{volume}{70}, \bibinfo{pages}{102002}.
\bibitem[{Ali et~al.(2022a)Ali, Ghatwary, Jha, Isik-Polat, Polat, Yang, Li, Galdran, Ballester, Thambawita et~al.}]{ali2022assessing}
\bibinfo{author}{Ali, S.}, \bibinfo{author}{Ghatwary, N.}, \bibinfo{author}{Jha, D.}, \bibinfo{author}{Isik-Polat, E.}, \bibinfo{author}{Polat, G.}, \bibinfo{author}{Yang, C.}, \bibinfo{author}{Li, W.}, \bibinfo{author}{Galdran, A.}, \bibinfo{author}{Ballester, M.{\'A}.G.}, \bibinfo{author}{Thambawita, V.}, et~al., \bibinfo{year}{2022}a.
\newblock \bibinfo{title}{Assessing generalisability of deep learning-based polyp detection and segmentation methods through a computer vision challenge}.
\newblock \bibinfo{journal}{arXiv preprint arXiv:2202.12031} .
\bibitem[{Ali et~al.(2022b)Ali, Jha, Ghatwary, Realdon, Cannizzaro, Salem, Lamarque, Daul, Anonsen, Riegler et~al.}]{ali2021polypgen}
\bibinfo{author}{Ali, S.}, \bibinfo{author}{Jha, D.}, \bibinfo{author}{Ghatwary, N.}, \bibinfo{author}{Realdon, S.}, \bibinfo{author}{Cannizzaro, R.}, \bibinfo{author}{Salem, O.E.}, \bibinfo{author}{Lamarque, D.}, \bibinfo{author}{Daul, C.}, \bibinfo{author}{Anonsen, K.V.}, \bibinfo{author}{Riegler, M.A.}, et~al., \bibinfo{year}{2022}b.
\newblock \bibinfo{title}{A multi-centre polyp detection and segmentation dataset for generalisability assessment}.
\newblock \bibinfo{journal}{Scientific Data} .
\bibitem[{Ali and Tomar(2021)}]{ali2021iterative}
\bibinfo{author}{Ali, S.}, \bibinfo{author}{Tomar, N.K.}, \bibinfo{year}{2021}.
\newblock \bibinfo{title}{Iterative deep learning for improved segmentation of endoscopic images}.
\newblock \bibinfo{journal}{Nordic Machine Intelligence} \bibinfo{volume}{1}, \bibinfo{pages}{38--40}.
\bibitem[{Ali et~al.(2020a)}]{ali2020objective}
\bibinfo{author}{Ali, S.}, et~al., \bibinfo{year}{2020}a.
\newblock \bibinfo{title}{An objective comparison of detection and segmentation algorithms for artefacts in clinical endoscopy}.
\newblock \bibinfo{journal}{Sci. Rep} , \bibinfo{pages}{1--21}.
\bibitem[{Ali et~al.(2020b)Ali, Khan, Haider, Ahmed, Khan and Tahir}]{ali2020depth}
\bibinfo{author}{Ali, S.M.F.}, \bibinfo{author}{Khan, M.T.}, \bibinfo{author}{Haider, S.U.}, \bibinfo{author}{Ahmed, T.}, \bibinfo{author}{Khan, Z.}, \bibinfo{author}{Tahir, M.A.}, \bibinfo{year}{2020}b.
\newblock \bibinfo{title}{Depth-wise separable atrous convolution for polyps segmentation in gastro-intestinal tract.}, in: \bibinfo{booktitle}{In Proceedings of the MediaEval}, pp. \bibinfo{pages}{1--3}.
\bibitem[{Angermann et~al.(2017)Angermann, Bernal, S{\'a}nchez-Montes, Hammami, Fern{\'a}ndez-Esparrach, Dray, Romain, S{\'a}nchez and Histace}]{angermann2017towards}
\bibinfo{author}{Angermann, Q.}, \bibinfo{author}{Bernal, J.}, \bibinfo{author}{S{\'a}nchez-Montes, C.}, \bibinfo{author}{Hammami, M.}, \bibinfo{author}{Fern{\'a}ndez-Esparrach, G.}, \bibinfo{author}{Dray, X.}, \bibinfo{author}{Romain, O.}, \bibinfo{author}{S{\'a}nchez, F.J.}, \bibinfo{author}{Histace, A.}, \bibinfo{year}{2017}.
\newblock \bibinfo{title}{Towards real-time polyp detection in colonoscopy videos: Adapting still frame-based methodologies for video sequences analysis}, in: \bibinfo{booktitle}{In Proceedings of the Computer Assisted and Robotic Endoscopy and Clinical Image-Based Procedures: 4th International Workshop, CARE 2017, and 6th International Workshop, CLIP 2017, Held in Conjunction with MICCAI 2017, Qu{\'e}bec City, QC, Canada, September 14, 2017, Proceedings 4}, pp. \bibinfo{pages}{29--41}.
\bibitem[{Asplund et~al.(2018)Asplund, Kauppila, Mattsson and Lagergren}]{asplund2018survival}
\bibinfo{author}{Asplund, J.}, \bibinfo{author}{Kauppila, J.H.}, \bibinfo{author}{Mattsson, F.}, \bibinfo{author}{Lagergren, J.}, \bibinfo{year}{2018}.
\newblock \bibinfo{title}{Survival trends in gastric adenocarcinoma: a population-based study in sweden}.
\newblock \bibinfo{journal}{Ann. Surgi. Oncol.} \bibinfo{volume}{25}, \bibinfo{pages}{2693--2702}.
\bibitem[{Bach et~al.(2015)Bach, Binder, Montavon, Klauschen, M{\"u}ller and Samek}]{bach2015pixel}
\bibinfo{author}{Bach, S.}, \bibinfo{author}{Binder, A.}, \bibinfo{author}{Montavon, G.}, \bibinfo{author}{Klauschen, F.}, \bibinfo{author}{M{\"u}ller, K.R.}, \bibinfo{author}{Samek, W.}, \bibinfo{year}{2015}.
\newblock \bibinfo{title}{On pixel-wise explanations for non-linear classifier decisions by layer-wise relevance propagation}.
\newblock \bibinfo{journal}{PloS one} \bibinfo{volume}{10}, \bibinfo{pages}{e0130140}.
\bibitem[{Ballesteros et~al.(2017)Ballesteros, Trujillo, Mazo, Chaves and Hoyos}]{inproceedings2}
\bibinfo{author}{Ballesteros, C.}, \bibinfo{author}{Trujillo, M.}, \bibinfo{author}{Mazo, C.}, \bibinfo{author}{Chaves, D.}, \bibinfo{author}{Hoyos, J.}, \bibinfo{year}{2017}.
\newblock \bibinfo{title}{Automatic classification of non-informative frames in colonoscopy videos using texture analysis}, in: \bibinfo{booktitle}{In Proceedings of the Lecture Notes in Computer Science}, pp. \bibinfo{pages}{401--408}.
\bibitem[{Banik and Bhattacharjee(2020)}]{banik2020deep}
\bibinfo{author}{Banik, D.}, \bibinfo{author}{Bhattacharjee, D.}, \bibinfo{year}{2020}.
\newblock \bibinfo{title}{Deep conditional adversarial learning for polyp segmentation.}, in: \bibinfo{booktitle}{In Proceedings of the MediaEval}, pp. \bibinfo{pages}{1--3}.
\bibitem[{Banik et~al.(2021)Banik, Roy and Bhattacharjee}]{Banik2021}
\bibinfo{author}{Banik, D.}, \bibinfo{author}{Roy, K.}, \bibinfo{author}{Bhattacharjee, D.}, \bibinfo{year}{2021}.
\newblock \bibinfo{title}{{EM-Net}: An efficient m-net for segmentation of surgical instruments in colonoscopy frames}.
\newblock \bibinfo{journal}{Nordic Machine Intelligence} \bibinfo{volume}{1}, \bibinfo{pages}{14--16}.
\bibitem[{Batchkala and Ali(2020)}]{batchkala2020real}
\bibinfo{author}{Batchkala, G.}, \bibinfo{author}{Ali, S.}, \bibinfo{year}{2020}.
\newblock \bibinfo{title}{Real-time polyp segmentation using u-net with iou loss.}, in: \bibinfo{booktitle}{In Proceedings of the MediaEval}, pp. \bibinfo{pages}{1--3}.
\bibitem[{Bernal and Aymeric(2017)}]{gianadataset2017}
\bibinfo{author}{Bernal, J.}, \bibinfo{author}{Aymeric, H.}, \bibinfo{year}{2017}.
\newblock \bibinfo{title}{{Gastrointestinal Image ANAlysis (GIANA) Angiodysplasia {D\&L} challenge}}.
\newblock \bibinfo{howpublished}{\url{https://endovissub2017-giana.grand-challenge.org/home/}}.
\newblock \bibinfo{note}{Accessed: 2017-11-20}.
\bibitem[{Bernal et~al.(2012)Bernal, S{\'a}nchez and Vilarino}]{bernal2012towards}
\bibinfo{author}{Bernal, J.}, \bibinfo{author}{S{\'a}nchez, J.}, \bibinfo{author}{Vilarino, F.}, \bibinfo{year}{2012}.
\newblock \bibinfo{title}{Towards automatic polyp detection with a polyp appearance model}.
\newblock \bibinfo{journal}{Pattern Recognition} \bibinfo{volume}{45}, \bibinfo{pages}{3166--3182}.
\bibitem[{Bernal et~al.(2017)Bernal, Tajkbaksh, Sanchez, Matuszewski, Chen, Yu, Angermann, Romain, Rustad, Balasingham et~al.}]{bernal2017comparative}
\bibinfo{author}{Bernal, J.}, \bibinfo{author}{Tajkbaksh, N.}, \bibinfo{author}{Sanchez, F.J.}, \bibinfo{author}{Matuszewski, B.J.}, \bibinfo{author}{Chen, H.}, \bibinfo{author}{Yu, L.}, \bibinfo{author}{Angermann, Q.}, \bibinfo{author}{Romain, O.}, \bibinfo{author}{Rustad, B.}, \bibinfo{author}{Balasingham, I.}, et~al., \bibinfo{year}{2017}.
\newblock \bibinfo{title}{Comparative validation of polyp detection methods in video colonoscopy: results from the miccai 2015 endoscopic vision challenge}.
\newblock \bibinfo{journal}{IEEE transactions on medical imaging} \bibinfo{volume}{36}, \bibinfo{pages}{1231--1249}.
\bibitem[{Bernal et~al.(2018)}]{bernal2018polyp}
\bibinfo{author}{Bernal, J.}, et~al., \bibinfo{year}{2018}.
\newblock \bibinfo{title}{Polyp detection benchmark in colonoscopy videos using gtcreator: A novel fully configurable tool for easy and fast annotation of image databases}, in: \bibinfo{booktitle}{Proceedings of the Comput. Assist. Radiol. Surg. (CARS)}.
\bibitem[{Bhattacharya et~al.(2021a)Bhattacharya, Betz, Eggert and Schlaefer}]{bhattacharya_betz_eggert_schlaefer_2021}
\bibinfo{author}{Bhattacharya, D.}, \bibinfo{author}{Betz, C.}, \bibinfo{author}{Eggert, D.}, \bibinfo{author}{Schlaefer, A.}, \bibinfo{year}{2021}a.
\newblock \bibinfo{title}{Dual parallel reverse attention edge network : {DPRA-EdgeNet}}.
\newblock \bibinfo{journal}{Nordic Machine Intelligence} \bibinfo{volume}{1}, \bibinfo{pages}{8–10}.
\bibitem[{Bhattacharya et~al.(2021b)Bhattacharya, Betz, Eggert and Schlaefer}]{bhattacharya2021self}
\bibinfo{author}{Bhattacharya, D.}, \bibinfo{author}{Betz, C.}, \bibinfo{author}{Eggert, D.}, \bibinfo{author}{Schlaefer, A.}, \bibinfo{year}{2021}b.
\newblock \bibinfo{title}{Self-supervised u-net for segmenting flat and sessile polyps}.
\newblock \bibinfo{journal}{arXiv preprint arXiv:2110.08776} .
\bibitem[{Cai and Vasconcelos(2019)}]{cai2019cascade}
\bibinfo{author}{Cai, Z.}, \bibinfo{author}{Vasconcelos, N.}, \bibinfo{year}{2019}.
\newblock \bibinfo{title}{Cascade r-cnn: High quality object detection and instance segmentation}.
\newblock \bibinfo{journal}{IEEE transactions on pattern analysis and machine intelligence} \bibinfo{volume}{43}, \bibinfo{pages}{1483--1498}.
\bibitem[{Chao et~al.(2019)Chao, Kao, Ruan, Huang and Lin}]{chao2019hardnet}
\bibinfo{author}{Chao, P.}, \bibinfo{author}{Kao, C.Y.}, \bibinfo{author}{Ruan, Y.S.}, \bibinfo{author}{Huang, C.H.}, \bibinfo{author}{Lin, Y.L.}, \bibinfo{year}{2019}.
\newblock \bibinfo{title}{Hardnet: A low memory traffic network}, in: \bibinfo{booktitle}{Proceedings of the IEEE/CVF international conference on computer vision}, pp. \bibinfo{pages}{3552--3561}.
\bibitem[{Chen et~al.(2018)Chen, Zhu, Papandreou, Schroff and Adam}]{chen2018encoder}
\bibinfo{author}{Chen, L.C.}, \bibinfo{author}{Zhu, Y.}, \bibinfo{author}{Papandreou, G.}, \bibinfo{author}{Schroff, F.}, \bibinfo{author}{Adam, H.}, \bibinfo{year}{2018}.
\newblock \bibinfo{title}{Encoder-decoder with atrous separable convolution for semantic image segmentation}, in: \bibinfo{booktitle}{Proceedings of the European conference on computer vision (ECCV)}, pp. \bibinfo{pages}{801--818}.
\bibitem[{Chen et~al.(2021)Chen, Kuo, Fang and Wang}]{Chen2021}
\bibinfo{author}{Chen, Y.H.}, \bibinfo{author}{Kuo, P.H.}, \bibinfo{author}{Fang, Y.Z.}, \bibinfo{author}{Wang, W.L.}, \bibinfo{year}{2021}.
\newblock \bibinfo{title}{More birds in the hand -medical image segmentation using a multi-model ensemble framework}.
\newblock \bibinfo{journal}{Nordic Machine Intelligence} \bibinfo{volume}{1}, \bibinfo{pages}{23--25}.
\bibitem[{Chou(2021)}]{Chou2021}
\bibinfo{author}{Chou, Y.}, \bibinfo{year}{2021}.
\newblock \bibinfo{title}{Automatic polyp and instrument segmentation in medai-2021}.
\newblock \bibinfo{journal}{Nordic Machine Intelligence} \bibinfo{volume}{1}, \bibinfo{pages}{17--19}.
\bibitem[{Dong et~al.(2021a)Dong, Wang, Fan, Li, Fu and Shao}]{dong2021polyp}
\bibinfo{author}{Dong, B.}, \bibinfo{author}{Wang, W.}, \bibinfo{author}{Fan, D.P.}, \bibinfo{author}{Li, J.}, \bibinfo{author}{Fu, H.}, \bibinfo{author}{Shao, L.}, \bibinfo{year}{2021}a.
\newblock \bibinfo{title}{Polyp-pvt: Polyp segmentation with pyramid vision transformers}.
\newblock \bibinfo{journal}{arXiv preprint arXiv:2108.06932} .
\bibitem[{Dong et~al.(2021b)Dong, Wang and Li}]{Dong2021}
\bibinfo{author}{Dong, B.}, \bibinfo{author}{Wang, W.}, \bibinfo{author}{Li, J.}, \bibinfo{year}{2021}b.
\newblock \bibinfo{title}{Transformer based multi-model fusion for medical image segmentation}.
\newblock \bibinfo{journal}{Nordic Machine Intelligence} \bibinfo{volume}{1}, \bibinfo{pages}{50--52}.
\bibitem[{Fan et~al.(2021)Fan, Ji, Cheng and Shao}]{fan2021concealed}
\bibinfo{author}{Fan, D.P.}, \bibinfo{author}{Ji, G.P.}, \bibinfo{author}{Cheng, M.M.}, \bibinfo{author}{Shao, L.}, \bibinfo{year}{2021}.
\newblock \bibinfo{title}{Concealed object detection}.
\newblock \bibinfo{journal}{IEEE Transactions on Pattern Analysis and Machine Intelligence} \bibinfo{volume}{44}, \bibinfo{pages}{6024--6042}.
\bibitem[{Fan et~al.(2020)Fan, Ji, Zhou, Chen, Fu, Shen and Shao}]{fan2020pranet}
\bibinfo{author}{Fan, D.P.}, \bibinfo{author}{Ji, G.P.}, \bibinfo{author}{Zhou, T.}, \bibinfo{author}{Chen, G.}, \bibinfo{author}{Fu, H.}, \bibinfo{author}{Shen, J.}, \bibinfo{author}{Shao, L.}, \bibinfo{year}{2020}.
\newblock \bibinfo{title}{Pranet: Parallel reverse attention network for polyp segmentation}, in: \bibinfo{booktitle}{Proceedings of the Medical Image Computing and Computer Assisted Intervention--MICCAI 2020: 23rd International Conference, Lima, Peru, October 4--8, 2020, Proceedings, Part VI 23}, pp. \bibinfo{pages}{263--273}.
\bibitem[{Foret et~al.(2020)Foret, Kleiner, Mobahi and Neyshabur}]{foret2020sharpness}
\bibinfo{author}{Foret, P.}, \bibinfo{author}{Kleiner, A.}, \bibinfo{author}{Mobahi, H.}, \bibinfo{author}{Neyshabur, B.}, \bibinfo{year}{2020}.
\newblock \bibinfo{title}{Sharpness-aware minimization for efficiently improving generalization}.
\newblock \bibinfo{journal}{arXiv preprint arXiv:2010.01412} .
\bibitem[{Galdran(2021)}]{galdran2021polyp}
\bibinfo{author}{Galdran, A.}, \bibinfo{year}{2021}.
\newblock \bibinfo{title}{Polyp and surgical instrument segmentation with double encoder-decoder networks}.
\newblock \bibinfo{journal}{Nordic Machine Intelligence} \bibinfo{volume}{1}, \bibinfo{pages}{5--7}.
\bibitem[{Haithami et~al.(2021)Haithami, Ahmed, Liao and Jalab}]{Haithami2021}
\bibinfo{author}{Haithami, M.}, \bibinfo{author}{Ahmed, A.}, \bibinfo{author}{Liao, I.Y.}, \bibinfo{author}{Jalab, H.}, \bibinfo{year}{2021}.
\newblock \bibinfo{title}{Employing {GRU} to combine feature maps in {DeeplabV}3 for a better segmentation model}.
\newblock \bibinfo{journal}{Nordic Machine Intelligence} \bibinfo{volume}{1}, \bibinfo{pages}{29--31}.
\bibitem[{He et~al.(2017)He, Gkioxari, Doll{\'a}r and Girshick}]{he2017mask}
\bibinfo{author}{He, K.}, \bibinfo{author}{Gkioxari, G.}, \bibinfo{author}{Doll{\'a}r, P.}, \bibinfo{author}{Girshick, R.}, \bibinfo{year}{2017}.
\newblock \bibinfo{title}{Mask r-cnn}, in: \bibinfo{booktitle}{Proceedings of the IEEE international conference on computer vision}, pp. \bibinfo{pages}{2961--2969}.
\bibitem[{He et~al.(2016)He, Zhang, Ren and Sun}]{he2016deep}
\bibinfo{author}{He, K.}, \bibinfo{author}{Zhang, X.}, \bibinfo{author}{Ren, S.}, \bibinfo{author}{Sun, J.}, \bibinfo{year}{2016}.
\newblock \bibinfo{title}{Deep residual learning for image recognition}, in: \bibinfo{booktitle}{In Proceedings of the IEEE Conf. Comput. Vis. Pattern Recognit. (CVPR)}, pp. \bibinfo{pages}{770--778}.
\bibitem[{Hicks et~al.(2021)Hicks, Jha, Thambawita, Halvorsen, Singstad, Gaur, Pettersen, Goodwin, Parasa, de~Lange and Riegler}]{medaireview}
\bibinfo{author}{Hicks, S.}, \bibinfo{author}{Jha, D.}, \bibinfo{author}{Thambawita, V.}, \bibinfo{author}{Halvorsen, P.}, \bibinfo{author}{Singstad, B.J.}, \bibinfo{author}{Gaur, S.}, \bibinfo{author}{Pettersen, K.}, \bibinfo{author}{Goodwin, M.}, \bibinfo{author}{Parasa, S.}, \bibinfo{author}{de~Lange, T.}, \bibinfo{author}{Riegler, M.}, \bibinfo{year}{2021}.
\newblock \bibinfo{title}{Medai: Transparency in medical image segmentation}.
\newblock \bibinfo{journal}{Nordic Machine Intelligence} \bibinfo{volume}{1}, \bibinfo{pages}{1--4}.
\bibitem[{Huang et~al.(2021)Huang, Wu and Lin}]{huang2021hardnet}
\bibinfo{author}{Huang, C.H.}, \bibinfo{author}{Wu, H.Y.}, \bibinfo{author}{Lin, Y.L.}, \bibinfo{year}{2021}.
\newblock \bibinfo{title}{Hardnet-mseg: A simple encoder-decoder polyp segmentation neural network that achieves over 0.9 mean dice and 86 fps}.
\newblock \bibinfo{journal}{arXiv preprint arXiv:2101.07172} .
\bibitem[{Hwang et~al.(2007a)Hwang, Oh, Tavanapong, Wong and De~Groen}]{hwang2007polyp}
\bibinfo{author}{Hwang, S.}, \bibinfo{author}{Oh, J.}, \bibinfo{author}{Tavanapong, W.}, \bibinfo{author}{Wong, J.}, \bibinfo{author}{De~Groen, P.C.}, \bibinfo{year}{2007}a.
\newblock \bibinfo{title}{Polyp detection in colonoscopy video using elliptical shape feature}, in: \bibinfo{booktitle}{In proceedings of the IEEE International Conference on Image Processing}, pp. \bibinfo{pages}{II--465}.
\bibitem[{Hwang et~al.(2007b)Hwang, Oh, Tavanapong, Wong and Groen}]{inproceedings3}
\bibinfo{author}{Hwang, S.}, \bibinfo{author}{Oh, J.}, \bibinfo{author}{Tavanapong, W.}, \bibinfo{author}{Wong, J.}, \bibinfo{author}{Groen, P.}, \bibinfo{year}{2007}b.
\newblock \bibinfo{title}{Polyp detection in colonoscopy video using elliptical shape feature}, in: \bibinfo{booktitle}{In Proceedings of the International Conference on Image Processing, ICIP}, pp. \bibinfo{pages}{II -- 465}.
\bibitem[{Jha et~al.(2021)Jha, Ali, Emanuelsen, Hicks, Thambawita, Garcia-Ceja, Riegler, de~Lange, Schmidt, Johansen et~al.}]{jha2021kvasir}
\bibinfo{author}{Jha, D.}, \bibinfo{author}{Ali, S.}, \bibinfo{author}{Emanuelsen, K.}, \bibinfo{author}{Hicks, S.A.}, \bibinfo{author}{Thambawita, V.}, \bibinfo{author}{Garcia-Ceja, E.}, \bibinfo{author}{Riegler, M.A.}, \bibinfo{author}{de~Lange, T.}, \bibinfo{author}{Schmidt, P.T.}, \bibinfo{author}{Johansen, H.D.}, et~al., \bibinfo{year}{2021}.
\newblock \bibinfo{title}{Kvasir-instrument: Diagnostic and therapeutic tool segmentation dataset in gastrointestinal endoscopy}, in: \bibinfo{booktitle}{MultiMedia Modeling: 27th International Conference, MMM 2021, Prague, Czech Republic, June 22--24, 2021, Proceedings, Part II 27}, pp. \bibinfo{pages}{218--229}.
\bibitem[{Jha et~al.(2020a)Jha, Hicks, Emanuelsen, Johansen, Johansen, de~Lange, Riegler and Halvorsen}]{jha2020medico}
\bibinfo{author}{Jha, D.}, \bibinfo{author}{Hicks, S.A.}, \bibinfo{author}{Emanuelsen, K.}, \bibinfo{author}{Johansen, H.}, \bibinfo{author}{Johansen, D.}, \bibinfo{author}{de~Lange, T.}, \bibinfo{author}{Riegler, M.A.}, \bibinfo{author}{Halvorsen, P.}, \bibinfo{year}{2020}a.
\newblock \bibinfo{title}{Medico multimedia task at mediaeval 2020: Automatic polyp segmentation}, in: \bibinfo{booktitle}{In Proceedings of the CEUR Worksh. Multim. Bench. Worksh. (MediaEval)}, pp. \bibinfo{pages}{1--2}.
\bibitem[{Jha et~al.(2019)Jha, Smedsrud, Riegler, Johansen, De~Lange, Halvorsen and Johansen}]{jha2019resunet++}
\bibinfo{author}{Jha, D.}, \bibinfo{author}{Smedsrud, P.H.}, \bibinfo{author}{Riegler, M.A.}, \bibinfo{author}{Johansen, D.}, \bibinfo{author}{De~Lange, T.}, \bibinfo{author}{Halvorsen, P.}, \bibinfo{author}{Johansen, H.D.}, \bibinfo{year}{2019}.
\newblock \bibinfo{title}{Resunet++: An advanced architecture for medical image segmentation}, in: \bibinfo{booktitle}{In Proceedings of the IEEE International Symposium on Multimedia (ISM)}, pp. \bibinfo{pages}{225--2255}.
\bibitem[{Jha et~al.(2020b)}]{jha2020kvasir}
\bibinfo{author}{Jha, D.}, et~al., \bibinfo{year}{2020}b.
\newblock \bibinfo{title}{Kvasir-seg: A segmented polyp dataset}, in: \bibinfo{booktitle}{In Proceedings of the Int. Conf. Multim. Model. (MMM)}, pp. \bibinfo{pages}{451--462}.
\bibitem[{Ji et~al.(2020)Ji, Fan, Zhou, Chen, Fu and Shao}]{ji2020automatic}
\bibinfo{author}{Ji, G.P.}, \bibinfo{author}{Fan, D.P.}, \bibinfo{author}{Zhou, T.}, \bibinfo{author}{Chen, G.}, \bibinfo{author}{Fu, H.}, \bibinfo{author}{Shao, L.}, \bibinfo{year}{2020}.
\newblock \bibinfo{title}{Automatic polyp segmentation via parallel reverse attention network.}, in: \bibinfo{booktitle}{In Proceedings of the MediaEval}, pp. \bibinfo{pages}{1--3}.
\bibitem[{Kang and Gwak(2020)}]{kang2020kd}
\bibinfo{author}{Kang, J.}, \bibinfo{author}{Gwak, J.}, \bibinfo{year}{2020}.
\newblock \bibinfo{title}{Kd-resunet++: Automatic polyp segmentation via self-knowledge distillation.}, in: \bibinfo{booktitle}{In Proceedings of the MediaEval}, pp. \bibinfo{pages}{1--3}.
\bibitem[{Keprate and Pandey(2021)}]{Keprate2021}
\bibinfo{author}{Keprate, A.}, \bibinfo{author}{Pandey, S.}, \bibinfo{year}{2021}.
\newblock \bibinfo{title}{Kvasir-instruments and polyp segmentation using {UNet}}.
\newblock \bibinfo{journal}{Nordic Machine Intelligence} \bibinfo{volume}{1}, \bibinfo{pages}{26--28}.
\bibitem[{Khadka(2020)}]{khadka2020transfer}
\bibinfo{author}{Khadka, R.}, \bibinfo{year}{2020}.
\newblock \bibinfo{title}{Transfer of knowledge: Fine-tuning for polyp segmentation with attention.}, in: \bibinfo{booktitle}{In Proceedings of the MediaEval}, pp. \bibinfo{pages}{1--3}.
\bibitem[{Kolesnikov et~al.(2020)Kolesnikov, Beyer, Zhai, Puigcerver, Yung, Gelly and Houlsby}]{kolesnikov2020big}
\bibinfo{author}{Kolesnikov, A.}, \bibinfo{author}{Beyer, L.}, \bibinfo{author}{Zhai, X.}, \bibinfo{author}{Puigcerver, J.}, \bibinfo{author}{Yung, J.}, \bibinfo{author}{Gelly, S.}, \bibinfo{author}{Houlsby, N.}, \bibinfo{year}{2020}.
\newblock \bibinfo{title}{Big transfer (bit): General visual representation learning}, in: \bibinfo{booktitle}{Computer Vision--ECCV 2020: 16th European Conference, Glasgow, UK, August 23--28, 2020, Proceedings, Part V 16}, pp. \bibinfo{pages}{491--507}.
\bibitem[{Krenzer and Puppe(2020)}]{krenzer2020bigger}
\bibinfo{author}{Krenzer, A.}, \bibinfo{author}{Puppe, F.}, \bibinfo{year}{2020}.
\newblock \bibinfo{title}{Bigger networks are not always better: Deep convolutional neural networks for automated polyp segmentation.}, in: \bibinfo{booktitle}{In Proceedings of the MediaEval}, pp. \bibinfo{pages}{1--3}.
\bibitem[{LeCun et~al.(2015)LeCun, Bengio and Hinton}]{article}
\bibinfo{author}{LeCun, Y.}, \bibinfo{author}{Bengio, Y.}, \bibinfo{author}{Hinton, G.}, \bibinfo{year}{2015}.
\newblock \bibinfo{title}{Deep learning} \bibinfo{volume}{521}, \bibinfo{pages}{436--44}.
\bibitem[{Levin et~al.(2008)}]{levin2008screening}
\bibinfo{author}{Levin, B.}, et~al., \bibinfo{year}{2008}.
\newblock \bibinfo{title}{Screening and surveillance for the early detection of colorectal cancer and adenomatous polyps, 2008: a joint guideline from the american cancer society, the us multi-society task force on colorectal cancer, and the american college of radiology}.
\newblock \bibinfo{journal}{CA: a Cancer Journ. Clinici.} \bibinfo{volume}{58}, \bibinfo{pages}{130--160}.
\bibitem[{Lin et~al.(2017)Lin, Doll{\'a}r, Girshick, He, Hariharan and Belongie}]{lin2017feature}
\bibinfo{author}{Lin, T.Y.}, \bibinfo{author}{Doll{\'a}r, P.}, \bibinfo{author}{Girshick, R.}, \bibinfo{author}{He, K.}, \bibinfo{author}{Hariharan, B.}, \bibinfo{author}{Belongie, S.}, \bibinfo{year}{2017}.
\newblock \bibinfo{title}{Feature pyramid networks for object detection}, in: \bibinfo{booktitle}{Proceedings of the IEEE conference on computer vision and pattern recognition}, pp. \bibinfo{pages}{2117--2125}.
\bibitem[{Lundberg and Lee(2017)}]{lundberg2017unified}
\bibinfo{author}{Lundberg, S.M.}, \bibinfo{author}{Lee, S.I.}, \bibinfo{year}{2017}.
\newblock \bibinfo{title}{A unified approach to interpreting model predictions}.
\newblock \bibinfo{journal}{Advances in neural information processing systems} \bibinfo{volume}{30}.
\bibitem[{Mahmud et~al.(2021)Mahmud, Paul and Fattah}]{mahmud2021polypsegnet}
\bibinfo{author}{Mahmud, T.}, \bibinfo{author}{Paul, B.}, \bibinfo{author}{Fattah, S.A.}, \bibinfo{year}{2021}.
\newblock \bibinfo{title}{Polypsegnet: A modified encoder-decoder architecture for automated polyp segmentation from colonoscopy images}.
\newblock \bibinfo{journal}{Computers in Biology and Medicine} \bibinfo{volume}{128}, \bibinfo{pages}{104119}.
\bibitem[{Maxwell~Hwang et~al.(2020)Maxwell~Hwang, Hwang, Xu and Wu}]{maxwell2020temporal}
\bibinfo{author}{Maxwell~Hwang, C.W.}, \bibinfo{author}{Hwang, K.S.}, \bibinfo{author}{Xu, Y.S.}, \bibinfo{author}{Wu, C.H.}, \bibinfo{year}{2020}.
\newblock \bibinfo{title}{A temporal-spatial attention model for medical image detection}, in: \bibinfo{booktitle}{In Proceedings of the MediaEval}, pp. \bibinfo{pages}{1--3}.
\bibitem[{Mehta and Sivaswamy(2017)}]{7950555}
\bibinfo{author}{Mehta, R.}, \bibinfo{author}{Sivaswamy, J.}, \bibinfo{year}{2017}.
\newblock \bibinfo{title}{M-net: A convolutional neural network for deep brain structure segmentation}, in: \bibinfo{booktitle}{2017 IEEE 14th International Symposium on Biomedical Imaging (ISBI 2017)}, pp. \bibinfo{pages}{437--440}.
\bibitem[{Mirza and Rajak(2021)}]{Mirza2021}
\bibinfo{author}{Mirza, A.}, \bibinfo{author}{Rajak, R.K.}, \bibinfo{year}{2021}.
\newblock \bibinfo{title}{Segmentation of polyp instruments using {UNet} based deep learning model}.
\newblock \bibinfo{journal}{Nordic Machine Intelligence} \bibinfo{volume}{1}, \bibinfo{pages}{44--46}.
\bibitem[{Moriyama et~al.(2015)Moriyama, Uraoka, Esaki and Matsumoto}]{moriyama2015advanced}
\bibinfo{author}{Moriyama, T.}, \bibinfo{author}{Uraoka, T.}, \bibinfo{author}{Esaki, M.}, \bibinfo{author}{Matsumoto, T.}, \bibinfo{year}{2015}.
\newblock \bibinfo{title}{Advanced technology for the improvement of adenoma and polyp detection during colonoscopy}.
\newblock \bibinfo{journal}{Digestive Endoscopy} \bibinfo{volume}{27}, \bibinfo{pages}{40--44}.
\bibitem[{Nathan and Ramamoorthy(2020)}]{nathan2020efficient}
\bibinfo{author}{Nathan, S.}, \bibinfo{author}{Ramamoorthy, S.}, \bibinfo{year}{2020}.
\newblock \bibinfo{title}{Efficient supervision net: Polyp segmentation using efficientnet and attention unit.}, in: \bibinfo{booktitle}{In Proceedings of the MediaEval}, pp. \bibinfo{pages}{1--3}.
\bibitem[{Nguyen et~al.(2020)Nguyen, Nguyen, Diep, Le, Nguyen-Dinh, Nguyen and Tran}]{nguyen2020hcmus}
\bibinfo{author}{Nguyen, T.P.}, \bibinfo{author}{Nguyen, T.C.}, \bibinfo{author}{Diep, G.H.}, \bibinfo{author}{Le, M.Q.}, \bibinfo{author}{Nguyen-Dinh, H.P.}, \bibinfo{author}{Nguyen, H.D.}, \bibinfo{author}{Tran, M.T.}, \bibinfo{year}{2020}.
\newblock \bibinfo{title}{Hcmus at medico automatic polyp segmentation task 2020: Pranet and resunet++ for polyps segmentation.}, in: \bibinfo{booktitle}{In Proceedings of the MediaEval}, pp. \bibinfo{pages}{1--3}.
\bibitem[{Oktay et~al.(2018)Oktay, Schlemper, Folgoc, Lee, Heinrich, Misawa, Mori, McDonagh, Hammerla, Kainz et~al.}]{oktay2018attention}
\bibinfo{author}{Oktay, O.}, \bibinfo{author}{Schlemper, J.}, \bibinfo{author}{Folgoc, L.L.}, \bibinfo{author}{Lee, M.}, \bibinfo{author}{Heinrich, M.}, \bibinfo{author}{Misawa, K.}, \bibinfo{author}{Mori, K.}, \bibinfo{author}{McDonagh, S.}, \bibinfo{author}{Hammerla, N.Y.}, \bibinfo{author}{Kainz, B.}, et~al., \bibinfo{year}{2018}.
\newblock \bibinfo{title}{Attention u-net: Learning where to look for the pancreas}.
\newblock \bibinfo{journal}{arXiv preprint arXiv:1804.03999} .
\bibitem[{Poudel and Lee(2020)}]{poudel2020automatic}
\bibinfo{author}{Poudel, S.}, \bibinfo{author}{Lee, S.W.}, \bibinfo{year}{2020}.
\newblock \bibinfo{title}{Automatic polyp segmentation using channel-spatial attention with deep supervision.}, in: \bibinfo{booktitle}{In Proceedings of the MediaEval}, pp. \bibinfo{pages}{1--3}.
\bibitem[{Poudel and Lee(2021)}]{Poudel2021}
\bibinfo{author}{Poudel, S.}, \bibinfo{author}{Lee, S.W.}, \bibinfo{year}{2021}.
\newblock \bibinfo{title}{Explainable {U-Net} model {forMedical} image segmentation}.
\newblock \bibinfo{journal}{Nordic Machine Intelligence} \bibinfo{volume}{1}, \bibinfo{pages}{41--43}.
\bibitem[{Qin et~al.(2020)Qin, Zhang, Huang, Dehghan, Zaiane and Jagersand}]{qin2020u2}
\bibinfo{author}{Qin, X.}, \bibinfo{author}{Zhang, Z.}, \bibinfo{author}{Huang, C.}, \bibinfo{author}{Dehghan, M.}, \bibinfo{author}{Zaiane, O.R.}, \bibinfo{author}{Jagersand, M.}, \bibinfo{year}{2020}.
\newblock \bibinfo{title}{U2-net: Going deeper with nested u-structure for salient object detection}.
\newblock \bibinfo{journal}{Pattern recognition} \bibinfo{volume}{106}, \bibinfo{pages}{107404}.
\bibitem[{Rauniyar et~al.(2021)Rauniyar, Jha, Jha, Jha and Rauniyar}]{Rauniyar2021}
\bibinfo{author}{Rauniyar, S.}, \bibinfo{author}{Jha, V.K.}, \bibinfo{author}{Jha, R.K.}, \bibinfo{author}{Jha, D.}, \bibinfo{author}{Rauniyar, A.}, \bibinfo{year}{2021}.
\newblock \bibinfo{title}{Improving polyp segmentation in colonoscopy using deep learning}.
\newblock \bibinfo{journal}{Nordic Machine Intelligence} \bibinfo{volume}{1}, \bibinfo{pages}{35--37}.
\bibitem[{Riegler et~al.(2016)}]{riegler2016eir}
\bibinfo{author}{Riegler, M.}, et~al., \bibinfo{year}{2016}.
\newblock \bibinfo{title}{Eir—efficient computer aided diagnosis framework for gastrointestinal endoscopies}, in: \bibinfo{booktitle}{In Proceedings of the Inter. Worksh. Content-Based Multime. Index. (CBMI)}, pp. \bibinfo{pages}{1--6}.
\bibitem[{Ronneberger et~al.(2015)Ronneberger, Fischer and Brox}]{ronneberger2015u}
\bibinfo{author}{Ronneberger, O.}, \bibinfo{author}{Fischer, P.}, \bibinfo{author}{Brox, T.}, \bibinfo{year}{2015}.
\newblock \bibinfo{title}{U-net: Convolutional networks for biomedical image segmentation}, in: \bibinfo{booktitle}{In Proceedings of the Medical Image Computing and Computer-Assisted Intervention--MICCAI 2015: 18th International Conference, Munich, Germany, October 5-9, 2015, Proceedings, Part III 18}, pp. \bibinfo{pages}{234--241}.
\bibitem[{Shrestha et~al.(2020)Shrestha, Khanal and Ali}]{shrestha2020ensemble}
\bibinfo{author}{Shrestha, S.}, \bibinfo{author}{Khanal, B.}, \bibinfo{author}{Ali, S.}, \bibinfo{year}{2020}.
\newblock \bibinfo{title}{Ensemble u-net model for efficient polyp segmentation.}, in: \bibinfo{booktitle}{In Proceedings of the MediaEval}, pp. \bibinfo{pages}{1--3}.
\bibitem[{Siegel et~al.(2023)Siegel, Miller, Wagle and Jemal}]{siegel2023cancer}
\bibinfo{author}{Siegel, R.L.}, \bibinfo{author}{Miller, K.D.}, \bibinfo{author}{Wagle, N.S.}, \bibinfo{author}{Jemal, A.}, \bibinfo{year}{2023}.
\newblock \bibinfo{title}{Cancer statistics, 2023}.
\newblock \bibinfo{journal}{CA: a cancer journal for clinicians} \bibinfo{volume}{73}, \bibinfo{pages}{17--48}.
\bibitem[{Somani et~al.(2021)Somani, Singh, Prasad and Horsch}]{Somani2021}
\bibinfo{author}{Somani, A.}, \bibinfo{author}{Singh, D.}, \bibinfo{author}{Prasad, D.}, \bibinfo{author}{Horsch, A.}, \bibinfo{year}{2021}.
\newblock \bibinfo{title}{T-{MIS}: Transparency adaptation in medical image segmentation}.
\newblock \bibinfo{journal}{Nordic Machine Intelligence} \bibinfo{volume}{1}, \bibinfo{pages}{11--13}.
\bibitem[{Tan and Le(2019a)}]{tan2019efficientnet}
\bibinfo{author}{Tan, M.}, \bibinfo{author}{Le, Q.}, \bibinfo{year}{2019}a.
\newblock \bibinfo{title}{Efficientnet: Rethinking model scaling for convolutional neural networks}, in: \bibinfo{booktitle}{Proceedings of the International conference on machine learning}, pp. \bibinfo{pages}{6105--6114}.
\bibitem[{Tan and Le(2019b)}]{DBLP:journals/corr/abs-1905-11946}
\bibinfo{author}{Tan, M.}, \bibinfo{author}{Le, Q.V.}, \bibinfo{year}{2019}b.
\newblock \bibinfo{title}{Efficientnet: Rethinking model scaling for convolutional neural networks}.
\newblock \bibinfo{journal}{CoRR} \bibinfo{volume}{abs/1905.11946}.
\bibitem[{Thambawita et~al.(2020)Thambawita, Hicks, Halvorsen and Riegler}]{thambawita2020pyramid}
\bibinfo{author}{Thambawita, V.}, \bibinfo{author}{Hicks, S.}, \bibinfo{author}{Halvorsen, P.}, \bibinfo{author}{Riegler, M.A.}, \bibinfo{year}{2020}.
\newblock \bibinfo{title}{Pyramid-focus-augmentation: Medical image segmentation with step-wise focus}.
\newblock \bibinfo{journal}{arXiv preprint arXiv:2012.07430} .
\bibitem[{Tomar(2021)}]{tomar2021automatic}
\bibinfo{author}{Tomar, N.K.}, \bibinfo{year}{2021}.
\newblock \bibinfo{title}{Automatic polyp segmentation using fully convolutional neural network}.
\newblock \bibinfo{journal}{arXiv preprint arXiv:2101.04001} .
\bibitem[{Tomar et~al.(2021)Tomar, Jha, Ali, Johansen, Johansen, Riegler and Halvorsen}]{tomar2021ddanet}
\bibinfo{author}{Tomar, N.K.}, \bibinfo{author}{Jha, D.}, \bibinfo{author}{Ali, S.}, \bibinfo{author}{Johansen, H.D.}, \bibinfo{author}{Johansen, D.}, \bibinfo{author}{Riegler, M.A.}, \bibinfo{author}{Halvorsen, P.}, \bibinfo{year}{2021}.
\newblock \bibinfo{title}{Ddanet: Dual decoder attention network for automatic polyp segmentation}, in: \bibinfo{booktitle}{Pattern Recognition. ICPR International Workshops and Challenges: Virtual Event, January 10-15, 2021, Proceedings, Part VIII}, pp. \bibinfo{pages}{307--314}.
\bibitem[{Tomar et~al.(2022)Tomar, Jha, Riegler, Johansen, Johansen, Rittscher, Halvorsen and Ali}]{tomar2022fanet}
\bibinfo{author}{Tomar, N.K.}, \bibinfo{author}{Jha, D.}, \bibinfo{author}{Riegler, M.A.}, \bibinfo{author}{Johansen, H.D.}, \bibinfo{author}{Johansen, D.}, \bibinfo{author}{Rittscher, J.}, \bibinfo{author}{Halvorsen, P.}, \bibinfo{author}{Ali, S.}, \bibinfo{year}{2022}.
\newblock \bibinfo{title}{Fanet: A feedback attention network for improved biomedical image segmentation}.
\newblock \bibinfo{journal}{IEEE Transactions on Neural Networks and Learning Systems} .
\bibitem[{Trinh et~al.(2020)Trinh, Nguyen, Huynh and Tran}]{trinh2020hcmus}
\bibinfo{author}{Trinh, Q.H.}, \bibinfo{author}{Nguyen, M.V.}, \bibinfo{author}{Huynh, T.G.}, \bibinfo{author}{Tran, M.T.}, \bibinfo{year}{2020}.
\newblock \bibinfo{title}{Hcmus-juniors 2020 at medico task in mediaeval 2020: Refined deep neural network and u-net for polyps segmentation.}, in: \bibinfo{booktitle}{In Proceedings of the MediaEval}, pp. \bibinfo{pages}{1--3}.
\bibitem[{Tzavara and Singstad(2021)}]{Tzavara2021}
\bibinfo{author}{Tzavara, N.P.}, \bibinfo{author}{Singstad, B.J.}, \bibinfo{year}{2021}.
\newblock \bibinfo{title}{Transfer learning in polyp and endoscopic tool segmentation from colonoscopy images}.
\newblock \bibinfo{journal}{Nordic Machine Intelligence} \bibinfo{volume}{1}, \bibinfo{pages}{32--34}.
\bibitem[{Wang et~al.(2022)Wang, Xie, Li, Fan, Song, Liang, Lu, Luo and Shao}]{wang2022pvt}
\bibinfo{author}{Wang, W.}, \bibinfo{author}{Xie, E.}, \bibinfo{author}{Li, X.}, \bibinfo{author}{Fan, D.P.}, \bibinfo{author}{Song, K.}, \bibinfo{author}{Liang, D.}, \bibinfo{author}{Lu, T.}, \bibinfo{author}{Luo, P.}, \bibinfo{author}{Shao, L.}, \bibinfo{year}{2022}.
\newblock \bibinfo{title}{Pvt v2: Improved baselines with pyramid vision transformer}.
\newblock \bibinfo{journal}{Computational Visual Media} \bibinfo{volume}{8}, \bibinfo{pages}{415--424}.
\bibitem[{Yang et~al.(2024)Yang, Xing and Zhu}]{yang2024vivim}
\bibinfo{author}{Yang, Y.}, \bibinfo{author}{Xing, Z.}, \bibinfo{author}{Zhu, L.}, \bibinfo{year}{2024}.
\newblock \bibinfo{title}{Vivim: a video vision mamba for medical video object segmentation}.
\newblock \bibinfo{journal}{arXiv preprint arXiv:2401.14168} .
\bibitem[{Yeung(2021)}]{Yeung2021}
\bibinfo{author}{Yeung, M.}, \bibinfo{year}{2021}.
\newblock \bibinfo{title}{Attention {U-Net} ensemble for interpretable polyp and instrument segmentation}.
\newblock \bibinfo{journal}{Nordic Machine Intelligence} \bibinfo{volume}{1}, \bibinfo{pages}{47--49}.
\bibitem[{Zhang et~al.(2022)Zhang, Wu, Zhang, Zhu, Lin, Zhang, Sun, He, Mueller, Manmatha et~al.}]{zhang2022resnest}
\bibinfo{author}{Zhang, H.}, \bibinfo{author}{Wu, C.}, \bibinfo{author}{Zhang, Z.}, \bibinfo{author}{Zhu, Y.}, \bibinfo{author}{Lin, H.}, \bibinfo{author}{Zhang, Z.}, \bibinfo{author}{Sun, Y.}, \bibinfo{author}{He, T.}, \bibinfo{author}{Mueller, J.}, \bibinfo{author}{Manmatha, R.}, et~al., \bibinfo{year}{2022}.
\newblock \bibinfo{title}{Resnest: Split-attention networks}, in: \bibinfo{booktitle}{Proceedings of the IEEE/CVF conference on computer vision and pattern recognition}, pp. \bibinfo{pages}{2736--2746}.
\bibitem[{Zhang et~al.(2021)Zhang, Liu and Hu}]{zhang2021transfuse}
\bibinfo{author}{Zhang, Y.}, \bibinfo{author}{Liu, H.}, \bibinfo{author}{Hu, Q.}, \bibinfo{year}{2021}.
\newblock \bibinfo{title}{Transfuse: Fusing transformers and cnns for medical image segmentation}, in: \bibinfo{booktitle}{Medical Image Computing and Computer Assisted Intervention--MICCAI 2021: 24th International Conference, Strasbourg, France, September 27--October 1, 2021, Proceedings, Part I 24}, pp. \bibinfo{pages}{14--24}.

\end{thebibliography}
\end{document}